\documentclass[12pt]{article}
\usepackage{mathptmx}       
\usepackage{helvet}         
\usepackage{courier}        
\usepackage{type1cm}        
\usepackage{makeidx}         
\usepackage{multicol}        
\usepackage[bottom]{footmisc}

\usepackage{times}
\usepackage{mathptm}
\usepackage{amsfonts}
\usepackage{amsmath}
\usepackage{theorem}
\usepackage{pifont}
\usepackage{mathrsfs}
\usepackage[dvips]{color}
\usepackage{graphics}
\usepackage{epsfig}

\usepackage{leqno}
\usepackage{graphicx}
\usepackage{caption}
\usepackage{subcaption}

\newcommand{\no}{\nonumber}

\newtheorem{definition}{Definition}[section]

\title{Reynolds number dependence of the structure functions in homogeneous turbulence
}
\author{John Kaminsky$^1$, Bj\"orn Birnir$^{1,3}$, Gregory P. Bewley$^2$, \\
and Michael Sinhuber$^4$\\
CNLS
and Department of Mathematics\\
UC Santa Barbara$^1$,
University of Iceland$^3$,\\
Department of Mechanical Engineering\\
Cornell University$^2$, and\\
Department of Civil and Enviromental Engineering\\
Stanford University$^4$.  
}

\date{}

\begin{document}

\maketitle
\begin{abstract}
We compare the predictions of stochastic closure theory (SCT) \cite{BB211} 
with experimental measurements of homogeneous turbulence 
made in the Variable Density Turbulence Tunnel (VDTT) \cite{BBNSX14} 
at the Max Planck Institute for Dynamics and Self-Organization in G\"ottingen.  
While the general form of SCT contains infinitely many free parameters, 
the data permit us to reduce the number to seven, only three of which are active over the entire inertial range.  
Of these three, one parameter characterizes the variance of the mean field noise in SCT and another characterizes the rate in the large deviations of the mean.  
The third parameter is the decay exponent of the Fourier variables in the Fourier expansion of the noise, 
which characterizes the smoothness of the turbulent velocity.  

SCT compares favorably with velocity structure functions measured in the experiment.  
We considered even-order structure functions ranging in order from two to eight 
as well as the third-order structure functions 
at five Taylor-Reynolds numbers ($R_\lambda$) between 110 and 1450.  
The comparisons highlight several advantages of the SCT, 
which include explicit predictions for the structure functions 
at any scale and for any Reynolds number.  
We observed that finite-$R_\lambda$ corrections, for instance, 
are important even at the highest Reynolds numbers produced in the experiments.  

SCT gives us the correct basis function to express all the moments of the velocity differences in turbulence in Fourier space. These turn out to be powers of the sine function indexed by the wavenumbers.  Here, the power of the sine function is the same as the order of the moment of the velocity differences (structure functions). The SCT produces the coefficients of the series and so determines the statistical quantities that characterize the small scales in turbulence. It also characterizes the random force acting on the fluid in the stochastic Navier-Stokes equation, as described in the paper. 
\end{abstract}

\section{Introduction}
\indent Let us begin with a brief history of wind tunnel research 
and of the effort to describe the structure of turbulence statistically.  
In Aeronautics, the design of airfoils and airplanes was a major challenge.  
The development of appropriate laboratory experiments facilitates progress to this day, 
including the invention of the wind tunnel.  
The first wind tunnel is credited to F. Wenham in Great Britain in $1871$.  
The Wright brothers also constructed their own wind tunnel in 1901 \cite{BC81}, 
but it was Ludwig Prandtl in $1917$ who designed the first ``modern'' wind tunnel.  
This 1917 tunnel was actually his second design, 
his first design in 1909 being a closed-loop wind tunnel which, by his own admission, was "of a temporary nature" \cite{Prandtl20}.  
Nonetheless, his second, more permanent design would become the model for many subsequent wind tunnels \cite{An99}.  
Prandtl's student, Max Munk, went on to design the first wind tunnel that allowed adjustment of the density of the working fluid \cite{BBNSX14}, 
and so for much higher Reynolds number flows in the tunnel.  
This tunnel was built at the Langley Research Center in Virgina in $1923$.  
Most of the early research done with wind tunnels was devoted to the study of airfoils and airplane shapes and Mach number \cite{BBNSX14}.  One interesting feature of these tunnels was the ability to adjust Mach number and Reynolds number independently.

Wind tunnels are essential tools to study not only airfoils and model airplanes, 
but also to study statistically homogeneous and isotropic turbulence, see Taylor \cite{Ta35}.  
Such flows limit turbulence to its essential ingredients: inertia, pressure and friction, 
minimize the effects of the boundaries on the flow, and do not exhibit a preferred orientation.  
It can be created by mechanically stirring a liquid or gas \cite{BBNSX14}.  
A close approximation of such flows are realized in a wind tunnel when 
a uniform free-stream flow is disturbed by a mesh or a grid, see \cite{Co61,CBC66}.  

Experiments to study turbulence were rare until the second half of the twentieth century.  
The $1940$s featured experiments on grid turbulence in California \cite{MSB48}, while another series of experiments were performed at the Nuclear Research Lab in J\"ulich in the $1970$s \cite{HSY92}.  
More recently, wind tunnels were built at the German Aerospace Center in G\"ottingen and at the Princeton Gas Dynamics Lab 
with similar goals.  
The experiments in this paper were performed in the facility 
at the Max Planck Institute for Dynamics and Self-Organization in G\"ottingen 
called the Variable Density Turbulence Tunnel (VDTT), which was completed in $2009$ \cite{BBNSX14}.  
It has achieved turbulent flow up to Taylor-Reynolds number $1600$, 
which is the highest recorded for a passive grid experiment until that time, with higher Reynolds number since being recorded with an active grid\cite{SBB15,SBB17}.  
Details about the VDTT can be found in \cite{BBNSX14}.  
One of Prandtl's original wind tunnels sits beside the VDTT in G\"ottingen, see \cite{BBNSX14}.  

The mathematical theory of turbulence has its roots in the work of Kolmogorov.  
In $1941$, Kolmogorov published his celebrated four-fifths law and postulated, with Obukhov, 
that the structure functions of turbulence should scale with the lag variable, $r$, so that 
\begin{displaymath}
S_p(x,y,t) = E(\vert u(x,t)-u(y,t)\vert^p) = C_p\ r^{p\slash 3}, 
\end{displaymath}
where $p$ is the order of the structure function and $r=\vert x-y\vert$.  
Lev Landau criticized the theory for neither taking into account the organization of the flow on large scales 
nor the influence of intermittency, which is the development of long tails in the velocity difference distributions at large Reynolds numbers.  
In $1962$, Kolmogorov and Obukhov revised their theory to address these criticisms.  
They introduced a correction to the exponent, such that 
\begin{displaymath}
S_p(x,y,t) = C_p<\epsilon^\frac{p}{3}>r^\frac{p}{3} = C^\prime_pr^{\frac{p}{3}+\tau_p} = C^\prime_pr^{\zeta_p}, 
\end{displaymath}
where $\epsilon$ is the dissipation rate and $\zeta_p=\frac{p}{3}+\tau_p$, 
and $\tau_p$ is the correction that needs to be determined.  
A prediction for the correction was found by She and Leveque in $1994$: 
\begin{align}
\label{eq:intcorr}
\tau_p=-\frac{2p}{9}+2(1-(\frac{2}{3})^\frac{p}{3}),
\end{align}
see \cite{SL94}.  
In \cite{BB211}, the log-Possonian processes of Dubrulle \cite{Du94} and She and Waymire  \cite{SW95}, 
responsible for the intermittency corrections, were derived from the stochastic Navier-Stokes equation.  

Kolmogorov and Obukhov considered the velocity in turbulent flow to be a stochastic process 
and their hypotheses include that the N-point probability distribution function 
(PDF) of turbulence does not depend on $x$ or $y$ individually but only on $r$, $\nu$ and $\epsilon$, 
where $r$ is the lag variable, $\nu$ is the kinematic viscosity, and $\epsilon$ is the kinetic energy dissipation rate of the turbulence per unit mass \cite{Po00}.  
Moreover, when $r >> \eta$, where $\eta$ is the Kolmogorov (or dissipation) scale, 
then the PDF depends on $\epsilon$ and $r$ alone, and not on $\nu$.  
Since the 2-point PDF determines the structure functions described above, 
the same statements that apply to the PDFs apply also to the structure functions.  

If the turbulent velocity is a stochastic process it must satisfy a stochastic Navier-Stokes equation.  
Such an equation was formulated by Landau and Lifschitz in their Fluid Dynamics book \cite{LL59} in 1959.  
They considered the noise in the stochastic Navier-Stokes equation to be the fluctuations in the velocity, 
which cannot be ignored in turbulence.  They argued that it should be white both in time and space, but this assumption cannot be true since the Navier-Stokes equation driven by noise that is white in space 
produces velocities the are not continuous \cite{Wa84}, and this is not observed in nature.  
Birnir \cite{BB211} argued that the noise has enough smoothness in space 
that the dissipation rate, $\epsilon$, is finite, 
and that the noise is of a generic nature that includes an additive term corresponding to a mean-field noise 
and another additive term corresponding to the large deviations of the mean-field.  
He also added a multiplicative noise term, modeling jumps in the gradient for the flow velocity, 
and showed that this term produced the log-Poisson process of Dubrulle, She and Waymire and their intermittency corrections $\tau_p$.  
These assumptions are the basis of the Stochastic Closure Theory (SCT) \cite{BB312} and are elaborated in the next section.  


\section{The Assumptions of SCT and Its Predictions}

The following assumptions produce the Stochastic Navier-Stokes equation (\ref{eq:sns}) given in the next section.  
The detailed arguments leading to the form of the noise are given in \cite{BB211} and \cite{BB312}.  
They follow the spirit of the argument in Landau and Lifschitz \cite{LL59}.  
We also list the predictions of the theory, 
which include a quantitative prediction 
of the She-Leveque intermittency corrections to the Kolmogorov-Obukhov '62 theory of turbulence.  \\

\noindent{\bf SCT Assumptions:}
\begin{enumerate}
\item 
The small scale flow in fully developed turbulence satisfies a stochastic Navier-Stokes (SNS) equation.
\item 
The noise in the SNS consists of both an additive and a multiplicative term. 
\item
The additive noise is in part a general mean field noise that is sufficiently smooth in space for the dissipation rate, 
\[
\epsilon = \nu\int_\Omega |\nabla u|^2 dx < \infty, 
\]
to be finite.  
In addition to this ``infinite-dimensional Brownian'' mean-field noise, 
there is a deterministic additive part that captures large-deviations in the mean-field. 
\item
The multiplicative noise models jumps in the velocity gradient, $\nabla u$.  This term is then multiplied by the velocity $u$.  
\item 
The most singular (having least spatial smoothness) structures in (3-d) turbulence are one-dimensional vortex lines. 
\end{enumerate}

\noindent{\bf SCT Predictions:}
\begin{enumerate}
\item
The structure functions of turbulence at finite Reynolds numbers are given by formulas that explicitly incorporate the Reynolds number dependence of the structure functions.  
\item
The N-point probability density function exists and can be computed.  
In the two-point case it is determined by the Kolmogorov-Hopf functional 
differential equation \cite{BB312} and has an explicit formula \cite{BB16}.  
\item
The PDF for the velocity distribution in turbulence is a Generalized-Hyperbolic distribution \cite{BN77} 
convolved with the Poisson distribution of the log-Poisson processes of Dubrulle, She and Waymire \cite{BB16}.  For large values of the lag variable and for the fluid velocities themselves, these distributions become (skewed and flat) Gaussians.
\end{enumerate}

The most important SCT prediction for this paper is (1), 
the explicit formulae for the structure functions with given Reynolds number dependencies.  
We use these formulas to fit the data measured in the VDTT, and this is the subject of the paper.  
A disadvantage of equation is that the noise has infinitely many undetermined coefficients $c_k, d_k$ and $h_k$.  
The last coefficients are a consequence of assumption 3 above.  
The vorticity lines are one-dimensional, and this implies that all the coefficients, $h_k$, are fixed \cite{BB312}.  
However, we are still left with infinitely many coefficients $c_k$ and $d_k$.  
What we find through comparison with the experimental data 
is that we can reduce the number of coefficients to only three.  
When the mean flow is given, we are left with one parameter that characterizes the infinite-dimensional Brownian, 
another parameter characterizing large deviations from the mean, and one exponent characterizing spatial smoothness.  
We find, as expected, that the mean flow and the three parameters depend on the Taylor-Reynolds number.  
They do not, however, depend on the order of the structure function.  
The upshot is a much improved stochastic closure model (\ref{eq:isns}) with only three parameters characterizing the noise.  

One can say that the SCT produces the correct basis, with basis functions that are functions of the lag variable indexed by the wavenumber, to represent all the statistical quantities of the velocity differences. This is a big improvement over previous theoretical result that only produce one or finitely many statistical quantities measurable in experiments.


\section{The Stochastic Closure Model}

In this section, we give a short derivation of the SCT model. We first explain how the form 
of the turbulent noise forcing, in the Navier-Stokes equation, is derived and then use some techniques from probability theory to transform the resulting stochastic Navier-Stokes equation, for the small scale velocity, to an integral equation.
The integral equation will be used in the next section to compute a sharp lower estimate for the structure functions. 
The reader is directed to \cite{BB312} for more details. 

The flow in the wind tunnel is governed by the Navier Stokes equation:
\begin{align}
\label{eq:NS}
	u_t+ (u\cdot\nabla) u&=\nu\Delta u-\nabla p,\no\\
	\mathrm{div}\, u & =0,\\
	u(x,0)& =u_0(x),\no
	\end{align}
where $u(x)$ is the fluid velocity, $x\in\mathbb{R}^3$, $p$ is pressure, and $\nu$ is the viscosity.  We also impose periodic boundary conditions upon the flow.  The second line in (\ref{eq:NS}) is the incompressibility condition.  Using this equation, we can eliminate the pressure to get 
\begin{align}
\label{eq:pressure}
	u_t+u\cdot\nabla u=\nu\Delta u+ \nabla(\Delta^{-1}[\mathrm{Trace}(\nabla u)]^2).
	\end{align}
This equation defines the evolution of the  velocity of the fluid in time.  We will impose periodic boundary conditions on the small scales below. 


Following the classical Reynolds decomposition \cite{Re85}, we decompose the velocity into mean flow $U$ and the fluctuations $u$. 
Then the velocity is written as $U+u$, where
$U$ describes the mean, or large scale flow and $u$ describes the velocity fluctuations.  These two terms describe the large scales and small scales of the flow, respectively.   If we also decompose the pressure
into mean pressure $P$ and the fluctuations $p$, then the equation for the large scale flow can be written as
\begin{equation}
\label{eq:RANS}
U_t + U\cdot \nabla U =\nu \Delta U -\nabla P -\nabla \cdot (\overline{u\otimes u}),
\end{equation}
where in coordinates 
\begin{displaymath}
\nabla \cdot (\overline{u\otimes u}) = \frac{\partial \overline{u_i u_j}}{\partial x_j},
\end{displaymath} 
that is $\nabla$ is dotted with the rows of $\overline{u_i u_j}$, and $R_{ij} = \overline{u\otimes u}$ is the Reynolds stress, see \cite{BW02}. The Reynolds stress has the interpretation of a turbulent momentum flux and the last term in (\ref{eq:RANS}) is also know as the eddy viscosity.  It describes how the small scales influence the large scales. In addition, from linearity, we get divergence free conditions for $U$, and $u$
\[
\nabla \cdot U = 0, \qquad \nabla \cdot u = 0.
\]
Together, (\ref{eq:RANS}) and the divergence free condition on $U$ give the Reynolds Averaged Navier-Stokes (RANS) that forms the basis for most contemporary simulations of turbulent flow. The large scale equation (\ref{eq:RANS}) is satisfied by the mean flow $U=$ constant in the measurement region of the VDTT.
Thus in our case (\ref{eq:RANS}) reduces to the pressure gradient balancing the eddy viscosity.

%
%


Finding a constitutive law for the Reynolds stress $\overline{u\otimes u}$ is the famous closure problem in turbulence and 
we will solve that by writing down a stochastic equation for the small scale velocity $u$. This was first done by Landau and Lifschitz in \cite{LL59}.

The consequence of the SCT hypothesis is that the fluctuating velocity $u$ in turbulence is a stochastic process that is determined by a stochastic partial differential equation (SPDE). It will be the  Navier-Stokes equation for the fluctuations driven by noise, see below. This is the point of view taken by Kolmogorov in \cite{K41a,K41b,K62}, but the question we have to answer is: what is the form of the noise? There is a wide array of literature on this question, trying to trace the form of the noise back to the fluid instabilities, but these attempts have proven to be unsuccessful. Any memory of the fluid instabilities is quickly forgotten in fully-developed turbulence and the noise seems to be of a general form. Thus it makes sense to try to put generic noise into the Navier-Stokes equations and see how the Navier-Stokes evolution colors generic noise. Below we will answer what generic noise in the Navier-Stokes equation must look like, see \cite{BB312} for more details. 

For fully developed turbulence, we close the model with a stochastic forcing term to account for the small scales in (\ref{eq:NS}) and (\ref{eq:pressure}).  This noise term models the dissipation in the flow.  We impose periodic boundary conditions and then discretize on the torus.  Let $p_k$ denote the dissipation process in the $j$-th box.  We assume these dissipation processes in the flow are weakly coupled and have mean $m$.  Thus, the average is given by
\begin{displaymath}
M_n=\frac{1}{n}\sum_{j=1}^n p_j.
\end{displaymath}
We now make use of the Central Limit Theorem, which implies $M_n$ will converge to a Gaussian distribution with mean zero and variance one.  For the statement and proof of the Central Limit Theorem, see page $194$ in \cite{GS01}. Then, define
\begin{displaymath}
x_t^n=\frac{S_{[tn]}-nm}{\sqrt{n}\sigma},
\end{displaymath}
where $S_n=\sum_{j=1}^n p_j$ and $[tn]$ denotes integer value.  We now apply the Functional Central Limit Theorem, as given by Theorem $8.1$ in \cite{BW90}, and so the processes $x_t^n$ must converge in distribution to a Brownian motion $b_t$ as $n\rightarrow\infty$.  This must occur in the direction of any Fourier component and so we get
\begin{displaymath}
\bar{D}=\sum_{k\neq0} c_k^\frac{1}{2}db_t^ke_k(x),
\end{displaymath}
where $e_k(x)=e^{2\pi ikx}$ are distinct Fourier compoents complete with its own Brownian motion $b_t^k$, and $c_k^\frac{1}{2}$ are coefficients that converge sufficiently fast to ensure convergence of the entire series, see \cite{BB312}.\\
\indent However, we also must measure the fluctuations in the dissipation which can be explained via the Large Deviation Principle.  To apply the Large Deviation Principle, we need to describe the rate function associated with the process, which depends on whether the fluctuations are random.  If they are, the fluctuations can be modeled by a Poisson process with rate $\lambda$ and furthermore, if there is bias in the fluctuations, then the deviations of $M_n$ are bounded above by a constant determining the direction of the bias times the rate $\eta$.  Cramer's Theorem, see \cite{BB312}, then gives that the rate function is bounded by $\eta=\lambda$, and so the second additive noise terms is
\begin{displaymath}
D^\prime=\sum_{k\neq0} d_k\eta_k dte_k(x).
\end{displaymath}
Here, $e_k(x)$ is defined as above, $d_k$ is defined similarly to $c_k^{0.5}$, and $\eta_k$ are the rates in the $k$-th direction.  We choose $\eta_k=\vert k\vert^\frac{1}{3}$ to line up with the scaling of the Central Limit Theorem term.  Thus, the Large Deviation Principle gives the term
\begin{displaymath}
D^\prime=\sum_{k\neq0} d_k\vert k\vert^\frac{1}{3} dte_k(x).
\end{displaymath}
These two terms defined the additive noise forcing term.  A more detailed description of these terms is given in \cite{BB312}.\\  
\indent A final forcing term comes from the multiplicative noise.  This noise models jumps in the velocity gradient or vorticity concentrations.  If we let $N_t^k$ denote the number of velocity jumps associated to the $k$-th wave number that have occured by time $t$.  This in turn implies that the differential
\begin{displaymath}
dN^k(t)=N^k(t+dt)-N^k(t)
\end{displaymath}
denotes the number of jumps in the time interval $(t,t+dt]$.  The multiplicative noise then has the form
\begin{displaymath}
J=\sum_{k\neq0}\int_{\mathbb{R}}h_k(t,z)\bar{N}^k(dt,dz),
\end{displaymath}
where $h_k$ measures the size of the jump and $\bar{N}^k$ is the compensated number of jumps.  For more information on the multiplicative noise, see \cite{BB312}.\\
\indent Thus, adding the terms $\bar D$, $D^\prime$ and, $J$ multiplied by $u$, to the Navier-Stokes equation, we get a stochastic PDE describing the fully developed turbulent small-scale flow in the wind tunnel:
\begin{align}
\label{eq:sns}
du+u\cdot\nabla u dt&=[\nu\Delta u+ \nabla(\Delta^{-1}[\mathrm{Trace}(\nabla u)])-u\cdot \nabla U-U\cdot \nabla u]dt\no\\&+\sum_{k\neq0} d_k\vert k\vert^\frac{1}{3} dte_k(x)+\sum_{k\neq0} c_k^\frac{1}{2}db_t^ke_k(x)\\
&+u\sum_{k\neq0}\int_{\mathbb{R}}h_k(t,z)\bar{N}^k(dt,dz).\no
\end{align}
 We drop the term $-u\cdot \nabla U$, in the equation above, since the mean flow $U$ is constant for homogeneous turbulence and approximately constant in the wind tunnel, see \cite{BBNSX14}.  
An application of Girsanov's Theorem allows us to eliminate the $(-u\cdot\nabla u-U\cdot\nabla u)dt$ term at the cost of adding an exponential martingale,
\begin{displaymath}
M_t=\mathrm{exp}(-\int (U+u(B_s,s))\cdot dB_s-\frac{1}{2}\int_0^t \vert U+u(B_s,s)\vert^2 ds),
\end{displaymath} 
where $B_t \in \mathbb{R}^3$ is an auxiliary Brownian motion,
to each term in the Navier Stokes equation:
\begin{align}
\label{eq:sns}
du&=[\nu\Delta u+ \nabla(\Delta^{-1}[\mathrm{Trace}(\nabla u)])]M_tdt\no\\&+\sum_{k\neq0} d_k\vert k\vert^\frac{1}{3}M_t dte_k(x)+\sum_{k\neq0} c_k^\frac{1}{2}M_tdb_t^ke_k(x)\\
&+u\sum_{k\neq0}\int_{\mathbb{R}}h_k(t,z)\bar{N}^kM_t(dt,dz).\no
\end{align}
For the statement and proof of Girsanov's Theorem, see pages $149-151$ of \cite{Ok98}.  The Feynman-Kac Formula allows us to eliminate the term 
\[
u \times \sum_{k\neq0}\int_{\mathbb{R}}h_k(t,z)\bar{N}^kM_t(dt,dz)
\]
at the cost of adding a log-Poisson process 
\begin{displaymath}
e^{\int_s^tdq}=\frac{1}{3}(\sum_{k\neq0}^m\{\int_0^t\int_\mathbb{R}\ln(1+h_k)\bar{N}^k(ds,dz)+\int_0^t\int_\mathbb{R}(\ln(1+h_k)-h_k)m_k(ds,dz)\})
\end{displaymath}
to each term in the Navier-Stokes equation.  For the statement and proof of Feynman-Kac Formula, see pages $128-129$ of \cite{Ok98}.  Thus, the new Navier Stokes equation becomes
\begin{align}
\label{eq:sns}
du&=[\nu\Delta u+ \nabla(\Delta^{-1}[\mathrm{Trace}(\nabla u)])]e^{\int_s^t dq}M_tdt\no\\&+\sum_{k\neq0} d_k\vert k\vert^\frac{1}{3}e^{\int_s^t dq}M_t dte_k(x)+\sum_{k\neq0} c_k^\frac{1}{2}e^{\int_s^t dq}M_tdb_t^ke_k(x)\\
\end{align}
Finally, we use the definition of mild (or Martingale) solutions of nonlinear stochastic partial differential equations (SPDE) in infinite-dimensional space:
\begin{definition} 
Consider the initial value SPDE problem
\begin{displaymath}
du=(Au+F(t,u))dt+G(t,u)dB_t, \,\,\,u(x,0)=u_0.
\end{displaymath}
A stochastic process $u(\omega,x,t)$ is a mild solution of this SPDE initial value problem (IVP) if
\begin{displaymath}
P(\int_0^t|u|^2_2(s)dt < \infty),\qquad {P}\:\: \mathrm{a.\ s.}
\end{displaymath}
and
\begin{displaymath}
u(t)=e^{At}u_0+\int_0^t e^{A(t-s)}F(s,u(s))ds+\int_0^t e^{A(t-s)}G(s,u(s))dB_s,\qquad {P}\:\: \mathrm{a.\ s.},
\end{displaymath}
where ${P}$ is the probability measure in the associated probability space $(\Omega, \mathcal{F}, { P})$. 
\end{definition}
For more information, see page $182$ in \cite{DZ14}. One can then state a theorem for the existence of unique mild local (in time) solutions, see page $186$ in \cite{DZ14}. Now, this theorem does not apply directly here, as the multiplicative noise concerns jumps and not only Brownian motion.  However, a slight alteration of the proof  gives local existence of solutions.
The mild solution of the stochastic Navier Stokes equation, governing fully developed turbulence, is given by
\begin{align}
\label{eq:NSintegral}
	u&=e^{K(t)}e^{\int_0^tdq}M_tu^0+\sum_{k\neq0}c_k^\frac{1}{2}\int_0^te^{K(t-s)}e^{\int_s^tdq}M_{t-s}db_s^ke_k(x)\\
	&+\sum_{k\neq0}d_k\int_0^te^{K(t-s)}e^{\int_s^tdq}M_{t-s}\vert k\vert^\frac{1}{3}dte_k(x),\no
	\end{align}
where $K$ is the operator
	
	\begin{align}
	K=\nu\Delta+\nabla\Delta^{-1}\mathrm{Trace}(\nabla u\nabla),
	\end{align}
$M_t$  is the above exponential martingale, $e_k(x)=e^{2\pi ikx}$ is a Fourier component complete with its own Brownian motion $b_t^k$, and the coefficients $c_k^\frac{1}{2}$ and $d_k$ decay fast enough so that the series converges, see \cite{BB312}, Chapter 1.  This is also the integral form of the stochastic Navier-Stokes equation (\ref{eq:sns}).

The integral equation (\ref{eq:NSintegral}) is equivalent to the stochastic Navier-Stokes initial value problem (\ref{eq:sns}) for the small scales. It will be our main tool in computing the structure functions.

\section{The Computation of the Structure Functions}
In this section, we describe the calculation of the structure functions of turbulence, which will be compared with the experimental data. 
Using the stochastic Navier-Stokes integral equation (\ref{eq:NSintegral}) from previous section, we have that
\begin{align}
&u(x,t)-u(y,t) =\\
&\sum_{k\neq0}\left [(c_k^\frac{1}{2}\int_0^t e^{K(t-s)}e^{\int_s^tdq}M_{t-s}db_s^k \right.\\
&+\left.d_k\int_0^t e^{K(t-s)}e^{\int_s^tdq}M_{t-s}\vert k\vert^\frac{1}{3}ds)(e_k(x)-e_k(y))\right],\no
\end{align}
where $u(x,t)$ and $u(y,t)$ are the flow velocities at two points $x$ and $y$ in the wind tunnel.  This permits us to describe the computation of the structure functions:
\begin{displaymath}
S_p(x-y,t)=E(\vert u(x,t)-u(y,t)\vert^p).
\end{displaymath}
First, we note that the expectation is actually a composition of three expectations, one for the Brownian motions in the Fourier series representation of the noise, denoted $E_b$, another for the log-Poisson process, denoted $E_p$, and the third $E_B$ for the auxiliary Brownian motion in the Martingale in the last section.  The log-Poisson expectation acts upon the term
\begin{displaymath}
e^{\int_s^tdq}=\mathrm{exp}\{\frac{\frac{2}{3} \mathrm{ln}\vert k\vert+N_k \mathrm{ln}(\frac{2}{3})}{3}\}=(\vert k\vert^\frac{2}{3} (\frac{2}{3})^{N_t^k})^\frac{1}{3},
\end{displaymath}
given by the Feynman-Kac formula, see \cite{BB312}.  Then, we get that
\begin{displaymath}
E_p([\vert k\vert^\frac{2}{3}(\frac{2}{3})^{N_t^k}]^\frac{p}{3})=\vert k\vert^{-(-\frac{2p}{9}+2(1-(\frac{2}{3})^\frac{p}{3}))},
\end{displaymath}
see \cite{BB312}.  Notice the exponent above is the She-Leveque intermittency correction (\ref{eq:intcorr}), denoted $\tau_p$.  Applying $E_p$ also eliminates all terms $(e_k(x)-e_k(y))(e_j(x)-e_j(y))$ for $k\neq j$.  Standard algebra and trigonometry gives
\begin{displaymath}
e_k(x)-e_k(y)=2e^{\pi ik(x+y)}\sin(\pi k\cdot(x-y)).
\end{displaymath}
Thus, we get that
\begin{align}
\label{eq:expect}
&E(\vert u(x,t)-u(y,t)\vert^p)= \\
&E(\big\vert\sum_{k\neq0}\left [(c_k^\frac{1}{2}\int_0^t e^{K(t-s)}e^{\int_s^tdq}M_{t-s}db_s^k\right.\\
&+ d_k\int_0^t \left. e^{K(t-s)}e^{\int_s^tdq}M_{t-s}\vert k\vert^\frac{1}{3}ds)(e_k(x)-e_k(y))\right]\big\vert^p)\no\\
&=E_b(\big\vert\sum_{k\neq0}\left [(c_k^\frac{1}{2}\int_0^t e^{K(t-s)}\vert k\vert^{-\tau_p}E_B(M_{t-s})db_s^k)\right.\\
&+ \left. d_k\int_0^t e^{K(t-s)}\vert k\vert^{-\tau_p}E_B(M_{t-s})\vert k\vert^\frac{1}{3}ds)\right]\no\times 2e^{\pi ki(x+y)}\sin(\pi k\cdot(x-y))\big\vert^p).\no
\end{align}
Now, we use the estimate for the action of the operator $K$ on the Fourier components, replacing it with $\lambda_k=C\vert k\vert^\frac{2}{3}+4\nu\pi^2\vert k\vert^2$, see \cite{BB312}. A lower estimate is $-C\vert k\vert^\frac{2}{3}+4\nu\pi^2\vert k\vert^2$. These estimates assume that the expectation of the norm of $u$ in the Sobolev space $H^{\frac{11}{6}^+}$ is finite, see Lemma 2.7 in \cite{BB312}. We will discuss this in more detail in a future paper. $M_t$ is the exponential martingale:
\begin{displaymath}
M_t=\mathrm{Exp}[\int (U+u)\cdot dB_s-\int \frac{\vert U+u\vert^2}{2} ds],
\end{displaymath}
where $B_t \in \mathbb{R}^3$ is an auxiliary Brownian motion and $U+u$ is the Reynolds decomposition of the flow.  A simple application of Ito's formula yields
\begin{displaymath}
M_t^p=1+\int_0^t (U+u) M_s\cdot dB_s+\frac{p(p-1)}{2}\int_0^t|U+u|^2 M_s^pds.
\end{displaymath}
Thus, we have
\begin{eqnarray*}
E_B[M_t^p]&=&1+\frac{p(p-1)}{2}\int_0^tE_B[|U+u|^2 M_s^p]ds \\
&\leq& 1+\frac{p(p-1)}{2}\int_0^t\sup_{x}|U+u|^2 E_B[M_s^{p}]ds.
\end{eqnarray*}
Thus by Gr\"onwall's inequality
\[
E_B[M_t^p] \leq e^{\frac{p(p-1)}{4}\int_0^t\sup_{x}(|U|^2+|u|^2)dt}. 
\]
The sup in $x$ frees the expectation from the expectation of the auxiliary Brownian motion, used to define the Martingale, and the 
exponent of the right hand side is easily estimated by the same methods as below 
\[
E_b[\sup_{x}(|U|^2+|u|^2)] \leq \sup_{x}|U|^2 + \frac{1}{C^2}\sum_{k\neq 0}\frac{(C/2)c_k+d_k^2}{|k|^{\zeta_2}} = \mathrm{Constant}.
\]
This implies that $E_B[M_t^p]$ only adds a constant, first to the exponents and then to the denominators, in the structure functions below and we will ignore it. 

Finally, we take the absolute value and expand the polynomial expression in (\ref{eq:expect}).  To ultimately compute the structure functions, we use Ito's Lemma
\begin{displaymath}
E[(\int_S^T f(t,w)dB_t)^2]=\int_S^T E[(f(t,w))^2] dt
\end{displaymath}
to turn any even power of the stochastic integral into a deterministic integral, which can then be solved for using standard calculus.  For odd powers, we use the fact that
\begin{displaymath}
E[\int_S^T f dB_t]=0
\end{displaymath}
to eliminate such terms.  We then find the first-order structure function is given by
\begin{align}
	E(\vert u(x,t)-u(y,t)\vert)&=S_1(x,y,t)\no\\
	 &=\frac{2}{C}\sum_{k\in\mathbb{Z}^3\backslash\{0\}}\frac{\vert d_k\vert(1-e^{-\lambda_kt})}{\vert k\vert^{\zeta_1}+\frac{4\pi^2\nu}{C}\vert k\vert^{\zeta_1+\frac{4}{3}}}\vert \sin(\pi k\cdot(x-y))\vert,
	\end{align}
where $|\cdot |$ denotes the vector norm in $\mathbb{R}^3$.
The second-order structure function is given by
\begin{align}
	E(\vert u(x,t)-u(y,t)\vert^2)&=S_2(x,y,t)\no\\
	&=\frac{4}{C^2}\sum_{k\,\in\,\mathbb{Z}^3}\Big[(\vert\sin^2(\pi k\cdot(x-y))\vert)\no\\
	&\Big\{\frac{\frac{C}{2}c_k(1-e^{-2\lambda_kt})}{\vert k\vert^{\zeta_2}+\frac{4\pi^2\nu}{C}\vert k\vert^{\zeta_2+\frac{4}{3}}}\\
	& \left.+\frac{|d_k|^2(1-e^{-\lambda_kt})}{\vert k\vert^{\zeta_2}+\frac{8\pi^2\nu}{C}\vert k\vert^{\zeta_2+\frac{4}{3}}+\frac{16\pi^4\nu^2}{C^2}\vert k\vert^{\zeta_2+\frac{8}{3}}}\Big\}\right]\no
	\end{align}
where $c_k = |c_k^{\frac{1}{2}}|^2$.  
The third-order structure function is given by
\begin{gather}
	E(\vert u(x,t)-u(y,t)\vert^3)=S_3(x,y,t)\no\\
	=\frac{8}{C^3}\sum_{k\in\mathbb{Z}^3}
	\Big[(\vert\sin^3(\pi k\cdot(x-y))\vert)\\
	\Big\{\frac{\frac{C}{2}c_k\vert d_k\vert(1-e^{-2\lambda_kt})(1-e^{-\lambda_kt})}{\vert k\vert^{\zeta_3}+\frac{8\pi^2\nu}{C}\vert k\vert^{\zeta_3+\frac{4}{3}}+\frac{16\pi^4\nu^2}{C^2}\vert k\vert^{\zeta_3+\frac{8}{3}}}\no\\
	+\frac{\vert d_k\vert^3(1-e^{-\lambda_kt})^3}{\vert k\vert^{\zeta_3}+\frac{12\pi^2\nu}{C}\vert k\vert^{\zeta_3+\frac{4}{3}}+\frac{48\pi^4\nu^2}{C^2} \vert k\vert^{\zeta_3+\frac{8}{3}}+\frac{64\pi^6\nu^3}{C^3}\vert k\vert^{\zeta_3+4}}\Big\}\Big]\no
	\end{gather}
The general $p$-th order structure function is given by
\begin{gather}
\label{eq:series}
S_p(x,y,t)=\frac{2^p}{C^p}\sum_{k\neq0} A_p\times \vert \sin^p[\pi k\cdot(x-y)]\vert,
\end{gather}
where
\begin{gather}
\label{eq:Ap}
A_p=\frac{2^\frac{p}{2}\Gamma(\frac{p+1}{2})\sigma_k^p{}_1F_1(-\frac{1}{2}p,\frac{1}{2},-\frac{1}{2}(\frac{M_k}{\sigma_k})^2)}{\vert k\vert ^{\zeta_p}+\frac{p_k\pi^2\nu}{C}\vert k\vert ^{\zeta_p+\frac{4}{3}}+\mathcal{O}(\nu^2)},
\end{gather}
where $\Gamma$ is the gamma function, ${}_1F_1$ is the hypergeometric function, $M_k=\vert d_k\vert(1-e^{-\lambda_kt})$,  $\sigma_k=\sqrt{(\frac{C}{2}c_k(1-e^{-2\lambda_kt}))}$, and $p_k$ is different for each denominator term in the series.  Note that the Reynolds number dependence is captured via the viscosity term $\nu$.  $C$ is a constant approximating the norm of the small-scale velocity of the flow. It will allowed to vary across structure functions to accommodate a relative change in the mean and the large deviations.

The equalities for the structure function above are really lower estimates because the action of $e^{Kt}$ on the $k$th Fourier component is estimated by $e^{-\lambda_k t}$, where $\lambda_k$ is an overestimate. We also get an overestimate of the structure functions by using the lower estimate $-C\vert k\vert^\frac{2}{3}+4\nu\pi^2\vert k\vert^2$ instead of $\lambda_k$. However, since the Kolmogorov-Obuhov cascade is a forward cascade in three dimensions, that is the energy flows from high to low wavenumbers, the lower estimate of the structure functions is sharp and we drop the inequality in favor of the equality. The other thing to notice is that we are not using the eigenfunctions of the operator $K$ and the $\lambda_k$s are not the eigenvalues of $K$. It turns out to be much simpler to work with the Fourier components instead of the eigenfunctions of $K$.

The above formulas clearly distinguish the stochastic closure theory (SCT) from previous theories on turbulence. SCT shows that the correct basis for the $p$th structure function, in Fourier space, is the collection of sine components of the lag variable, indexed by different wavenumbers and raised to the power $p$. The coefficients for this basis are given by the formula (\ref{eq:Ap}). This allows us to represent all structure functions as functions of the lag variable for all Reynolds numbers and all power $p$. Such a result has been unattainable until now. It permits a complete characterization of the experimental data for the structure functions in homogeneous turbulence.

\subsection{The One-dimensional Structure Functions}

We want to fit the structure functions (\ref{eq:series}) to the experimental data collected in the VDTT.  To do this we have to reduce the three-dimensional structure functions to one-dimensional ones. 
We will perform the reduction in this subsection. 

We consider structure functions where the measurements are taken at two distinct points along the length of the tunnel, in the direction of the mean velocity. These are called the longitudinal structure functions, $S_p(r,t)$, where $r = x-y$, is a vector along the main axis of the tunnel. One can also consider the transversal structure functions, $S_p(q,t)$, where $q=x-y$, is a vector in the radial direction of the tunnel, perpendicular to $r$. In homogeneous turbulence these two structure functions are not independent.  Their correlation matrix is given by \cite{Po00}:
\[
D_{ij}=E[(u_i(x,t)-u_i(y,t)) (u_j(x,t)-u_j(y,t))]=S_2(r,t)I +(S_2(r,t)-S_2(q,t))\frac{r_ir_j}{r^2},
\]
where $I$ is the identity matrix in $\mathbb{R}^3 \times \mathbb{R}^3$, and 
\[
S_2(q,t)= S_2(r,t)+ r \frac{\partial}{\partial r}S_2(r,t), 
\]
with $r=|r|$, $|\cdot|$ denoting the vector norm in $\mathbb{R}^3$. For $\eta << r$, $D_{ij}$ is expected to reduce to 
\[
D_{ij}=C_2 (\epsilon r)^{2/3}(\frac{4}{3} I - \frac{1}{3} \frac{r_i r_j}{r^2}).
\]
Thus in $\mathbb{R}^3$ the correlation matrix is determined by longitudinal structure function $S_2(r,t)$ alone and we will restrict our attention to the longitudinal structure functions. 

Consider the longitudinal third-order structure function given by the SCT,
\begin{gather}
S_3(r,t)
	=\frac{8}{C^3}\sum_{k\in\mathbb{Z}^3}
	\Big[(\vert\sin^3(\pi k\cdot r)\vert)
	\Big\{\frac{\frac{C}{2}c_k\vert d_k\vert(1-e^{-2\lambda_kt})(1-e^{-\lambda_kt})}{\vert k\vert^{\zeta_3}+\frac{8\pi^2\nu}{C}\vert k\vert^{\zeta_3+\frac{4}{3}}+\frac{16\pi^4\nu^2}{C^2}\vert k\vert^{\zeta_3+\frac{8}{3}}}\no\\
	+\frac{\vert d_k\vert^3(1-e^{-\lambda_kt})^3}{\vert k\vert^{\zeta_3}+\frac{12\pi^2\nu}{C}\vert k\vert^{\zeta_3+\frac{4}{3}}+\frac{48\pi^4\nu^2}{C^2} \vert k\vert^{\zeta_3+\frac{8}{3}}+\frac{64\pi^6\nu^3}{C^3}\vert k\vert^{\zeta_3+4}}\Big\}\Big],\no
\end{gather}
where $c_k= c_1+c_2+c_2$, $|d_k|= \sqrt{d_1^2+d_2^2+d_2^2}$ and $|k|=  \sqrt{k_1^2+k_2^2+k_2^2}$, and $r=x-y$. If we take $r=(r,0,0)$ to lie along the axis of the VDTT (cylinder), then $r\cdot k = (rk_1,0,0)$ and if we ignore $k_2$ and $k_3$ in the denominator of $S_2$, and take $t \to \infty$, we get the inequality
\begin{gather}
S_3(r,t)
	\leq \frac{8}{C^3}\sum_{k_1\neq 0}
	\Big[(\vert\sin^3(\pi k_1 r)\vert)
	\Big\{\frac{\frac{C}{2}{\tilde c}_{k_1}\vert {\tilde d}_{k_1}\vert}{\vert k_1\vert^{\zeta_3}+\frac{8\pi^2\nu}{C}\vert k_1\vert^{\zeta_3+\frac{4}{3}}+\frac{16\pi^4\nu^2}{C^2}\vert k_1\vert^{\zeta_3+\frac{8}{3}}}\no\\
	+\frac{\vert {\tilde d}_{k_1}\vert^3}{\vert k_1\vert^{\zeta_3}+\frac{12\pi^2\nu}{C}\vert k_1\vert^{\zeta_3+\frac{4}{3}}+\frac{48\pi^4\nu^2}{C^2} \vert k_1\vert^{\zeta_3+\frac{8}{3}}+\frac{64\pi^6\nu^3}{C^3}\vert k_1\vert^{\zeta_3+4}}\Big\}\Big],\no
\end{gather}
because $\zeta_3 =1$, where ${\tilde c}_{k_1}=\sum_{k_2\neq0}\sum_{k_3\neq0} c_{(k_1,k_2,k_3)}$,
$\vert{\tilde d}_{k_1}\vert=\sum_{k_2\neq0}\sum_{k_3\neq0} \vert d_{(k_1,k_2,k_3)}\vert$. We have used the convexity of the functions $f(x)=x^p,\ p \geq 1$ to take the sum into the powers, here $p = 1, 3$. This upper estimate, that is supposed to be close, reduces the three dimensional $S_3$ to the one dimensional one. 
The argument for all the structure functions $S_p,\ p \geq 3$, is similar but the argument does not hold for 
$p = 1,$ or $2$, because $\zeta_1=0.37$ and $\zeta_2 = 0.696$ in the She-Leveque model, so both are less than one. This means that the upper estimate does not hold for all $|k|$s, only the ones that are big enough so that the second term in the denominators of $S_1$ and $S_2$ dominates the first. We will use the upper estimate with this understanding. 
We will compare the one-dimensional structure function with experimental data and drop the subscript $1$ on 
$k_1$. Thus the general $p$-th one-dimensional longitudinal structure function, in the stationary state, is given by
\begin{gather}
\label{eq:series1}
S_p(r,\infty) \leq \frac{2^p}{C^p}\sum_{k\neq0} \frac{2^\frac{p}{2}\Gamma(\frac{p+1}{2})\sigma_k^p{}_1F_1(-\frac{1}{2}p,\frac{1}{2},-\frac{1}{2}(\frac{M_k}{\sigma_k})^2)}{\vert k\vert ^{\zeta_p}+\frac{p_k\pi^2\nu}{C}\vert k\vert ^{\zeta_p+\frac{4}{3}}+\mathcal{O}(\nu^2)} \vert \sin^p[\pi kr]\vert,
\end{gather}
where $\Gamma$ is the gamma function, ${}_1F_1$ is the hypergeometric function, $M_k=\vert {\tilde d}_k\vert$,  $\sigma_k^2=\frac{C}{2}{\tilde c}_k$, and $p_k$ is different for each denominator term in the series.  Note that the Taylor-Reynolds number dependence is captured via the viscosity term $\nu$, as the Taylor Reynolds number is given by $U\lambda\slash\nu$.  $C$ is a constant approximating the mean velocity fluctuation of the flow. The upper estimate is understood to hold for $p=1,2,$ when $|k|$ is large enough. 

We can think about the triple sum as an integral 
\begin{displaymath}
\sum_{k \in \mathbb{Z}^3\setminus\{0\}}c_k\sim \int_0^\infty \int_\omega   c_{k} d\omega |k|^2 d|k| = \int_0^\infty {\tilde c}_k dk \sim \sum_{k\neq0} {\tilde c}_{k},
\end{displaymath}
where $|k|$ is the radius of the three-vector $k$, and $d\omega$ is the angular part of the volume element. This means that 
\[
{\tilde c}_{k}=\int_\omega   c_{k} d\omega |k|^2,
\]
is the integral of $c_k$ over a sphere of radius $k$ is Fourier space, analogous to the 
energy shell in the Kolmogorov-Obukhov cascade. Thus, whereas $c_k \sim \frac{1}{k^{3+\epsilon}}$, where $k=|k|$, in order for the sum to converge, ${\tilde c}_k \sim \frac{1}{k^{1+\epsilon}}$. A similar argument applies to $\sum_{k \in \mathbb{Z}^3\setminus\{0\}}\vert d_k\vert$ For this reason we expect the exponent $m$ of $k$ below to satisfy $m>1$.
We will in fact make the ansatz,
\begin{align}
\label{eq:ans}
{\tilde c}_k=\sqrt{\frac{2}{\pi}}\frac{b}{b^2+k^m},\,\,{\tilde d}_k=\sqrt{\frac{2}{\pi}}\frac{a}{a^2+k^m},
\end{align}
where ${\tilde c}_k$ and ${\tilde d}_k$ are the one-dimensional versions of the coefficients in the structure functions, to approximate the experimental data. 
Provided that $m$ is greater than $1$, the series determining the one dimensional restriction of the structure functions (\ref{eq:series1}) will converge.  The thinking here is that there is a universal coefficient $m$ for each Reynolds number that will determine how fast the sine series converges, and thus the spatial smoothness of the structure functions.  
Thus for $k$ large, ${\tilde c}_k$ and ${\tilde d}_k \sim \frac{1}{k^m}$. Moreover, we are (optimistically) assuming that the two contributions ${\tilde c}_k$ and ${\tilde d}_k$, to the large eddies, also scale with the order of the structure functions and can be characterized by a number $b$, respectively $a$, for each Taylor-Reynolds number. Thus for $k$ small, ${\tilde c}_k\sim\frac{1}{b}$ and ${\tilde d}_k \sim \frac{1}{a}$.  This turns out to work reasonably well, see Table 5.\\

In summary, we reduce the coefficients for the three-dimensional structure functions, $c_k$ and $d_k$, 
to the ones for the one-dimensional structure functions, $\tilde{c_k}$ and $\tilde{d_k}$.  
We then fit the formulas for the one-dimensional structure functions to the data, 
and propose a simple ansatz (\ref{eq:ans}), 
for the coefficients' dependence on the Taylor-Reynolds number and the wavenumber $1/k$.

\section{Comparison of the Model with the Data}
\indent  The VDTT is capable of using pressurized inert gases as working fluids.  Specifically, the use of pressurized Sulfur Hexafluoride with a low kinematic viscosity enables classical grid experiments at $R_\lambda$ up to $1600$.  The turbulence in the VDTT was generated by a fixed grid of crossed bars, 
and is called classical grid turbulence \cite{Co61,CBC66}.  
The classical grid disturbed the free flow mechanically 
at the upstream end of the test section.  
In the wake of the grid, the turbulence evolved along the length of the tunnel 
without the middle region being substantially influenced by the walls of the tunnel \cite{BBNSX14}.  
The measurements were made with a Dantec StreamLine hot-wire anemometry system, using NSTAPs developed at Princeton University, see \cite{VS14}.  
The hot-wire probes were at a distance of $7.1$ meters downstream from the $186.6$ mm classical grid, 
so that the turbulence evolved through at least one eddy turnover time.  
Taylor's frozen flow hypothesis is used to extract $x$ and $r$ from the time series of the probe as in \cite{BBNSX14}.  
Measurements were taken for Taylor Reynolds Numbers $110$, $264$, $508$, $1000$, and $1450$.  
The pertinent parameters for the data are given Table $1$.  
For more information about the experiments, see \cite{SBB15,SBB17}.  Each measurement was taken over five minutes and sampled at $60$kHz, giving $1.8(10^7)$ data points.

The longitudinal velocity differences are 
\[
\delta u(r,t)=u(y,t)-u(x,t)= u(x+r,t)-u(x,t)
\]
where $u$, $x$ and $r$ are parallel vectors along the $x$-axis.  
The system length in the tunnel is an important value when fitting the data since we scale the lag variable, $r$, 
$\frac{r/\eta}{ \mathrm{system\ length}}=(x-y)$, with the system length.  
The system length in our case is the mesh size of the grid, 
and not the square root of the cross sectional area of the tunnel, for instance. 
\begin{table}
\begin{center}
    \begin{tabular}{ | l | l | l | p{2.5cm} |}
    \hline
    Taylor Reynolds Number & $\eta$ & L & $\nu$\\ \hline
110&1025&165.1&1.55$(10^{-5})$\\ \hline
264&162&102.5&2.34$(10^{-6})$\\ \hline
508&91&123.9&1.00$(10^{-6})$\\ \hline
1000&36&136.6&2.91$(10^{-7})$\\ \hline
1450&22&129.5&1.50$(10^{-7})$ \\
    \hline
 \end{tabular}
\caption{Here, $\eta$ is the Kolmogorov length scale given in micrometers, 
L is the integral length scale given in millimeters, and $\nu$ is the viscosity given in ${m^2}/{s}$.  The data here is at the furthest measured point downstream of the grid, as $\eta$ and $L$ evolve downstream.}
\end{center}
\end{table}
\indent The structure functions were plotted against ${r}/{\eta}$, 
where $r$ is the distance between positions $x$ and $y$ as given by the Taylor Frozen Flow Hypothesis 
and $\eta$ is the Kolmogorov length scale.  
In order for our sine series formula to capture the entire data set, 
we divided ${r}/{\eta}$ by its maximum value for which we computed structure functions, which was $\frac{r}{\eta}=19540$.  
We also introduced a variable, $D$, so that 
we substituted 
\begin{displaymath}
\frac{r}{\eta}/(19540(D))
\end{displaymath}
for $x-y$ in the formulas.  
The fitted values for $D$ are given in Table 5.  Note that $D$ is one of the four parameters which are not active over the range of Taylor-Reynolds numbers, which is shown in Figure \ref{fig:1450Length} and will be justified later.\\

\indent Fitting was done in Mathematica using the built-in ``findfit'' function.  
To bound computational time reasonably, the series given in Section $2$ were limited to one thousand terms.  
Because initial fitting to the formulas given in Section $2$ proved not to be effective, 
we permitted the first two sine terms in the expansion to have free coefficients 
to allow for variation in the nonuniversal largest scales of the flow \cite{BBBGGMVXY2011}.  
In other words, the new model used to fit the data is given by 
\begin{align}
\label{eq:SpFull}
S_p=A_1\vert\sin[(\pi\times r)/(19540.3(D))]\vert&+A_2\vert\sin^2[(2\pi\times r)/(19540.3(D))]\vert\\
&+\sum_{k=3}\frac{2^p}{C^p}A_p\vert \sin^p[\pi k\cdot(x-y)]\vert, 
\end{align}
where $A_p$ is given by $(\ref{eq:Ap})$.  
This was done for all structure function fits.\\
\indent As it stands right now, that leaves us with seven parameters for the fits, namely $a$, $b$, $m$, $C$, $D$, $A_1$, and $A_2$.  However, only three parameters are active over the entire inertial range, specifically $a$, $b$, and $m$, in the sense that they are changing the relative weights of the Fourier components of the solution $u(x,t)$.  The parameter $D$ measures the system length correction for large Reynolds numbers.  This correction serves to place the transition from the dissipative range to the inertial range.  The parameter $C$ measures the root-mean squared velocity whereas $A_1$ and $A_2$ measure the influence of the large eddies upon the grid.  These three parameters measure the transition out of the inertial range.  This is shown in Figures \ref{fig:1450Length}, \ref{fig:1450A}, and \ref{fig:1450CC}, and will be justified later.  Through experimentation, we found the best result when using the fourth-order structure functions for each Taylor Reynolds Number 
to fix the coefficients $a$, $b$, and $D$.  

\begin{table}
\begin{center}
    \begin{tabular}{| l | l | l | l | l |p{1.9cm}|}
    \hline
    Re Lambda&  110 & 264 &508& 1000& 1450 \\ \hline
Second-Order&0.00744&.0153&.0169&.0183&.0195\\ \hline
Third-Order&.00154&.00162&.00484&.00564&.00664\\ \hline
Fourth-Order&.000384&.00189&.00228&.00251&.00305\\ \hline
Sixth-Order&.0000341&.000431&.000566&.000552&.000691\\ \hline
Eighth-Order&3.12($10^{-6}$)&.0000839&.000122&.000144&.000204\\
    \hline
 \end{tabular}
\caption{The fitted values for $A_1$ in eq. $(\ref{eq:SpFull})$}   
\end{center}
\end{table}
\begin{table}
\begin{center}
{\small
    \begin{tabular}{| l | l | l | l | l |p{2.1cm}|}
    \hline
Re Lambda &  110 & 264 &508 & 1000& 1450\\ \hline
Second-Order&.00285&.00583&.00653&.00697&.00666 \\ \hline
Third-Order&.000872&.00124&.00526&.00488&.00395\\ \hline
Fourth-Order&.000174&.000746&.000804&.001&.0006\\ \hline
Sixth-Order&4.24($10^{-6}$)&-.0000756&-.00011&.0000919&.0000654\\ \hline
Eighth-Order&1.04($10^{-6}$)&.0000127&.0000147&.0000264&-4.71($10^{-7}$)\\ \hline
 \end{tabular}}
\caption{The fitted values for $A_2$ in eq. $(\ref{eq:SpFull})$}   
\end{center}
\end{table}

\begin{figure}
\begin{subfigure}{.5\textwidth}
  \centering
  \includegraphics[scale=.5]{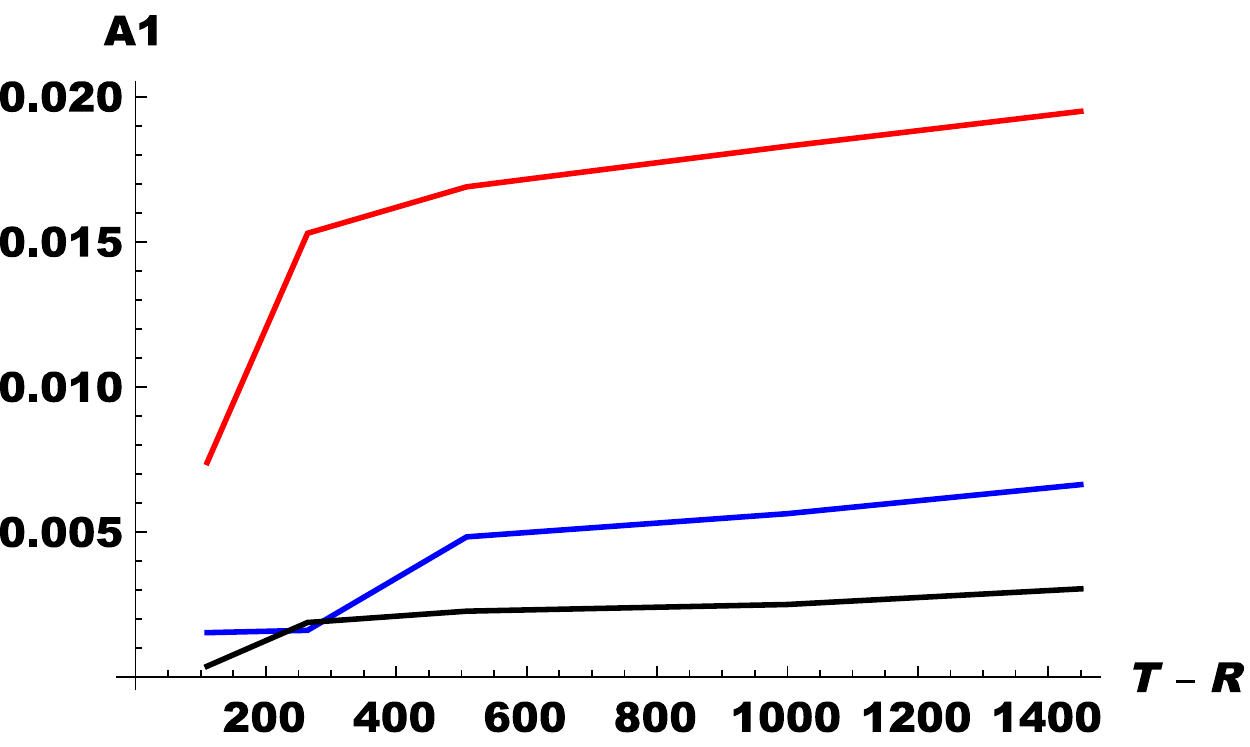}
  \caption{}
  \label{fig:A1}
\end{subfigure}%
\begin{subfigure}{.5\textwidth}
  \centering
  \includegraphics[scale=.5]{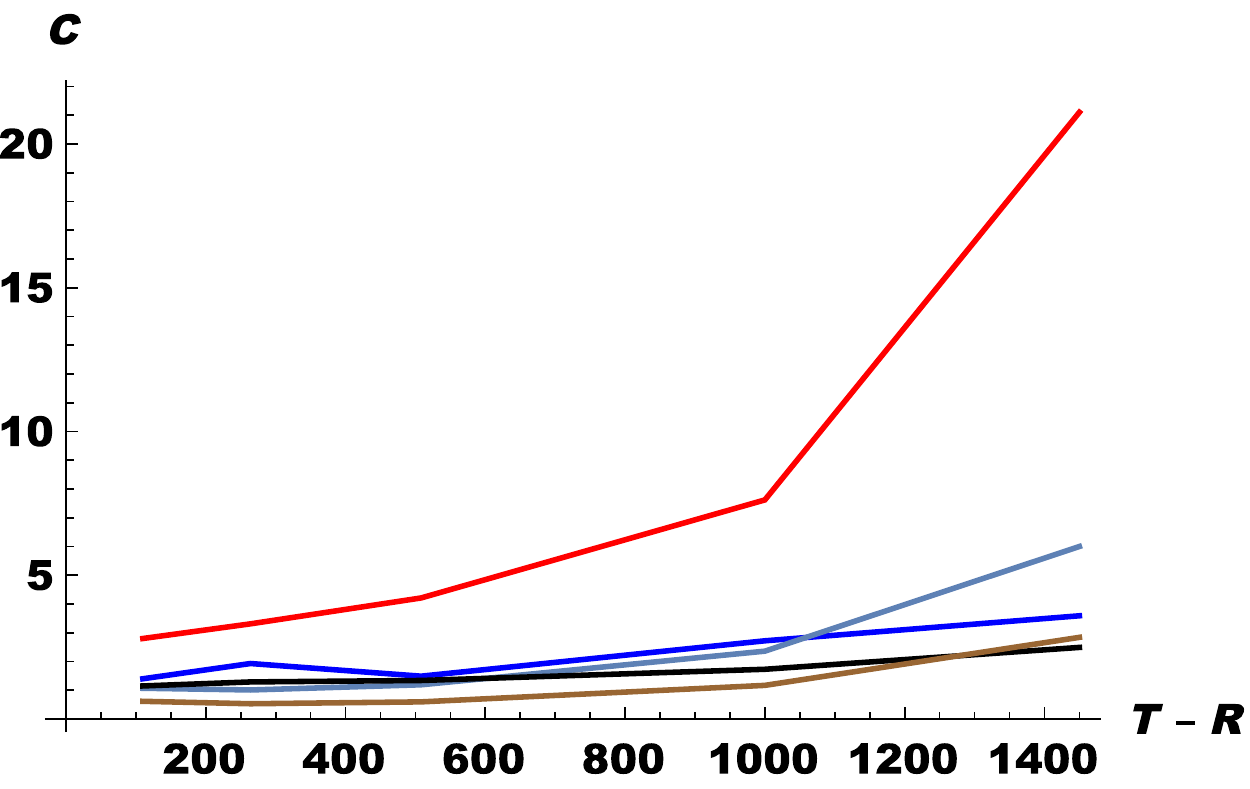}
  \caption{}
  \label{fig:C}
\end{subfigure}
\caption{(a) The values of the coefficient $A_1$ of the first Fourier component, as a function of the Taylor-Reynolds number, from Table 2. Top (red) 2nd moment, second from top (blue) 3rd moment, third from top (black) 4th moments. The higher moments give negligible correction, whereas the first three change little, notice the scale, over the range of T-R numbers in the experiment.\\
(b) The values of the coefficient $C$, as a function of the Taylor-Reynolds number,  from Table 4, top (red) 2nd moment (root means square size), lower, the higher moments. Notice that neither (a) or (b) are plotted on a log scale.}
 \end{figure}
 
\begin{table}
\begin{center}
    \begin{tabular}{| l | l | l | l | l |p{1.5cm}|}
    \hline
    Taylor Reynolds Number&  110 & 264 &508& 1000& 1450 \\ \hline
Second-Order&2.8&3.31&4.21&7.62&21.1\\ \hline
Third-Order&1.4&1.93&1.49&2.72&3.59\\ \hline
Fourth-Order&1.07&1.01&1.19&2.36&6\\ \hline
Sixth-Order&1.15&1.29&1.34&1.73&2.49\\ \hline
Eighth-Order&0.616&.531&.596&1.17&2.84\\
    \hline
 \end{tabular}
\caption{The fitted values for $C$ in eq. $(\ref{eq:SpFull})$}   
\end{center}
\end{table}
\begin{table}
\begin{center}
    \begin{tabular}{| l | l | l | l | l |p{1.5cm}|}
    \hline
    Taylor Reynolds Number&  a & b & D \\ \hline
110&11.64&0.01612&1.569\\ \hline
264&9.581&0.05236&1.769\\ \hline
508&8.314&0.06504&1.518\\ \hline
1000&3.792&0.09247&1.32\\ \hline
1450&2.684&0.4092&1.3\\
    \hline
 \end{tabular}
\caption{The fitted values for $a$, $b$, and $D$ in eq. $(\ref{eq:SpFull})$.  Note that these parameters are grouped as they are independent of the order of the structure functions.}   
\end{center}
\end{table}
\begin{table}
\begin{center}
    \begin{tabular}{| l | l | l | l | l |p{1.5cm}|}
    \hline
    Taylor Reynolds Number&  110 & 264 &508& 1000& 1450 \\ \hline
Second-Order&1.563&1.16&1.069&.8965&.9148\\ \hline
Third-Order&1.408&1.185&0.922&.6488&.5262\\ \hline
Fourth-Order&1.269&.8751&.7936&.5554&.4865\\ \hline
Sixth-Order&.98607&.5055&.5192&.4339&.3398\\ \hline
Eighth-Order&.9711&.5924&.5755&.3771&.2482\\
    \hline
 \end{tabular}
\caption{The fitted values for $m$ in eq. \ref{eq:SpFull}}   
\end{center}
\end{table}
Tables $2$, $3$, and $4$ contain the fitted values of $A_1$, $A_2$ and $C$ respectively, as described in $(\ref{eq:SpFull})$.  Note that $A_1$, $A_2$, $C$, and $m$ are given their own tables as they change with the order of the structure function, whereas $a$, $b$, and $D$ are placed in the same table as they do not. Consider Figure \ref{fig:A1} that shows the values of the coefficient $A_1$ as a function of the Taylor-Reynolds number taken from Table 2. We show that the values are small and do not change much over the range of T-R numbers in the experiment. The figure shows the values based on the first three structure functions do not change much of the whole over the range of T-R numbers in the experiment and the higher order structure functions give negligible correction. One can think of the second-order structure function as the root-mean square size and the higher order ones measure the roughness. Table 3 shows that the same analysis applies to the coefficient $A_2$ except that its values are even smaller.  Consequently we omit the plot of $A_2$. Table 4 shows that the parameter $C$ increases over the range of T-R numbers in the experiment and Figure \ref{fig:C} shows its plot corresponding to the increasing structure functions. We see that only the 2nd moment (red), measuring the root-mean square size of the mean fluctuation velocity, increases significantly over the range of T-R numbers in the experiment.  The plots corresponding to the higher moments increase significantly less, although they indicate that the mean velocity is getting spatially rougher. However, this is not influencing much the balance of the Fourier components in the Fourier representation of $u(x,t)$ compared to the significant changes in the parameters $a, b$ and
$m$ discussed below. It simply measures the increase of the mean turbulent velocity as the turbulence increases. Table 5 shows the parameter $D$ does not change with the order of the structure functions and does not change much over the range of T-R numbers in the experiment.

\section{Evaluation of the Model}

\begin{figure}
\centering
\begin{subfigure}{.6\textwidth}
  \centering
  \includegraphics[scale=.6]{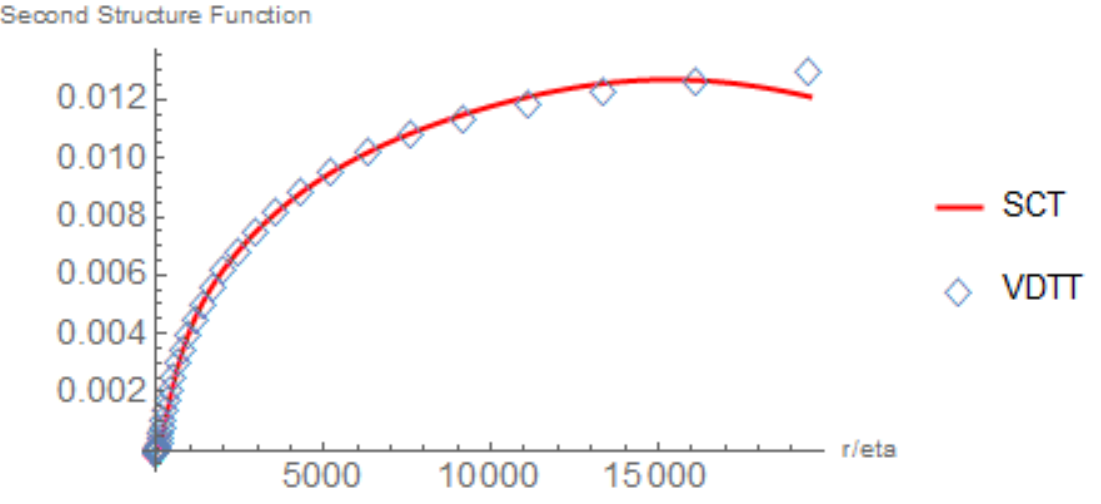}
  \caption{Second-Order Structure Function, Normal Scale}
  \label{fig:1102NLL}
\end{subfigure}%
\begin{subfigure}{.6\textwidth}
  \centering
  \includegraphics[scale=.6]{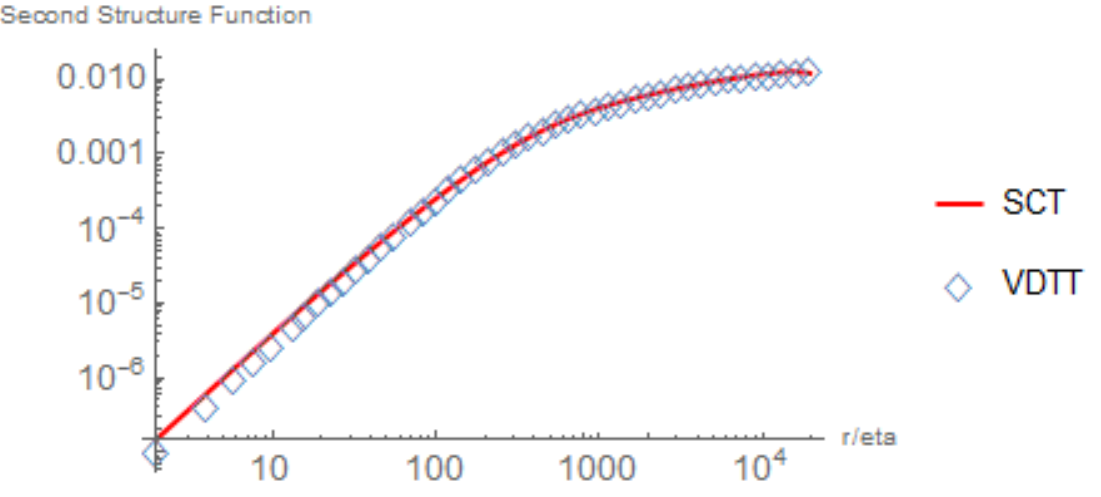}
  \caption{Second-Order Structure Function}
  \label{fig:1102}
\end{subfigure}\\
\begin{subfigure}{.6\textwidth}
  \centering
  \includegraphics[scale=.6]{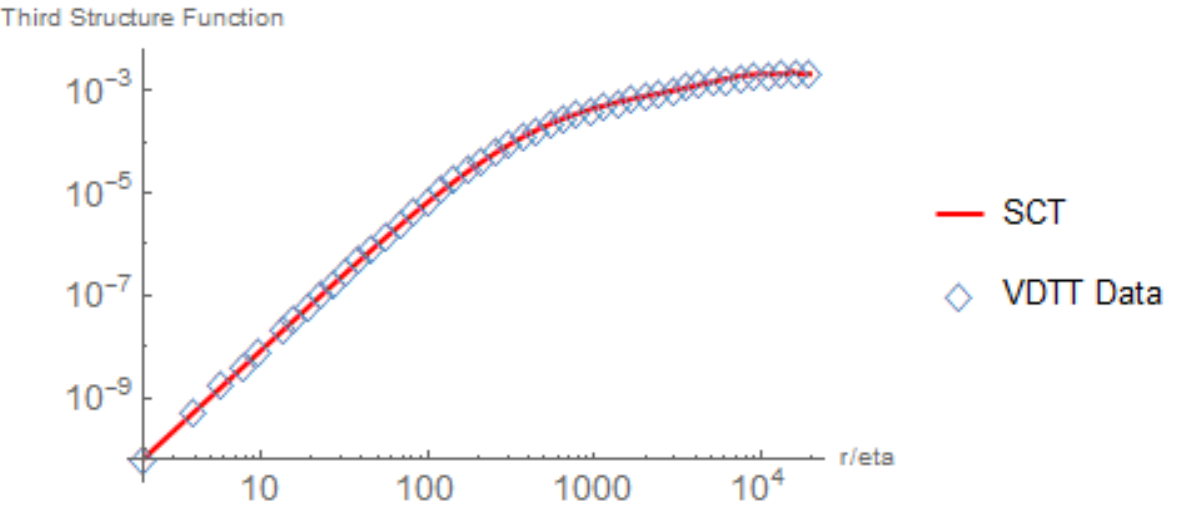}
  \caption{Third-Order Structure Function}
  \label{fig:1103}
\end{subfigure}%
\begin{subfigure}{.6\textwidth}
 \centering
 \includegraphics[scale=.6]{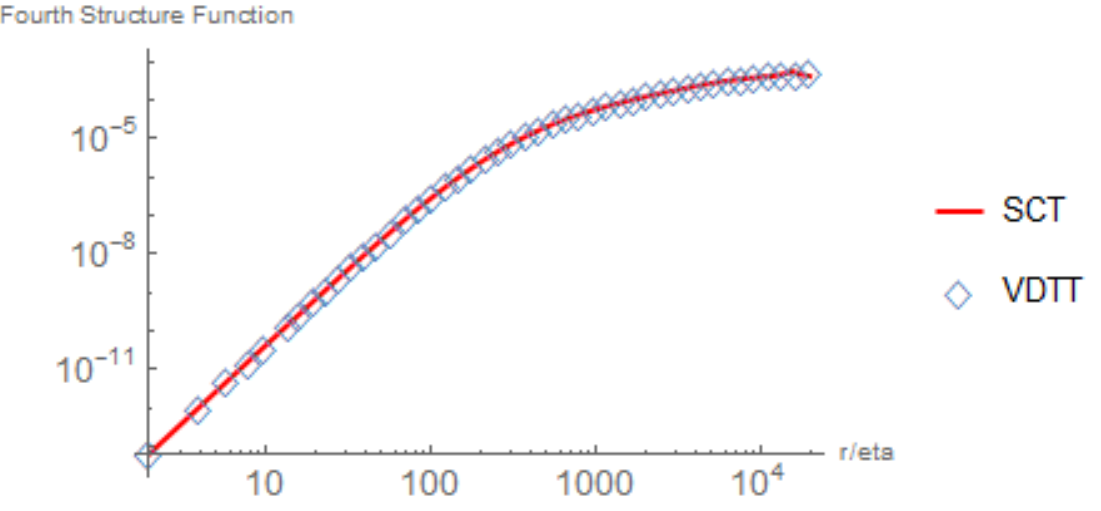}
 \caption{Fourth-Order Structure Function}
 \label{fig:1104}
\end{subfigure}\\
\begin{subfigure}{.6\textwidth}
  \centering
  \includegraphics[scale=.6]{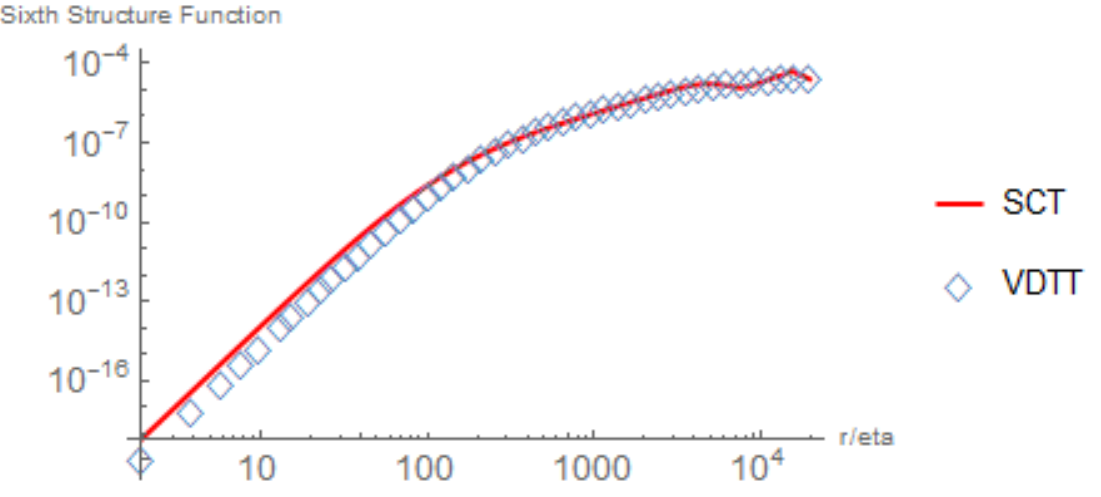}
  \caption{Sixth-Order Structure Function}
  \label{fig:1106}
\end{subfigure}%
\begin{subfigure}{.6\textwidth}
  \centering
  \includegraphics[scale=.6]{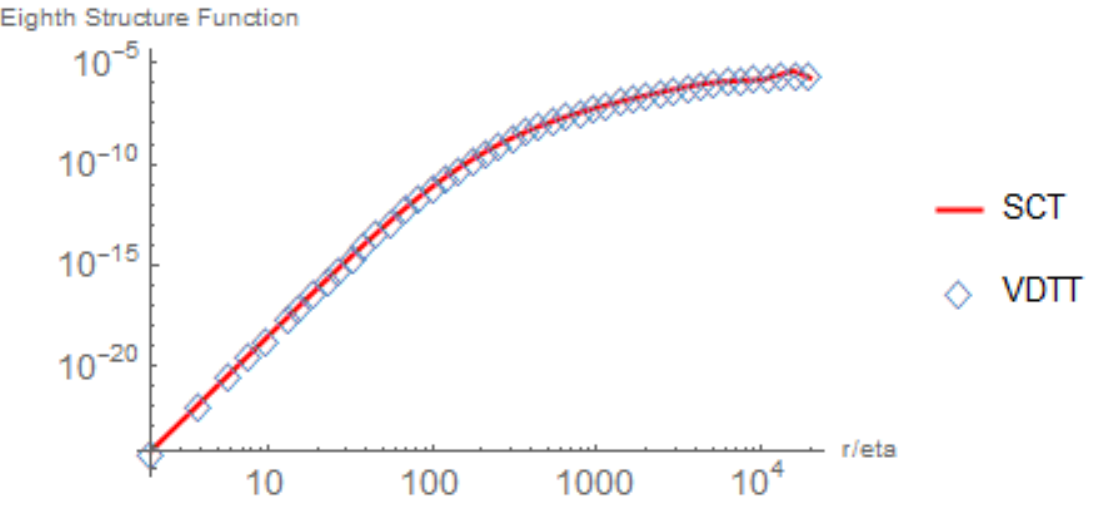}
  \caption{Eighth-Order Structure Function}
  \label{fig:1108}
\end{subfigure}
\caption{Taylor Reynolds Number 110.  Note that the plots (b)-(f) are made on a log-log scale.  The blue diamonds correspond to the data from the VDTT whereas the red line is the fitted SCT}
\label{fig:110}
\end{figure}

\begin{figure}
\centering
\begin{subfigure}{.6\textwidth}
  \centering
  \includegraphics[scale=.6]{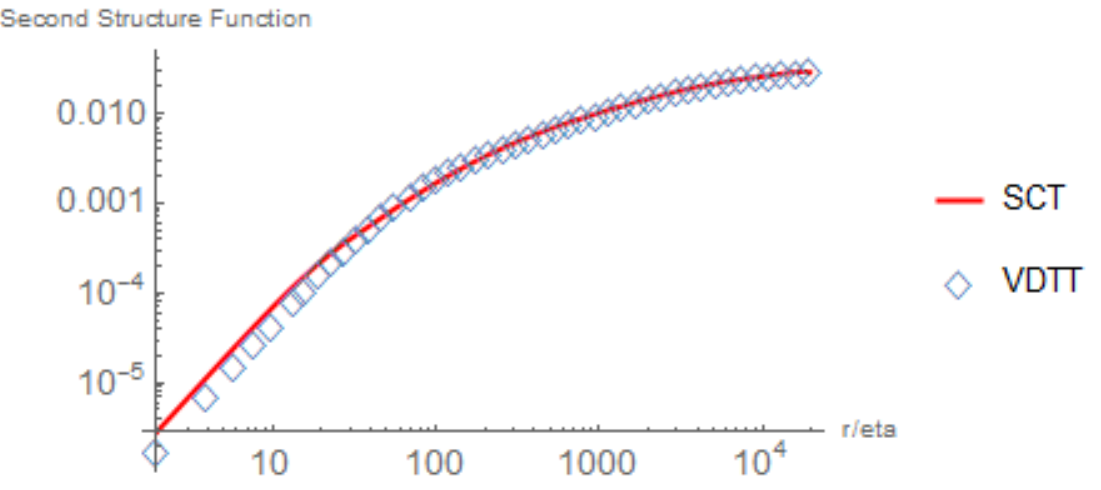}
  \caption{Second-Order Structure Function}
  \label{fig:2642}
\end{subfigure}%
\begin{subfigure}{.6\textwidth}
  \centering
  \includegraphics[scale=.6]{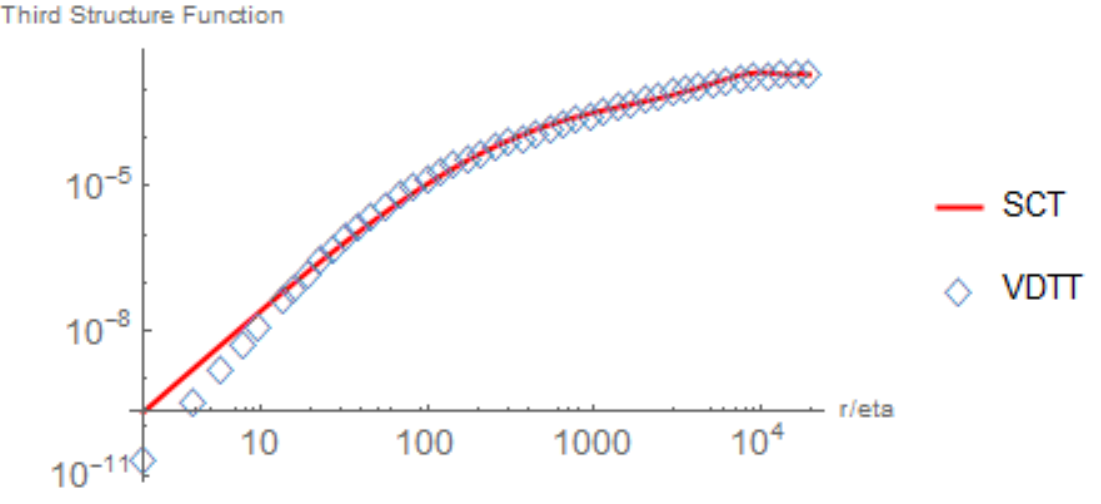}
  \caption{Third-Order Structure Function}
  \label{fig:2643}
\end{subfigure}\\
\begin{subfigure}{.6\textwidth}
 \centering
 \includegraphics[scale=.6]{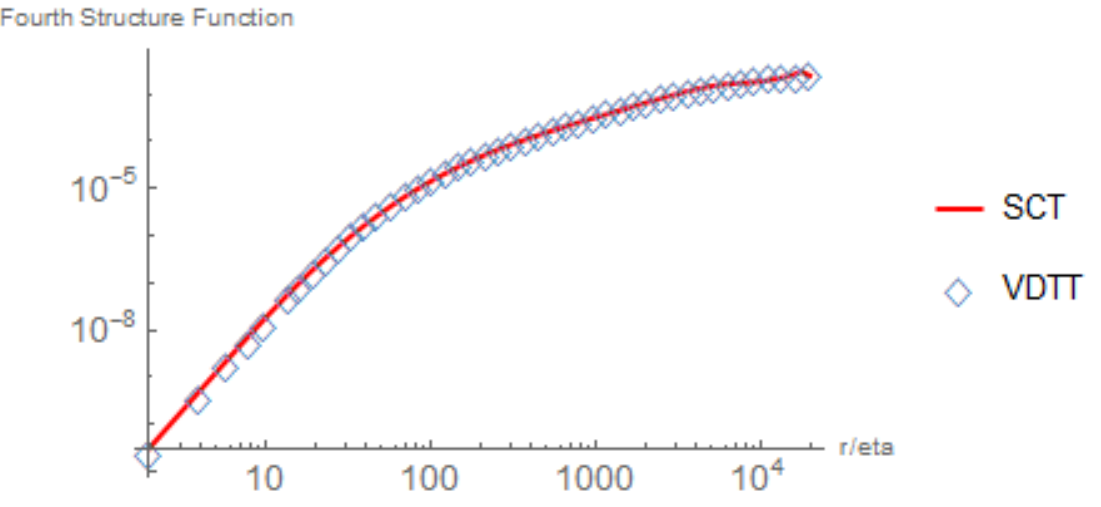}
 \caption{Fourth-Order Structure Function}
 \label{fig:2644}
\end{subfigure}%
\begin{subfigure}{.6\textwidth}
  \centering
  \includegraphics[scale=.6]{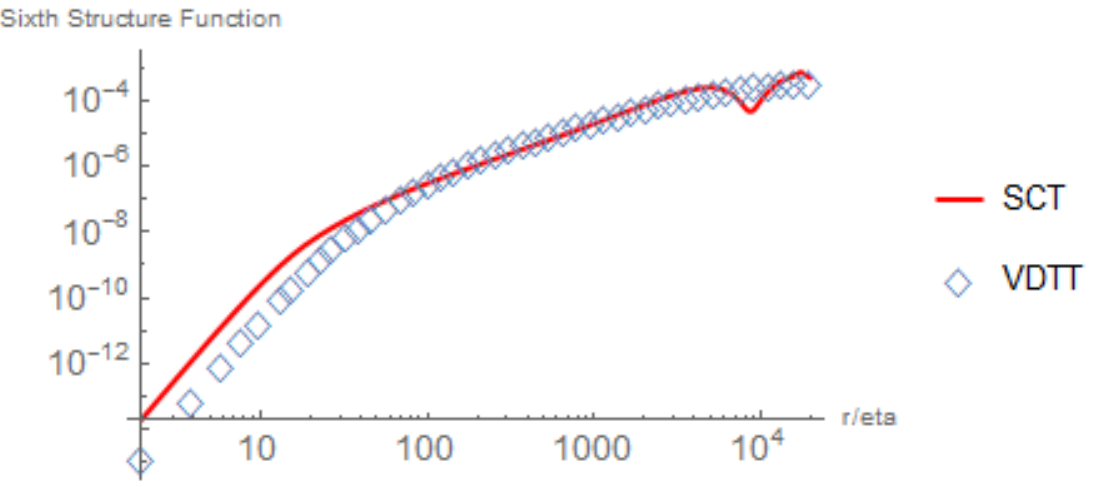}
  \caption{Sixth-Order Structure Function}
  \label{fig:2646}
\end{subfigure}
\begin{subfigure}{.6\textwidth}
  \centering
  \includegraphics[scale=.6]{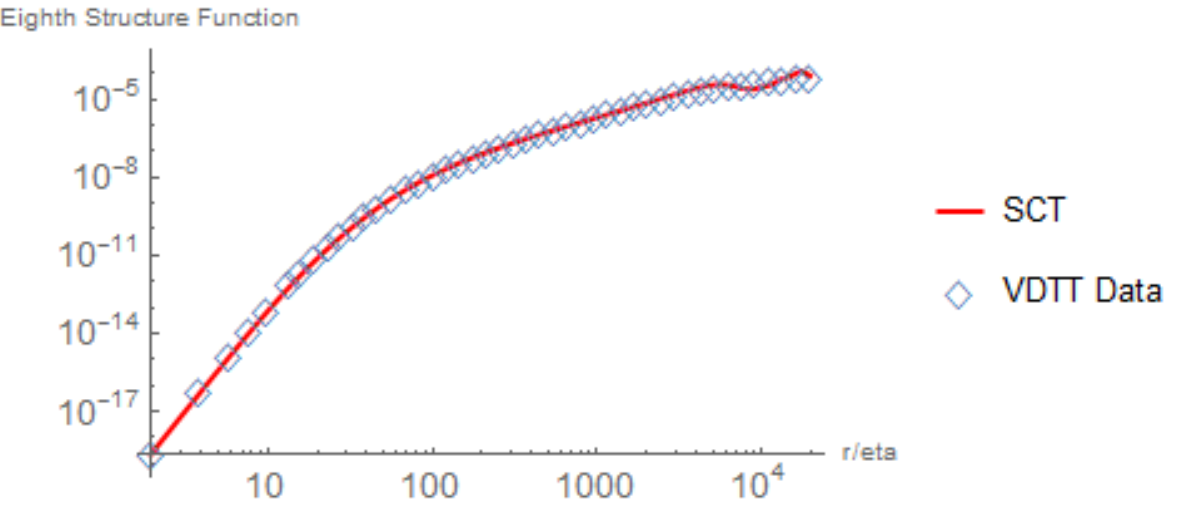}
  \caption{Eighth-Order Structure Function}
  \label{fig:2648}
\end{subfigure}
\caption{Taylor Reynolds Number 264.  Note that the plots are made on a log-log scale.  The blue diamonds correspond to the data from the VDTT whereas the red line is the fitted SCT}
\label{fig:264}
\end{figure}

\begin{figure}
\centering
\begin{subfigure}{.6\textwidth}
  \centering
  \includegraphics[scale=.6]{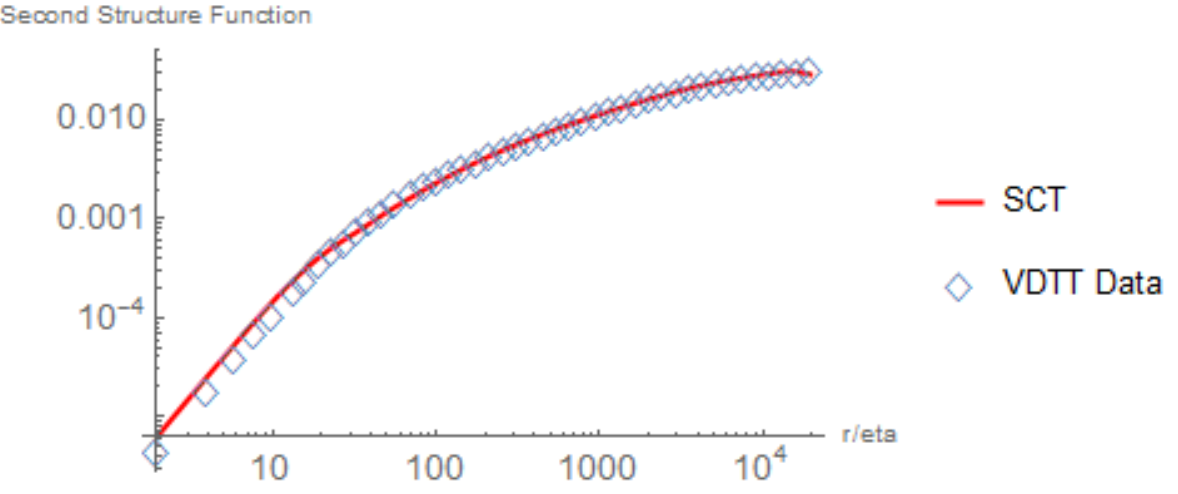}
  \caption{Second-Order Structure Function}
  \label{fig:5082}
\end{subfigure}%
\begin{subfigure}{.6\textwidth}
  \centering
  \includegraphics[scale=.6]{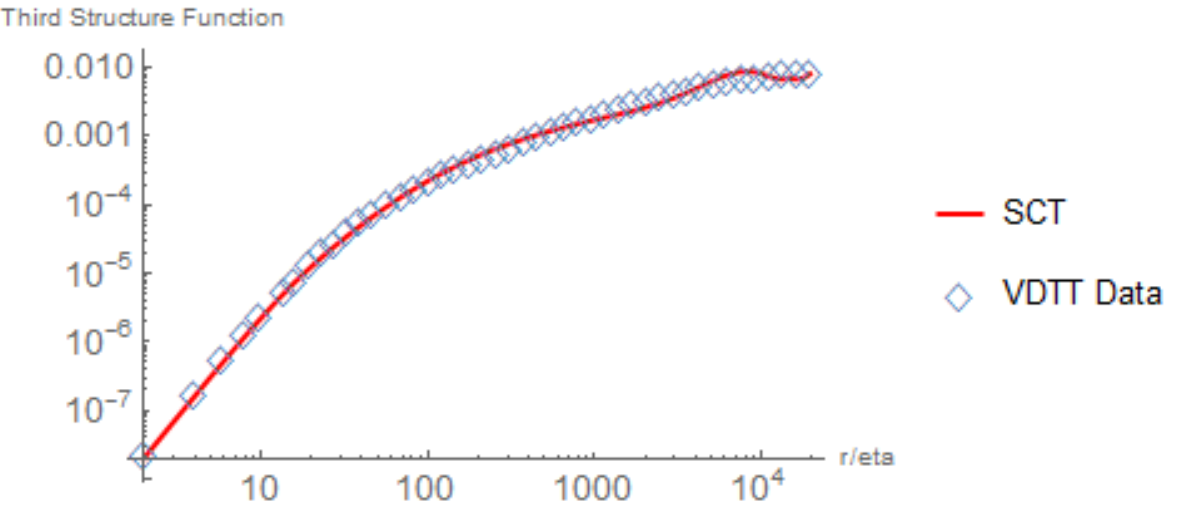}
  \caption{Third-Order Structure Function}
  \label{fig:5083}
\end{subfigure}\\
\begin{subfigure}{.6\textwidth}
 \centering
 \includegraphics[scale=.6]{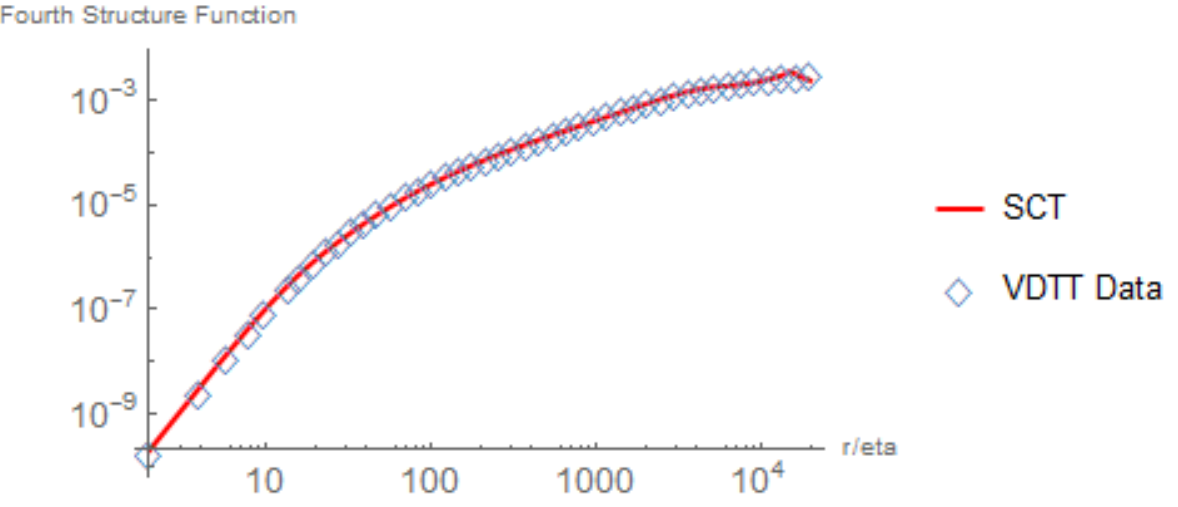}
 \caption{Fourth-Order Structure Function}
 \label{fig:5084}
\end{subfigure}%
\begin{subfigure}{.6\textwidth}
  \centering
  \includegraphics[scale=.6]{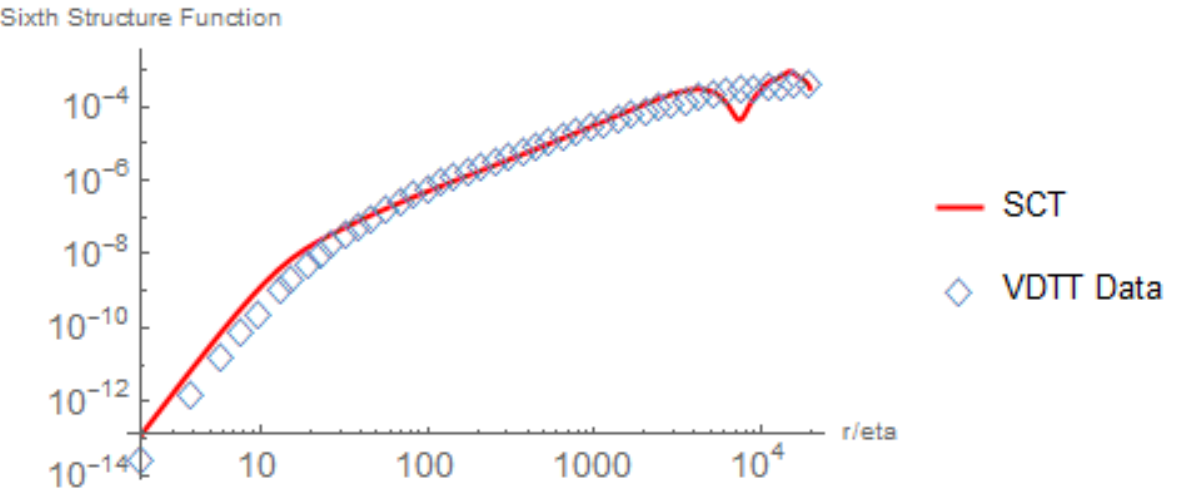}
  \caption{Sixth-Order Structure Function}
  \label{fig:5086}
\end{subfigure}
\begin{subfigure}{.6\textwidth}
  \centering
  \includegraphics[scale=.6]{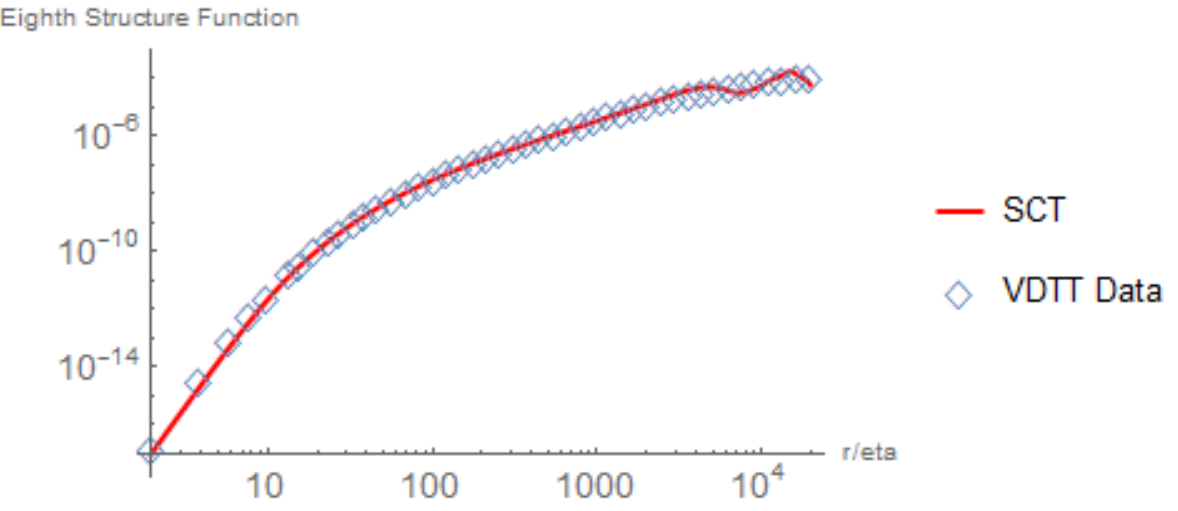}
  \caption{Eighth-Order Structure Function}
  \label{fig:5088}
\end{subfigure}
\caption{Taylor Reynolds Number 508.  Note that the plots are made on a log-log scale.  The blue diamonds correspond to the data from the VDTT whereas the red line is the fitted SCT}
\label{fig:508}
\end{figure}

\begin{figure}
\centering
\begin{subfigure}{.6\textwidth}
  \centering
  \includegraphics[scale=.6]{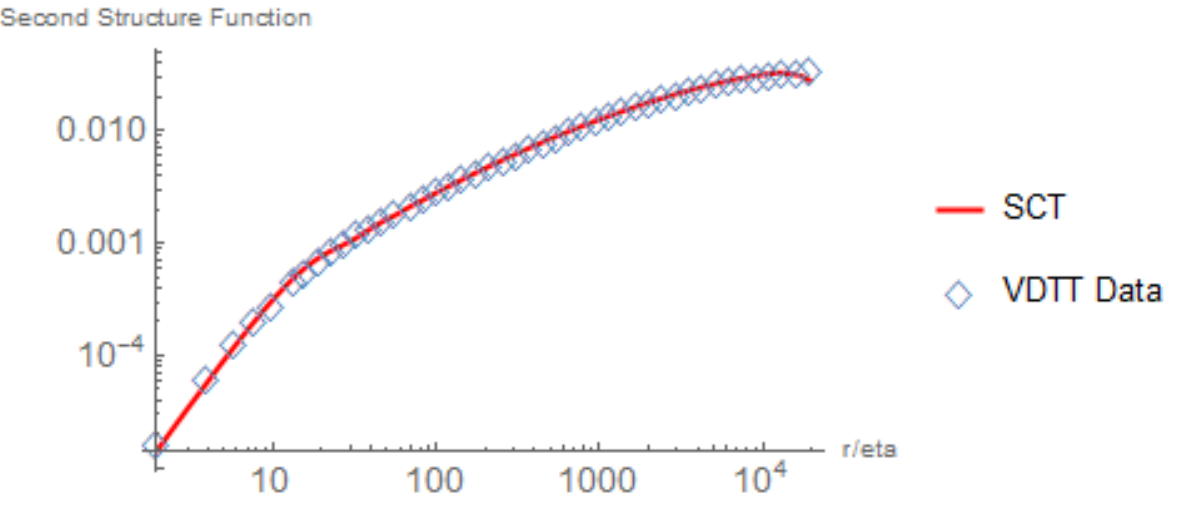}
  \caption{Second-Order Structure Function}
  \label{fig:10002}
\end{subfigure}%
\begin{subfigure}{.6\textwidth}
  \centering
  \includegraphics[scale=.6]{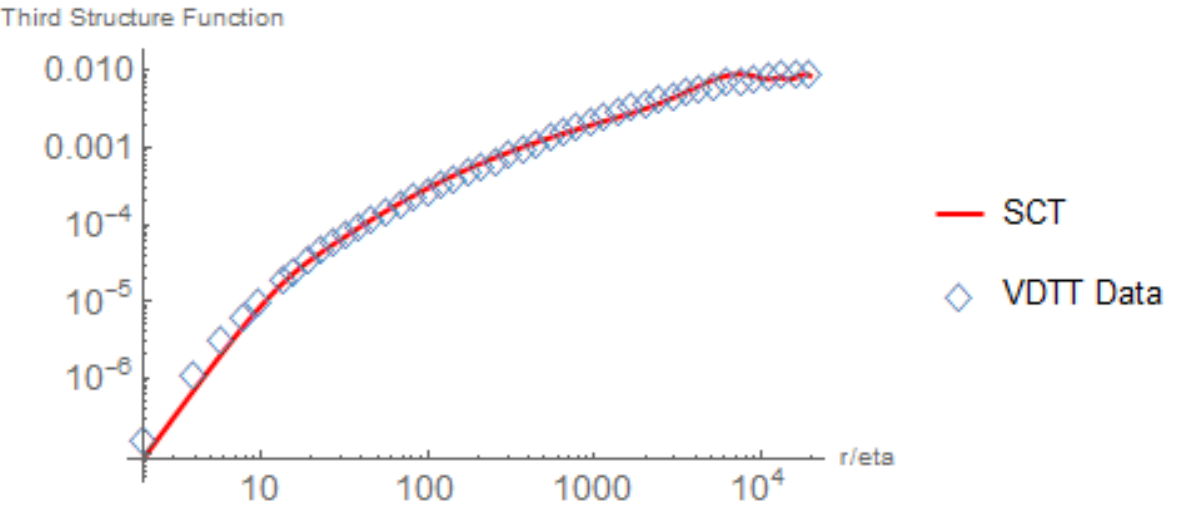}
  \caption{Third-Order Structure Function}
  \label{fig:10003}
\end{subfigure}\\
\begin{subfigure}{.6\textwidth}
 \centering
 \includegraphics[scale=.6]{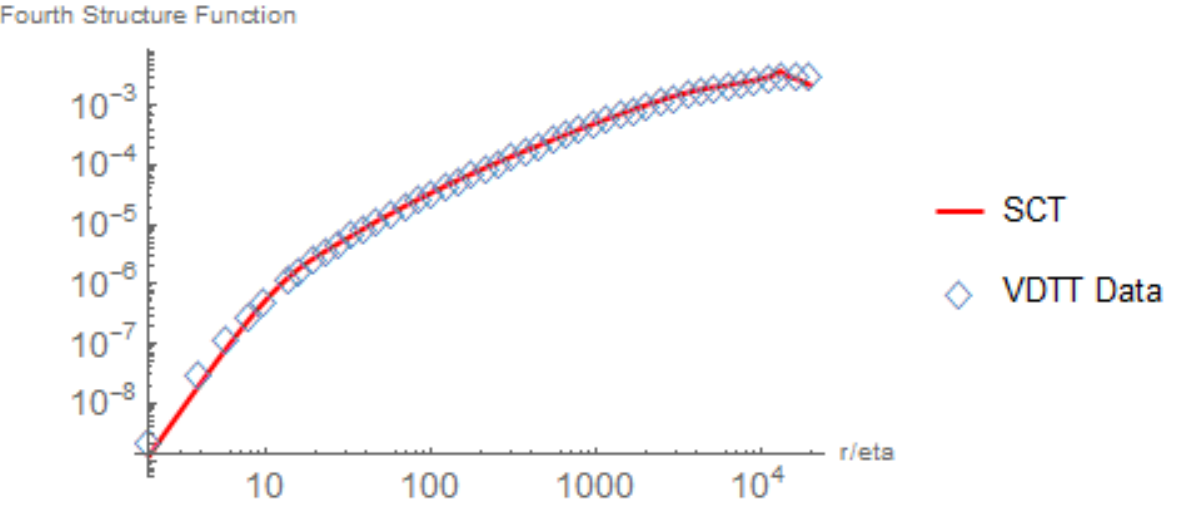}
 \caption{Fourth-Order Structure Function}
 \label{fig:10004}
\end{subfigure}%
\begin{subfigure}{.6\textwidth}
  \centering
  \includegraphics[scale=.6]{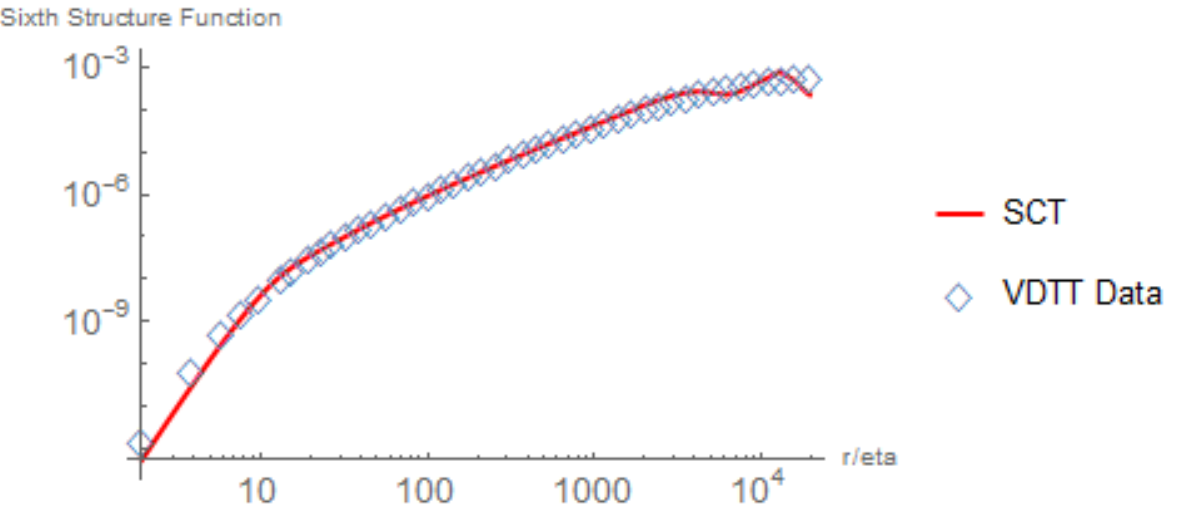}
  \caption{Sixth-Order Structure Function}
  \label{fig:10006}
\end{subfigure}
\begin{subfigure}{.6\textwidth}
  \centering
  \includegraphics[scale=.6]{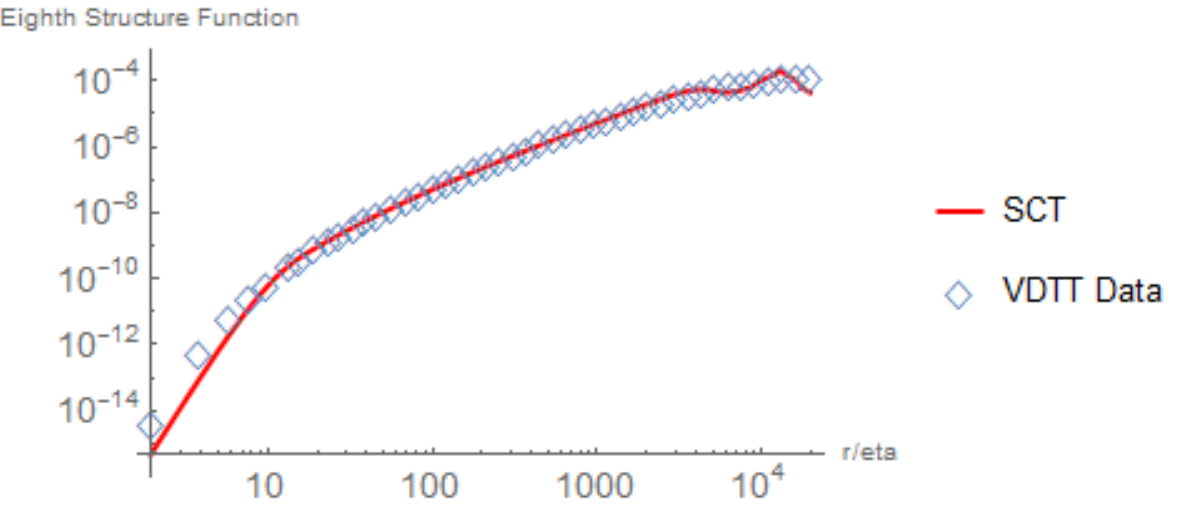}
  \caption{Eighth-Order Structure Function}
  \label{fig:10008}
\end{subfigure}
\caption{Taylor Reynolds Number 1000.  Note that the plots are made on a log-log scale.  The blue diamonds correspond to the data from the VDTT whereas the red line is the fitted SCT}
\label{fig:1000}
\end{figure}

\begin{figure}
\centering
\begin{subfigure}{.6\textwidth}
  \centering
  \includegraphics[scale=.6]{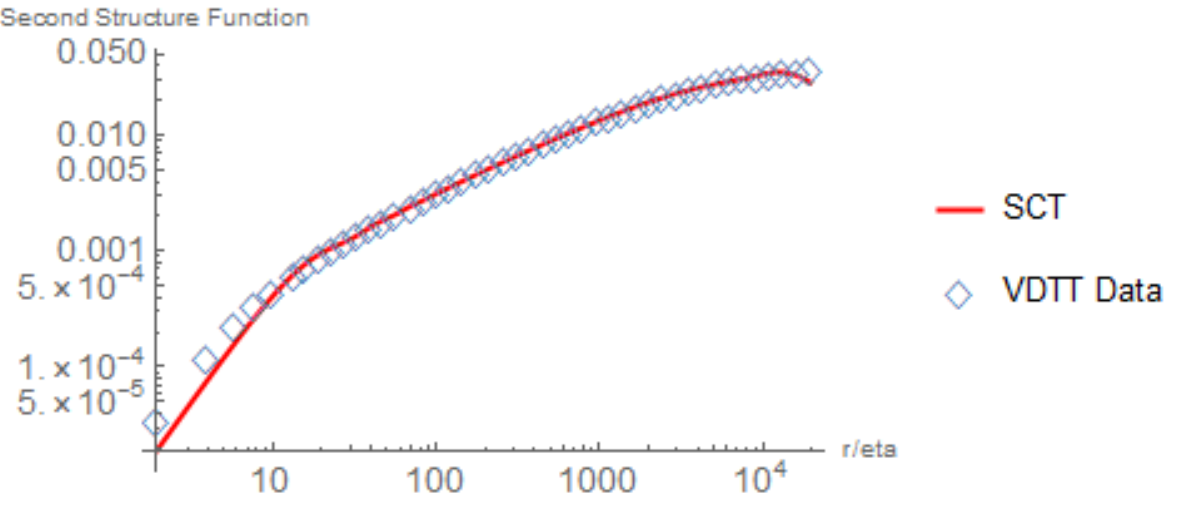}
  \caption{Second-Order Structure Function}
  \label{fig:14502}
\end{subfigure}%
\begin{subfigure}{.6\textwidth}
  \centering
  \includegraphics[scale=.6]{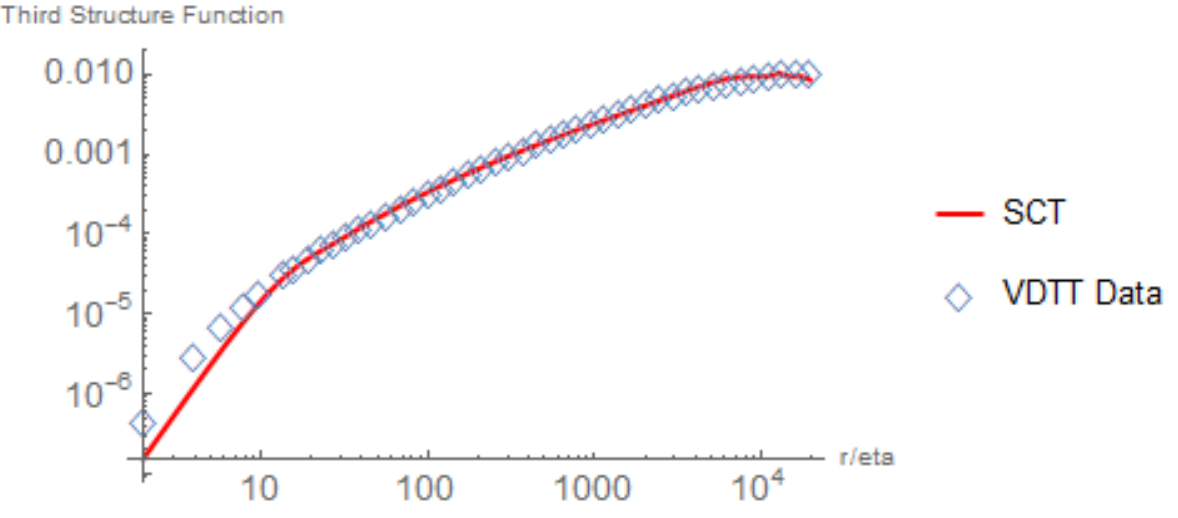}
  \caption{Third-Order Structure Function}
  \label{fig:14503}
\end{subfigure}\\
\begin{subfigure}{.6\textwidth}
 \centering
 \includegraphics[scale=.6]{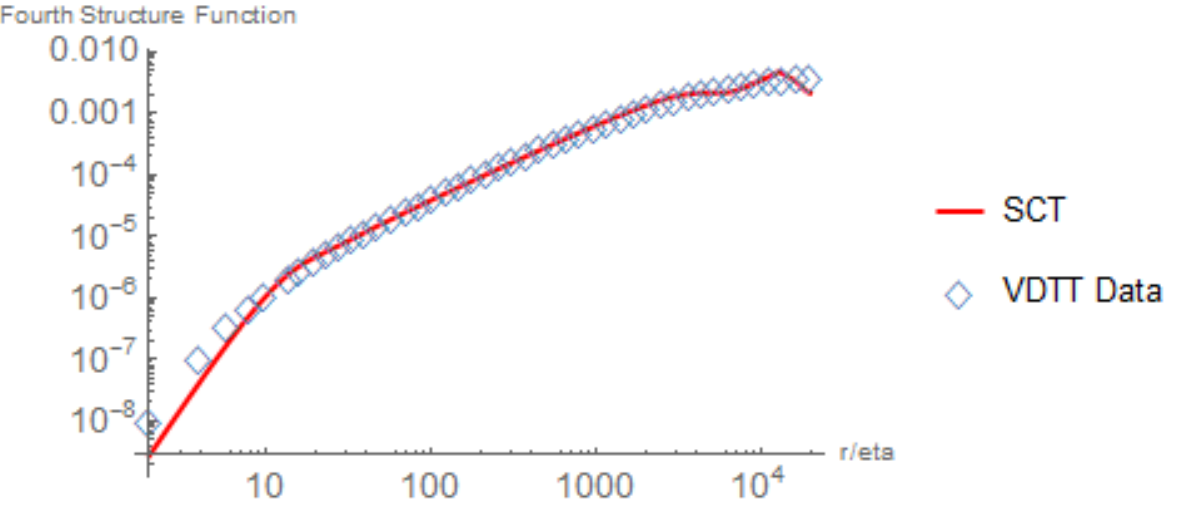}
 \caption{Fourth-Order Structure Function}
 \label{fig:14504}
\end{subfigure}%
\begin{subfigure}{.6\textwidth}
  \centering
  \includegraphics[scale=.6]{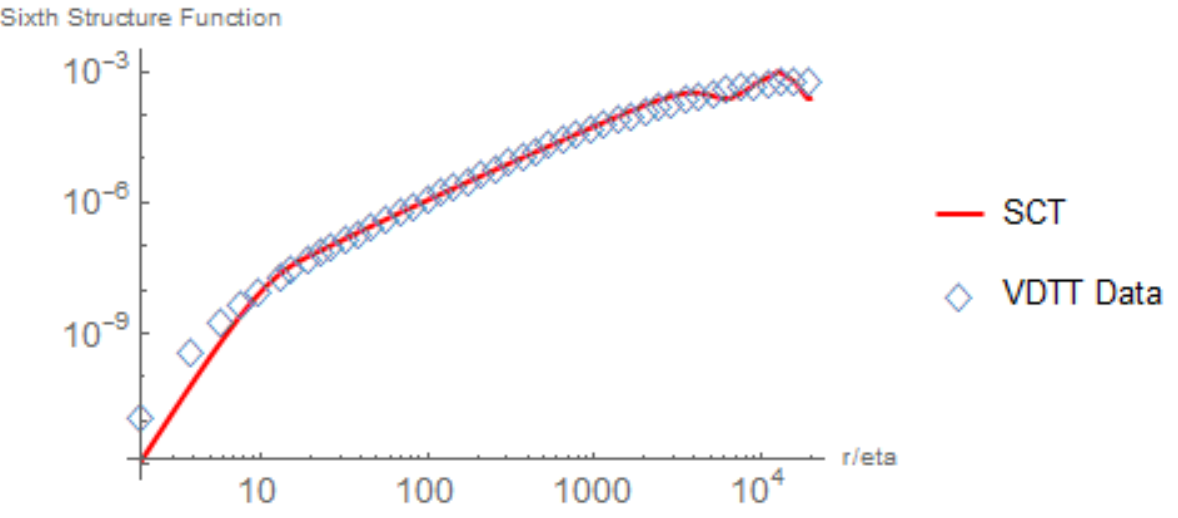}
  \caption{Sixth-Order Structure Function}
  \label{fig:14506}
\end{subfigure}
\begin{subfigure}{.6\textwidth}
  \centering
  \includegraphics[scale=.6]{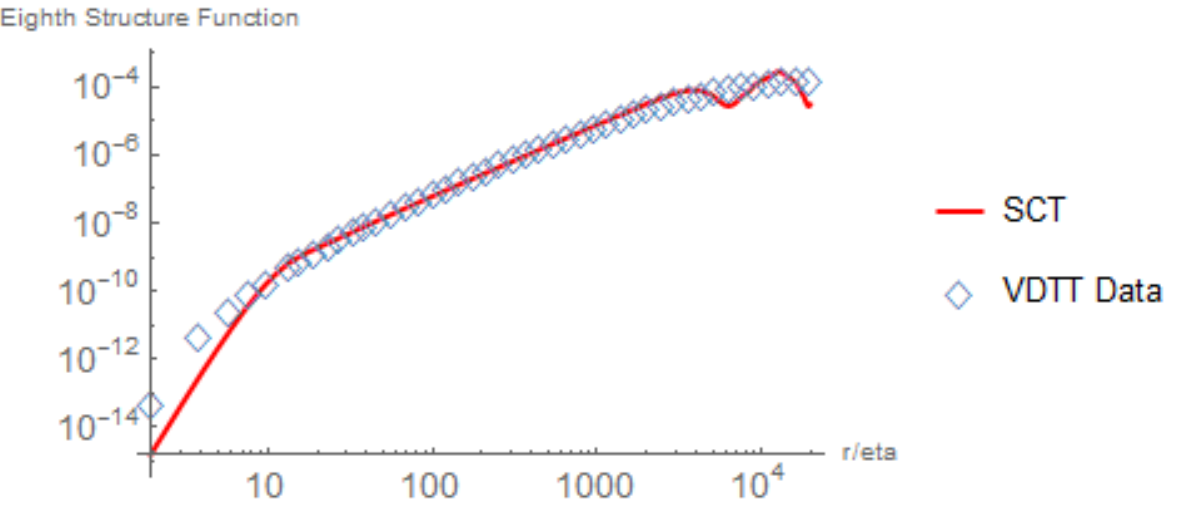}
  \caption{Eighth-Order Structure Function}
  \label{fig:14508}
\end{subfigure}
\caption{Taylor Reynolds Number 1450.  Note that the plots are made on a log-log scale.  The blue diamonds correspond to the data from the VDTT whereas the red line is the fitted SCT}
\label{fig:1450}
\end{figure}

In this section we present the results of the fits to the data.  
In the figures \ref{fig:110}, \ref{fig:264}, \ref{fig:508}, \ref{fig:1000}, and \ref{fig:1450}, the blue diamonds are the data from the experiment while the red lines are the SCT theory predictions.  
All the plots are on a log-log scale except for a single plot of the second-order structure function at Taylor Reynolds Number 110, 
this latter plot is included for perspective.  
The agreement between the theory is satisfactory for most orders of the structure functions and for most Reynolds numbers.  
For the highest Reynolds numbers and highest order (sixth- and eighth-order) structure functions, 
we see differences between the theory and experiment at the smallest scales.    
In general, we note that the fits become less accurate as we increase the order of the structure functions.  
The fits for the second-, third-, and fourth-order structure functions are generally better than 
the fits for the sixth- and eighth-order structure function fits, which are rougher.  
This is expected from the theory given by in \cite{BB312}, and will be explored further in a future paper.  

Table $5$ gives the fitted values for $a$ and $b$ that change significantly over the range of T-R numbers in the experiment, see Figure \ref{fig:a-b}.  This table shows that the Central Limit Theorem term,
\begin{displaymath}
\bar{D}=\sum_{k\neq0} c_k^\frac{1}{2}db_t^ke_k(x),\,\,c_k^\frac{1}{2}=\sqrt{\frac{2}{\pi}}\frac{b}{b^2+k^m},
\end{displaymath}
as given by $b$ has a greater influence for smaller Taylor Reynolds numbers than the Large Deviation Principle term:
\begin{displaymath}
D^\prime=\sum_{k\neq0} d_k\eta_k dte_k(x),\,\,d_k=\sqrt{\frac{2}{\pi}}\frac{a}{a^2+k^m},
\end{displaymath}
given by $a$, as for small values of $k$, these terms essentially become $\frac{1}{a^2}$ and $b$, respectively, because $b$ is small.  As the Reynolds number goes up, we do see an increasing influence of $b$ dominating the increase of $\frac{1}{a^2}$, see the plot in Figure \ref{fig:a-b} (b). Thus the contribution of the Central Limit Theorem is greater.

The values of the exponent $m$ of the wavenumber $k=k_1$ are given in Table 6. Their change
over the inertial range seems small, but since $m$ is and exponent the influence on the weight of
the Fourier components of $u(x,t)$ is highly significant.  
In general the exponents are larger or very close to $1$, at least near the top of the table.  
The first (top) line in Table 6, corresponding to the second-order structure function, 
verifies the hypothesis concerning the coefficients $c_k$ and $d_k$ in Sections 2 and 4.  
The energy shell coefficient $\tilde c_k$ and $\tilde d_k$ should decay as $|k|^{-m}$, $m >1$.  
All the exponents in the first line in Table 6 satisfy this except the last two.  
However, both still lie within the fitting uncertainty and may be explained by the Reynolds number corrections absorbing the weight of the power.  
Thus the exponents $m(R_\lambda)$ in the first line depend on $R_\lambda$, but approach $1$ as $R_\lambda$ becomes large.  We would expect the exponents to remain above one for the rest of the lines on the table, but this is not the case.  We will seek to explain this result in a future paper.

\section{The Improved SCT Model}

\begin{figure}
\begin{subfigure}{.5\textwidth}
  \includegraphics[scale=.5]{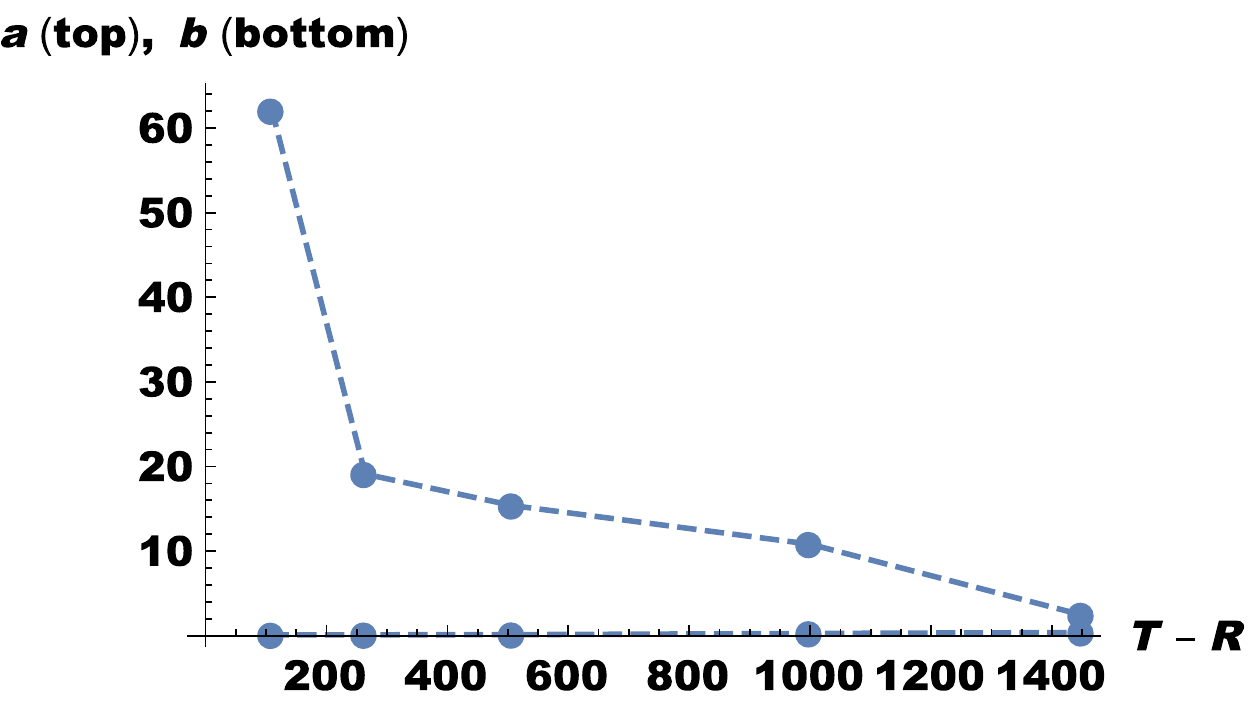}
  \caption{The coefficients a and b, from Table 5.}
  \label{fig:sub1ab}
\end{subfigure}%
\begin{subfigure}{.5\textwidth}
  \centering
  \includegraphics[scale=.5]{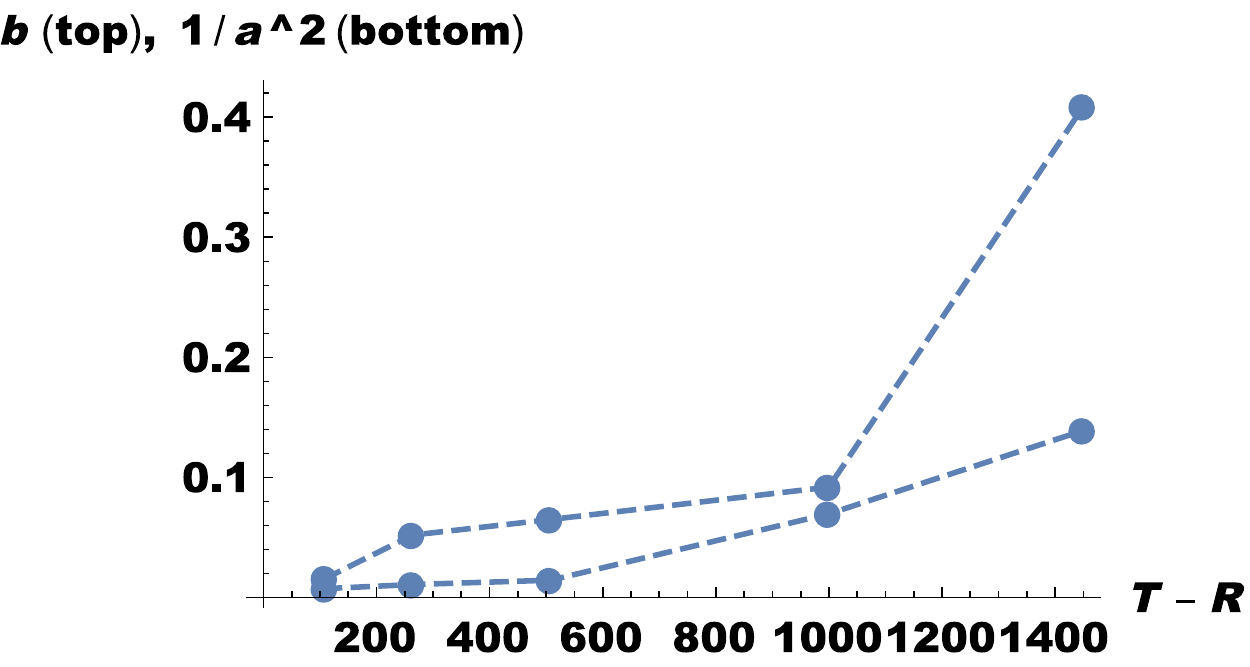}
  \caption{The coefficients b and $\frac{1}{a^2}$. $b$ changes by a factor of 40 over the Taylor-Reynolds number range in the experiment.}
  \label{fig:sub21b1a}
\end{subfigure}
\caption{The dependence of the coefficients, in the improved SCT model (\ref{eq:isns}), on the Taylor-Reynolds number. The coefficient $\frac{1}{a^2}$ makes the large deviation contribution in (\ref{eq:isns}) so it is plotted separately against $b$.  Note that $a$ versus $b$ is included for completeness but due to the nature of the Fourier coefficients $c_k$ and $d_k$ as defined in $(\ref{eq:ans})$, the ideal comparison is $b$ against $\frac{1}{a^2}$.}
\label{fig:a-b}
\end{figure}

The comparison of theory and data for homogeneous turbulence now produces a much improved Stochastic Closure Model, removing the infinitely many coefficients $c_k$, $d_k$, and $h_k$ from Equation (\ref{eq:sns}).  
What we find is that the large scales satisfy equation (\ref{eq:RANS}), 
whereas the small scale flow satisfies the stochastic Navier-Stokes equation, 
\begin{align}
\label{eq:isns}
du+u\cdot\nabla u dt&=(\nu\Delta u+ \nabla(\Delta^{-1}[\mathrm{Trace}(\nabla u)]))dt-u\cdot \nabla U-U\cdot \nabla u\no\\&+\sum_{k\neq0} \left(\frac{\textbf{a}}{|a|^2+|k|^m}\right)\vert k\vert^{-\frac{5}{3}} dte_k(x)+\sum_{k\neq0} \frac{\textbf{b}^{1/2}}{(|b|^2+|k|^m)^{1/2}}\vert k\vert^{-2}db_t^ke_k(x)\\
&-u\frac{1}{3}\sum_{k\neq0}\bar{N}_t^kdt, \no
\end{align}
where $\textbf{a},\textbf{b}^{1/2}, k \in \mathbb{R}^3$, $a=\vert\textbf{a}\vert$, and $b=\vert\textbf{b}\vert$.  
The improved SCT model depends on three parameters $a$, $b$ and $m$, 
which are all function of the Taylor-Reynolds number $R_\lambda$.  A plot of $a$ and $b$ from Table 5 are shown in Figure \ref{fig:a-b} (a). It shows that the Large Deviation coefficient $a$ is larger than the Central Limit Theorem coefficient $b$. But this is deceiving since the right comparison is between $1/a^2$ and $b$ for small wavenumber $k$, because of the form of the coefficients $c_k,\ d_k$ in (\ref{eq:ans}). This comparison is shown in Figure \ref{fig:a-b} (b). We see that $b$ is larger than $1/a^2$ and dominates for large Reynolds numbers. For large wavenumbers $k$, $b$ dominates even more because now it is compared with $a^2/k^2$. The conclusion is that the Central Limit Theorem term is the main contributor to the noise in the velocity differences, and the bias given by the Large Deviation term is only significant for small wavenumbers $k$ and small Reynolds numbers. 


The coefficient $C$ that appears in the computation of the structure functions (\ref{eq:series1}) is not constant for each Taylor-Reynolds number, see Table 4, because it measures both the size of the velocity fluctuations and the relative strength of the Center Limit Theorem term and the Large Deviation term in the noise. However, it does not vary much over the center part of Table 4 as a function of the Taylor-Reynolds number. The exponent $m$ also varies with Taylor-Reynolds number.  However, it also does not vary much with the Taylor-Reynolds number above the diagonal, as indicated by the bold numbers, in Table 6.

\begin{figure}
\centering
\begin{subfigure}{.6\textwidth}
  \centering
  \includegraphics[scale=.6]{14504-eps-converted-to.pdf}
  \caption{Fourth Structure Function as shown in figure $5$}
  \label{fig:145042}
\end{subfigure}%
\begin{subfigure}{.6\textwidth}
  \centering
  \includegraphics[scale=.6]{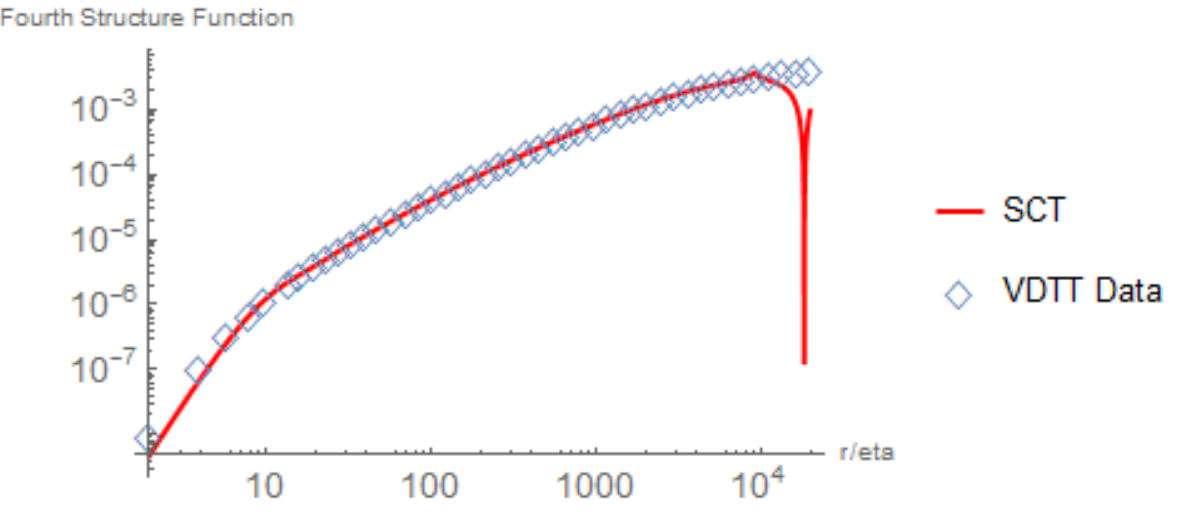}
  \caption{Fourth Structure Function at $D=0.920864$}
  \label{fig:1450atlowlength}
\end{subfigure}
\caption{The two different fits for Taylor Reynolds number $1450$.  Note the downward peak resulting from the Sine series wanting to return to zero before the last data point.}
\label{fig:1450Length}
\end{figure}

\begin{figure}
\centering
\begin{subfigure}{.6\textwidth}
  \centering
  \includegraphics[scale=.6]{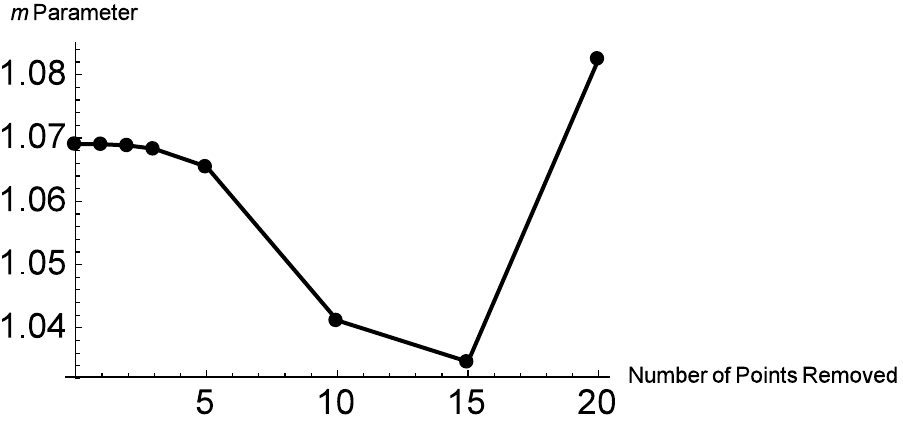}
  \caption{Second-order Structure Function}
  \label{fig:robust2}
\end{subfigure}%
\begin{subfigure}{.6\textwidth}
  \centering
  \includegraphics[scale=.6]{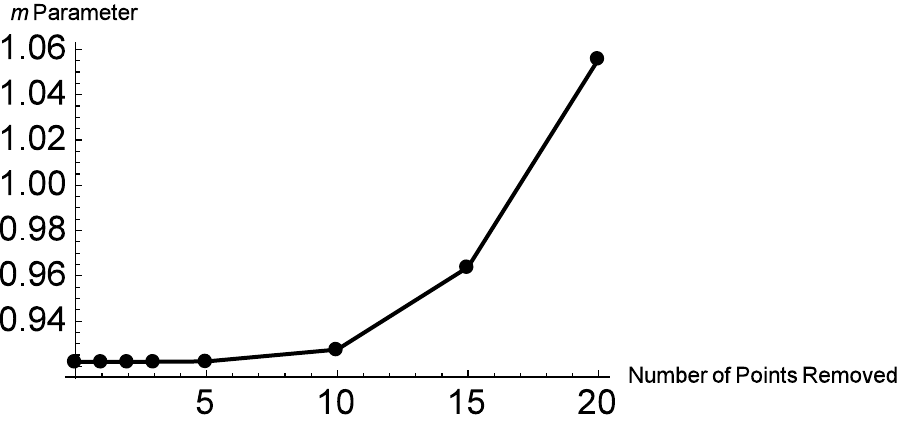}
  \caption{Third-order Structure Function}
  \label{fig:robust3}
\end{subfigure}\\
\begin{subfigure}{.6\textwidth}
 \centering
 \includegraphics[scale=.6]{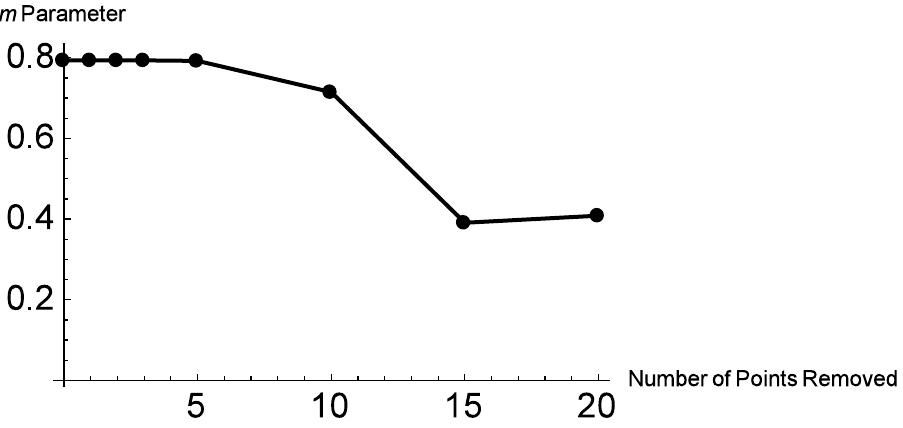}
 \caption{Fourth-order StructureFunction}
 \label{fig:robust4}
\end{subfigure}%
\begin{subfigure}{.6\textwidth}
  \centering
  \includegraphics[scale=.6]{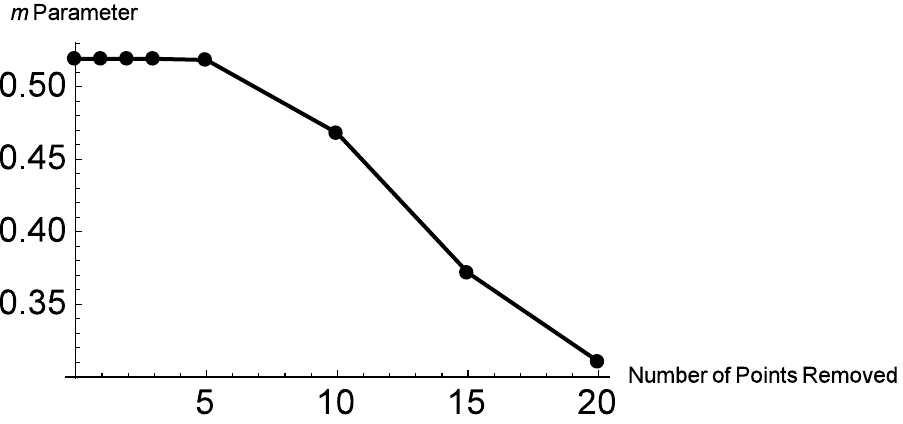}
  \caption{Sixth-order Structure Function}
  \label{fig:robust6}
\end{subfigure}
\begin{subfigure}{.6\textwidth}
  \centering
  \includegraphics[scale=.6]{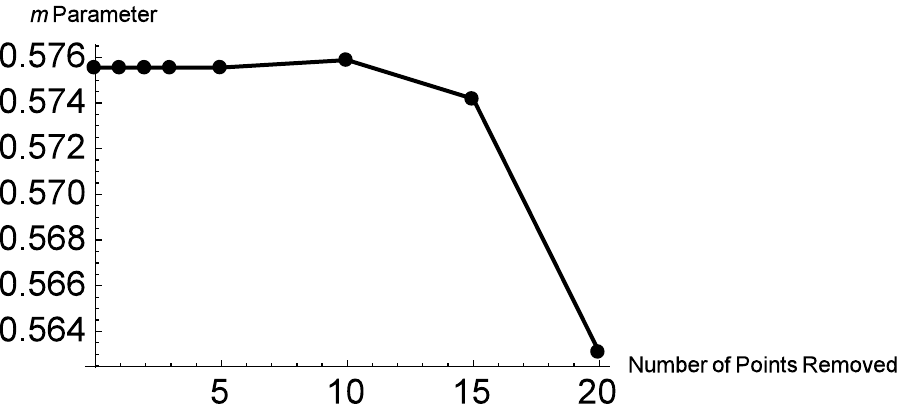}
  \caption{Eighth-order Structure Function}
  \label{fig:robust8}
\end{subfigure}
\caption{Robustness test for a Reynolds number of 508.  Note that the x-axis is the number of data points removed from the fitting.  We see very little change in the $m$ parameter until we remove enough data points to eliminate the dissipative range completely.  Note the scales on the y-axis.  As a result, we are convinced our fits are not dependent on the probe size.}
\label{fig:robust}
\end{figure}

\subsection{The Characterization of the Noise}

We will now answer the question: "What is the noise in homogeneous turbulence?" based on the improvements of the SCT model. This is the question that was stated in Section 3 and partially answered by the original SCT model. We can completely answer the question and characterize the noise appearing in the stochastic Navier-Stokes equation (\ref{eq:sns}). Recall that the original conjecture by Landau and Lifschitz \cite{LL59} was that the noise was white or uncorrelated. The question can be rephrased to ask what the noise forcing is that the fluid velocity is subjected to in fully developed turbulence. In the stochastic Navier-Stokes equation  the noise was modeled (SCT) as a Fourier series with infinitely many coefficients, but now these coefficients have been determined by the experimental data in Section 5. The following observations can be made:\\

\begin{figure}
\label{fig:corr}
\begin{subfigure}{.5\textwidth}
  \centering
  \includegraphics[scale=.5]{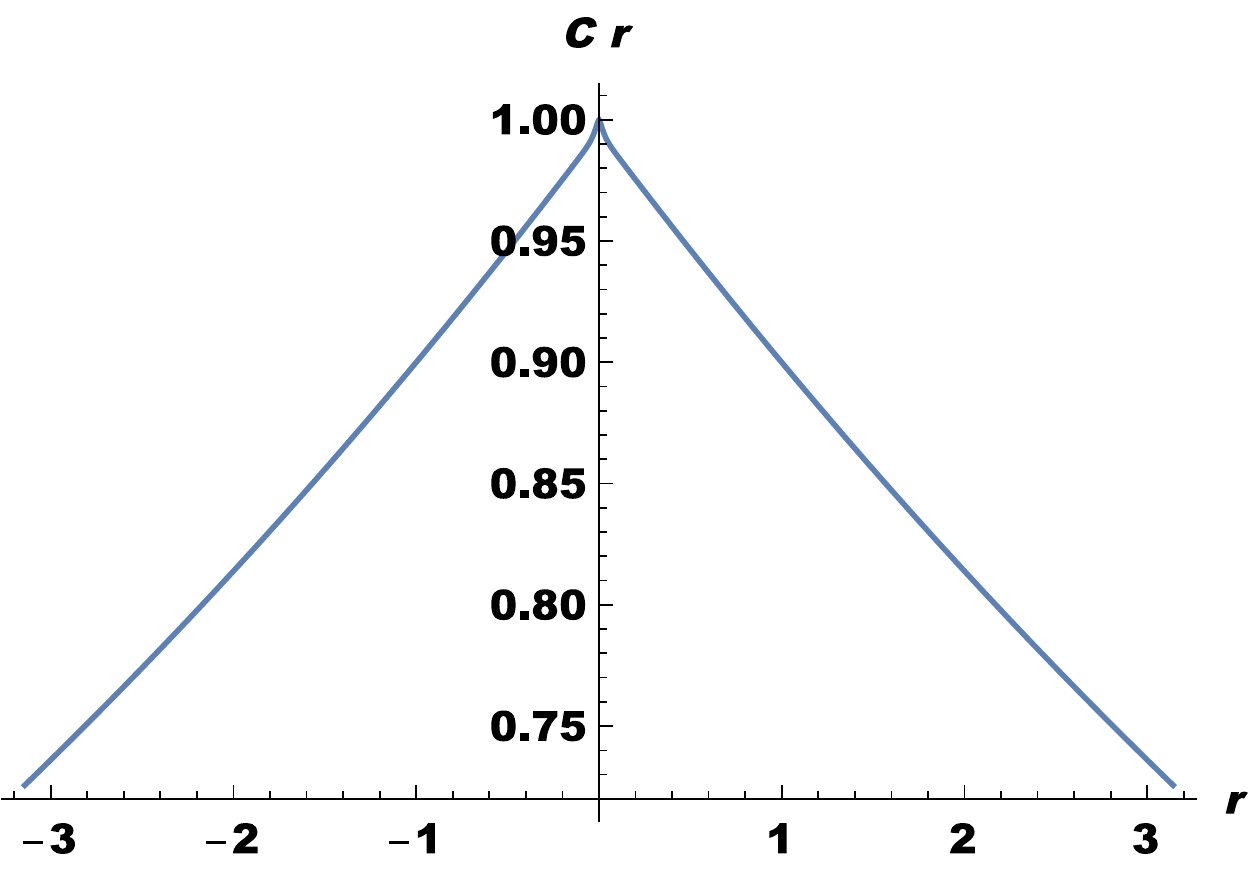}
  \caption{}
  \label{fig:corr1}
\end{subfigure}%
\begin{subfigure}{.5\textwidth}
  \centering
  \includegraphics[scale=.5]{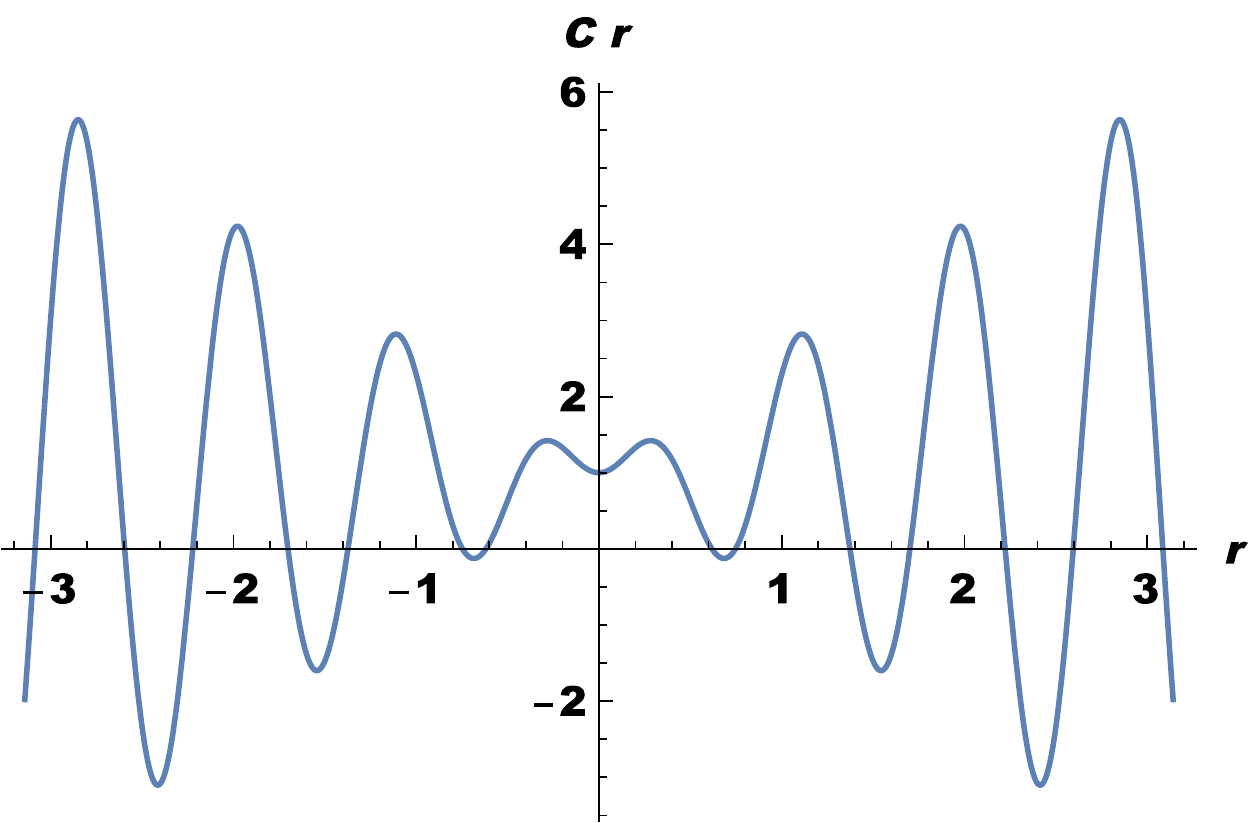}
  \caption{}
  \label{fig:corr2}
\end{subfigure}
\caption{(a) The normalized (Pearson's coefficient) two point correlation, of the noise in the Navier-Stokes equation (\ref{eq:isns}), for Taylor-Reynolds number 110, with values of a and b from the first line in Table 5, and C from the first column, first line of Table 4.\\
(b) The normalized (Pearson's coefficient) two point correlation, of the noise in the Navier-Stokes equation (\ref{eq:isns}), for Taylor-Reynolds number 1450, with values of a and b from the last line in Table 5, and C from the last column, first line of Table 4.}
 \end{figure}

\noindent{\bf The Noise in Homogeneous Turbulence:}\\
\begin{enumerate}
\item
The color of the noise, in the stochastic Navier-Stokes equation (\ref{eq:sns}) depends on the Reynolds number through the coefficients $a, b$ and the exponent $m$.
\item
For small Reynolds number the mean-field noise, or the central limit theorem and large deviation noise, is
exponentially correlated (decaying), with correlation:
\[
C_r= \frac{C}{2}e^{-2\pi b r}+ \frac{1}{2}e^{-2\pi a r}(r+\frac{1}{2\pi a}),
\] 
up to a multiplicative factor of $\frac{1}{C^2}$, where the values of $b(Re_\lambda)$ (central limit theorem), $a(Re_\lambda)$ (large deviation), and $m(Re_\lambda)$ (spatial smoothness) are taken from Tables 4, 5 and 6, for low values of $Re_\lambda$, and $r=|x-y|$ is the correlation distance between two points $x$ and $y$ in the fluid, see  Figure \ref{fig:corr1}. 
\item
For large Reynolds number the mean-field noise become oscillatory, approaching the correlation:
\[
C_r= \frac{C}{2}b\cos( 2\pi b^2 r) +2\pi a^2 r \sin( 2\pi a^2 r), 
\]
up to a multiplicative factor of $\frac{1}{C^2}$, again taking the values from Tables 4, 5 and 6, for high values of $Re_\lambda$, see Figure \ref{fig:corr2}. 
\end{enumerate}
These correlations are computed using the variance of the velocity and the second-order structure function above, using the well-known formula
$
S_2(r)=2(\sigma-C_r) 
$
and taking the limit of a very large spatial period. The exact formulas of the correlations depend on the ansatz that we made for the coefficients $c_k$ and $d_k$ in Section 4.1, but the above statements, about the nature of the correlations, are true in general. 

Once we put in the values of $b$ and $a$ from Table 5, we see that the slow decay of $e^{-br}$ (central limit theorem) dominates for  
small Reynolds numbers, but the rapid oscillations of $\sin(a^2 r)$ (large deviation) dominate for large Reynolds number. Recall, however, that these are the correlations of the noise in the stochastic Navier-Stokes equation, not the correlations of the turbulent velocity itself, see section $7$. 

\section{Sensitivity Analysis}

\noindent Now that we have compared the experimental data from the VDTT to formulas computed by the SCT, it is desirable to check how robust these results are. In particular, we want to know if the formulas with the Reynolds number corrections do better than the formulas without them? Also do the results depend on the probe size used in the experiments or are they independent of it? In this section we perform a sensitivity analysis to test the results and answer these questions.

\begin{figure}
\centering
\begin{subfigure}{.6\textwidth}
  \centering
  \includegraphics[scale=.6]{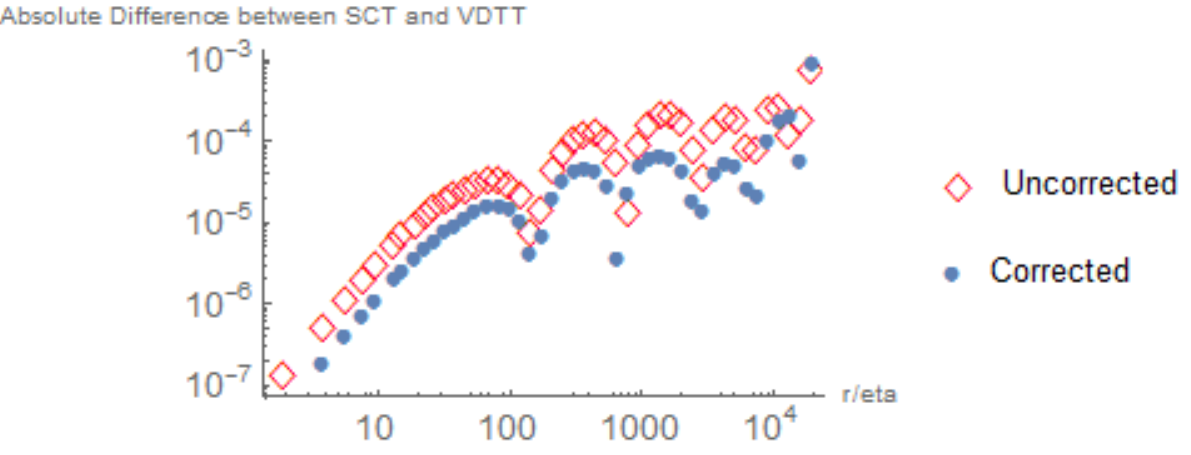}
  \caption{Second-Order Structure Function Error}
  \label{fig:110error2}
\end{subfigure}\\
\begin{subfigure}{.6\textwidth}
  \centering
  \includegraphics[scale=.6]{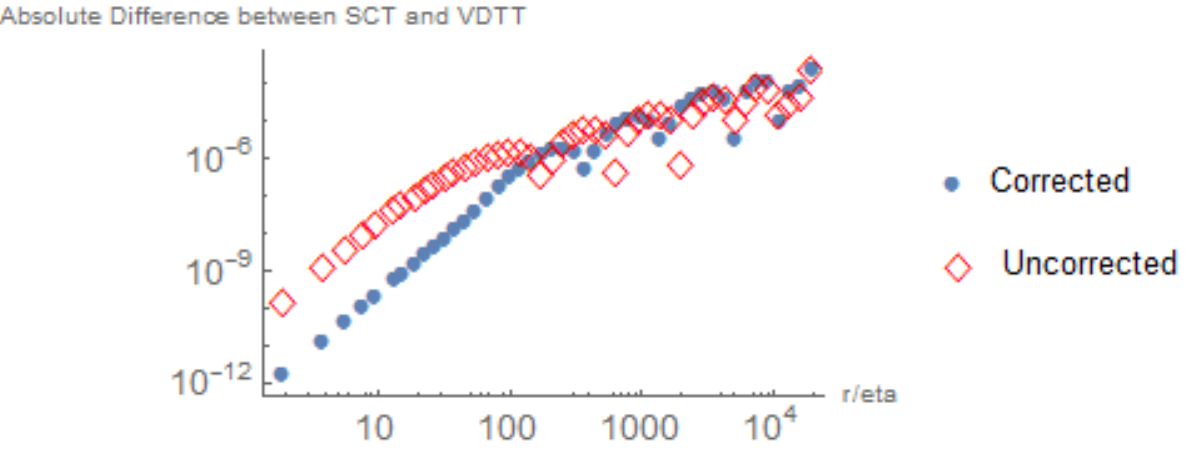}
  \caption{Third-Order Structure Function Error}
  \label{fig:110error3}
\end{subfigure}%
\begin{subfigure}{.6\textwidth}
 \centering
 \includegraphics[scale=.6]{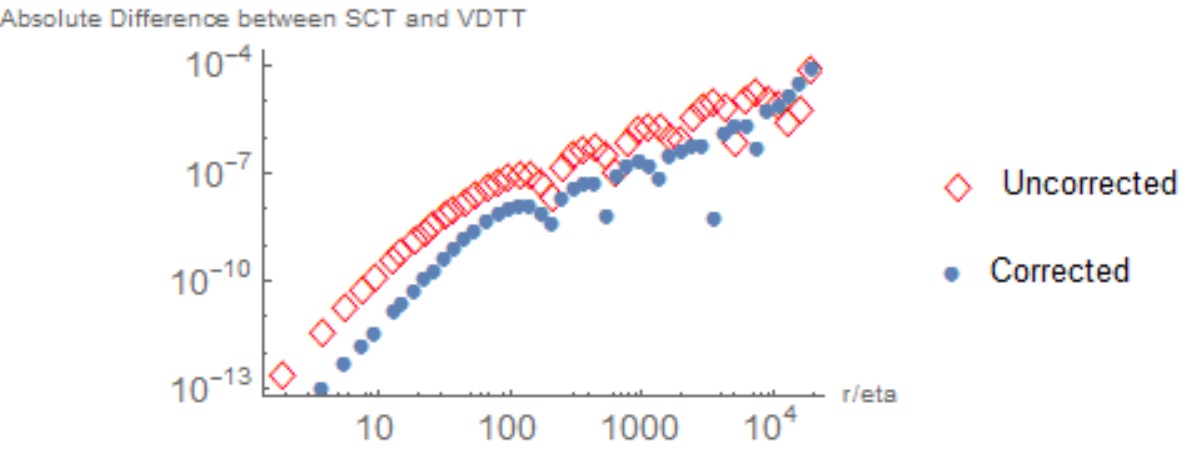}
 \caption{Fourth-Order Structure Function Error}
 \label{fig:110error4}
\end{subfigure}\\
\begin{subfigure}{.6\textwidth}
  \centering
  \includegraphics[scale=.6]{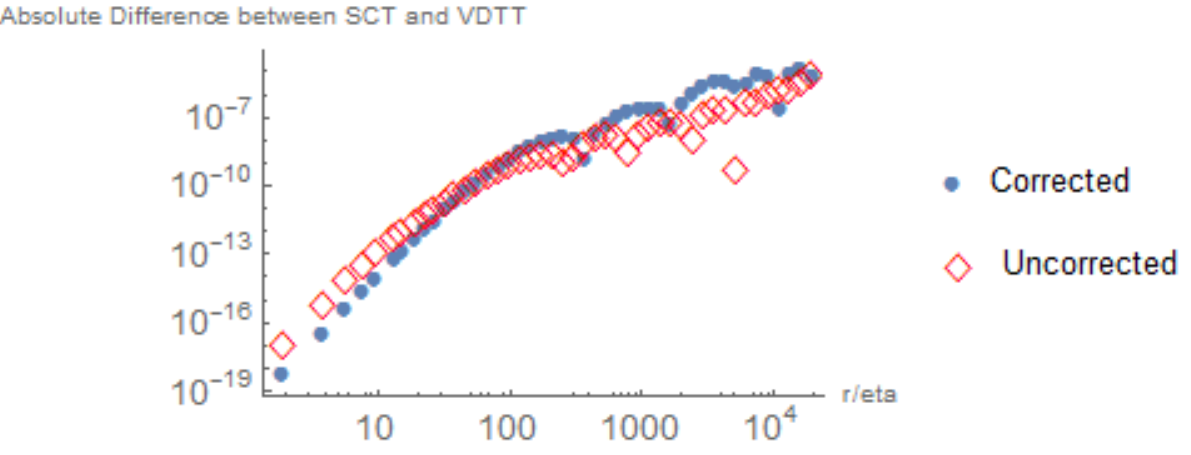}
  \caption{Sixth-Order Structure Function Error}
  \label{fig:110error6}
\end{subfigure}%
\begin{subfigure}{.6\textwidth}
  \centering
  \includegraphics[scale=.6]{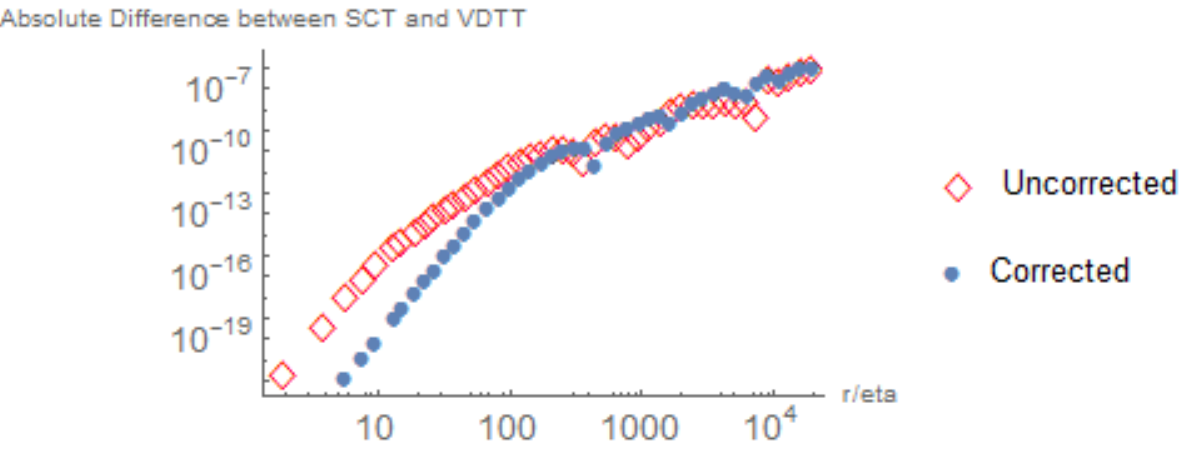}
  \caption{Eighth-Order Structure Function Error}
  \label{fig:110error8}
\end{subfigure}
\caption{Error for Taylor Reynolds Number 110.  Note that the plots are made on a log-log scale.  The blue dots are fits to the structure function formula featuring the Reynolds correction whereas the red diamonds are fits to the structure function formula without the Reynolds number correction}
\label{fig:110error}
\end{figure}

\begin{figure}
\centering
\begin{subfigure}{.6\textwidth}
  \centering
  \includegraphics[scale=.6]{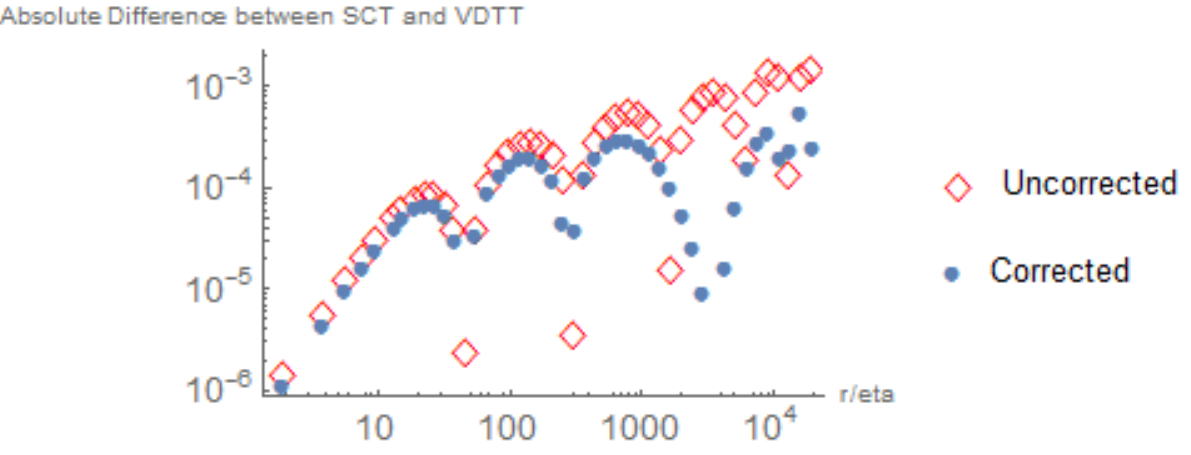}
  \caption{Second-Order Structure Function Error}
  \label{fig:264error2}
\end{subfigure}%
\begin{subfigure}{.6\textwidth}
  \centering
  \includegraphics[scale=.6]{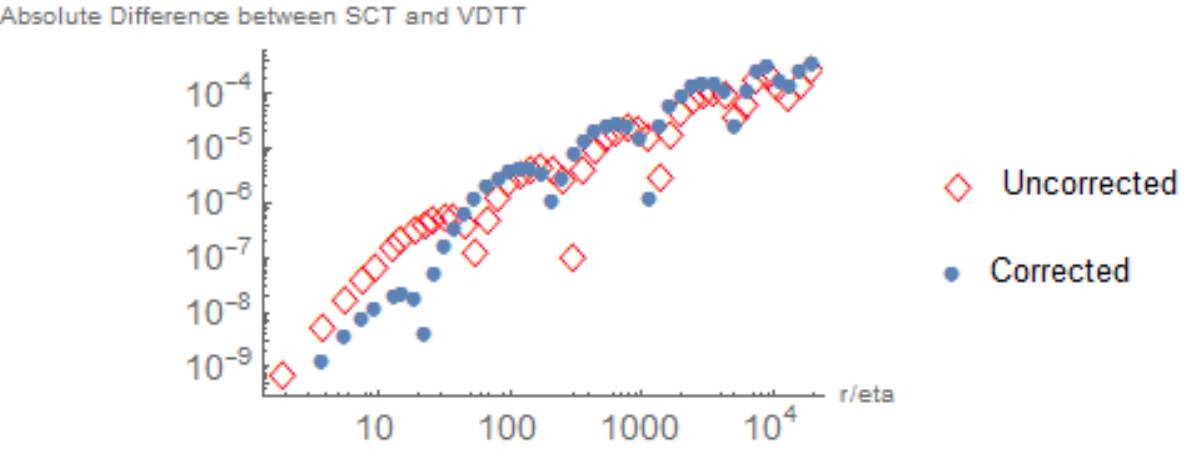}
  \caption{Third-Order Structure Function Error}
  \label{fig:264error3}
\end{subfigure}\\
\begin{subfigure}{.6\textwidth}
 \centering
 \includegraphics[scale=.6]{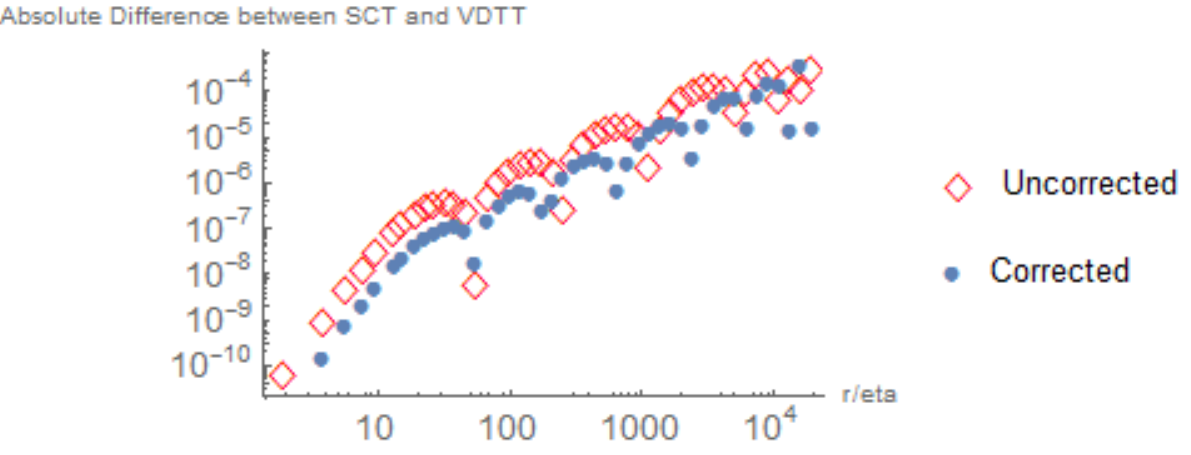}
 \caption{Fourth-Order Structure Function Error}
 \label{fig:264error4}
\end{subfigure}%
\begin{subfigure}{.6\textwidth}
  \centering
  \includegraphics[scale=.6]{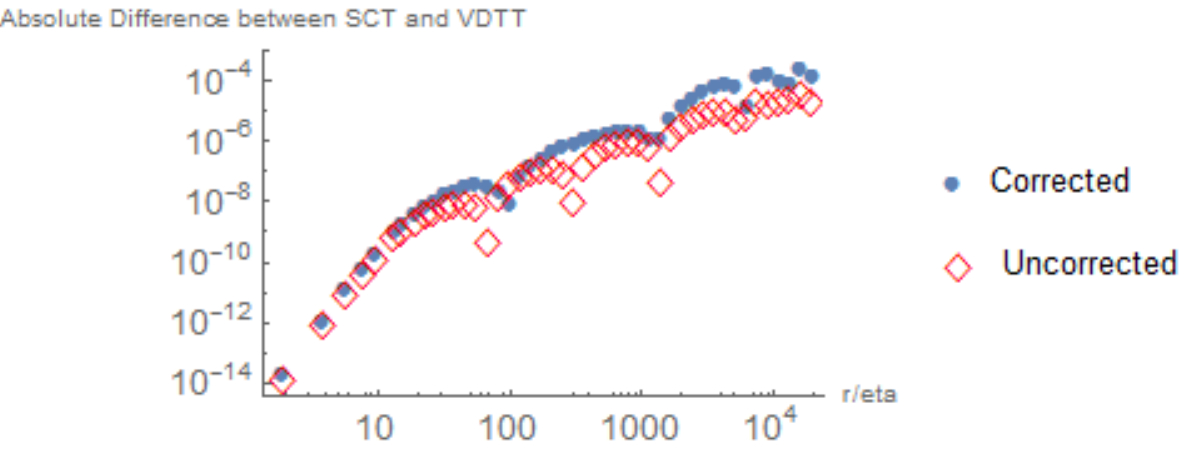}
  \caption{Sixth-Order Structure Function Error}
  \label{fig:264error6}
\end{subfigure}
\begin{subfigure}{.6\textwidth}
  \centering
  \includegraphics[scale=.6]{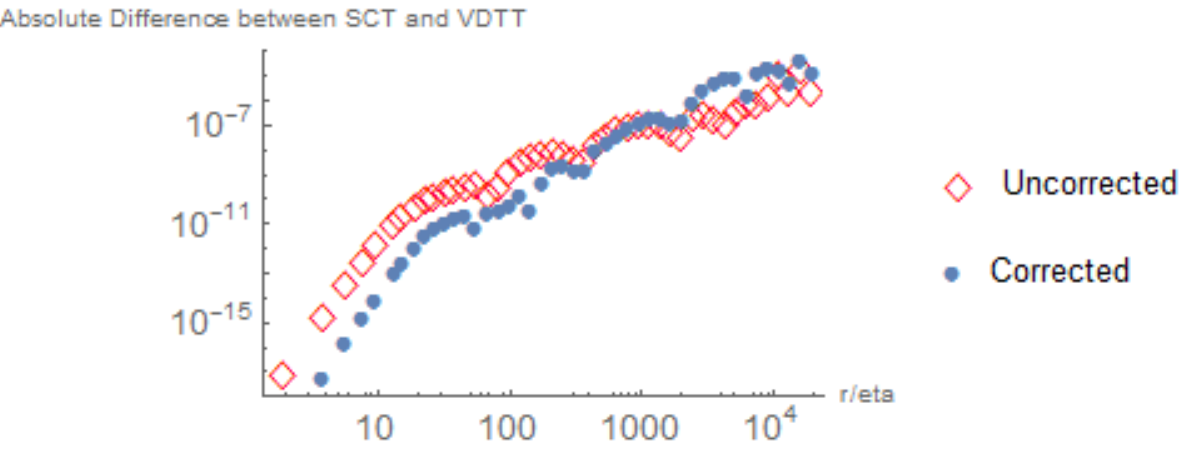}
  \caption{Eighth-Order Structure Function Error}
  \label{fig:264error8}
\end{subfigure}
\caption{Error for Taylor Reynolds Number 264.  Note that the plots are made on a log-log scale.  The blue dots are fits to the structure function formula featuring the Reynolds correction whereas the red diamonds are fits to the structure function formula without the Reynolds number correction}
\label{fig:264error}
\end{figure}

\begin{figure}
\centering
\begin{subfigure}{.6\textwidth}
  \centering
  \includegraphics[scale=.6]{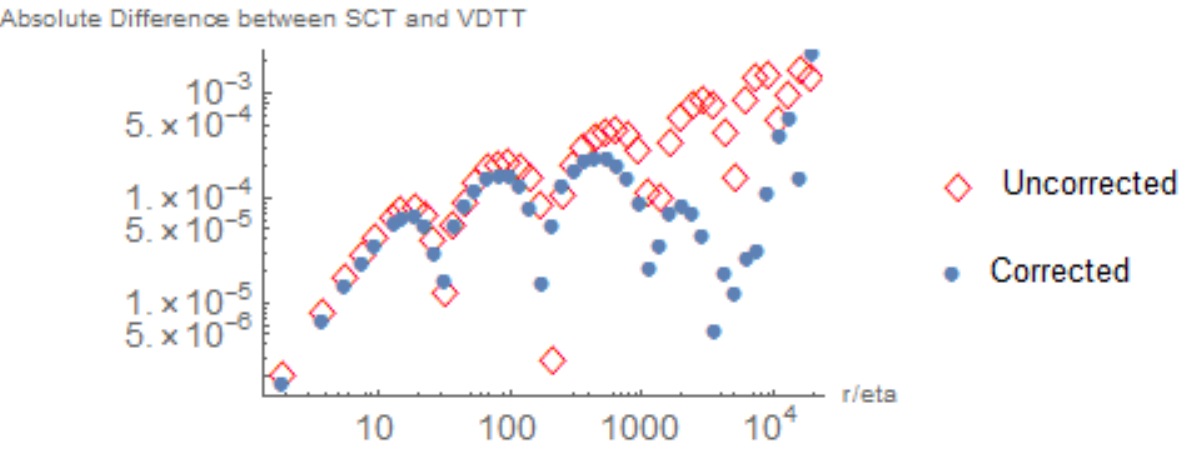}
  \caption{Second-Order Structure Function Error}
  \label{fig:508error2}
\end{subfigure}%
\begin{subfigure}{.6\textwidth}
  \centering
  \includegraphics[scale=.6]{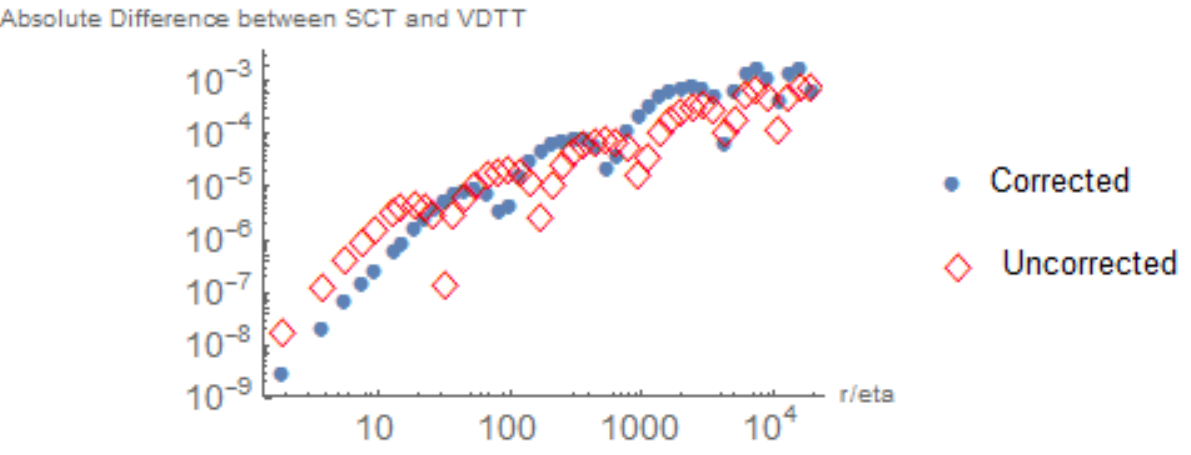}
  \caption{Third-Order Structure Function Error}
  \label{fig:508error3}
\end{subfigure}\\
\begin{subfigure}{.6\textwidth}
 \centering
 \includegraphics[scale=.6]{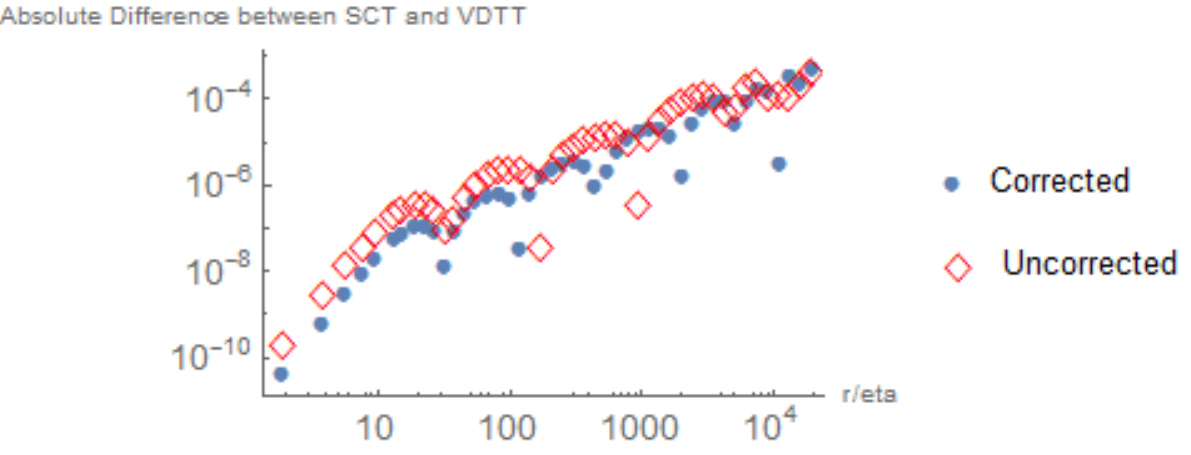}
 \caption{Fourth-Order Structure Function Error}
 \label{fig:508error4}
\end{subfigure}%
\begin{subfigure}{.6\textwidth}
  \centering
  \includegraphics[scale=.6]{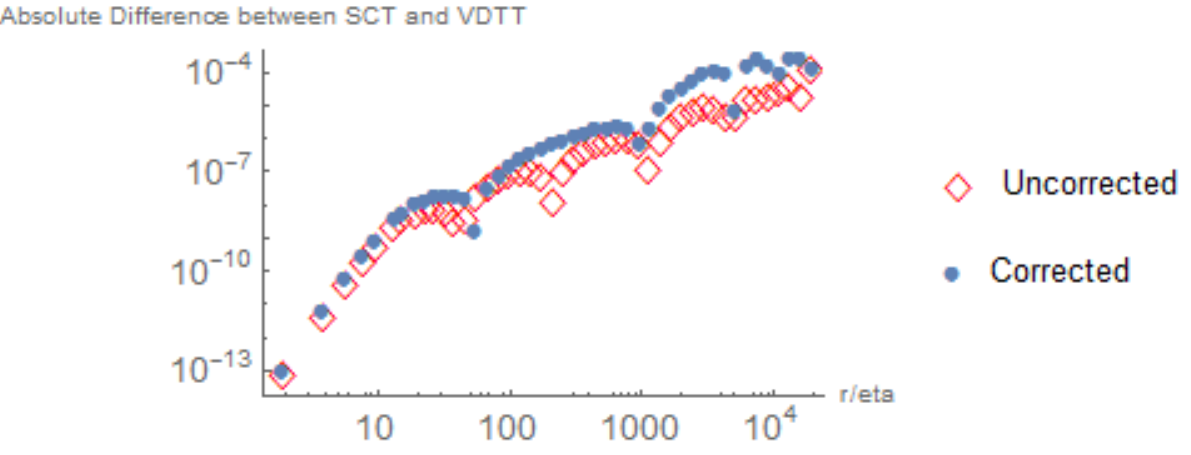}
  \caption{Sixth-Order Structure Function Error}
  \label{fig:508error6}
\end{subfigure}
\begin{subfigure}{.6\textwidth}
  \centering
  \includegraphics[scale=.6]{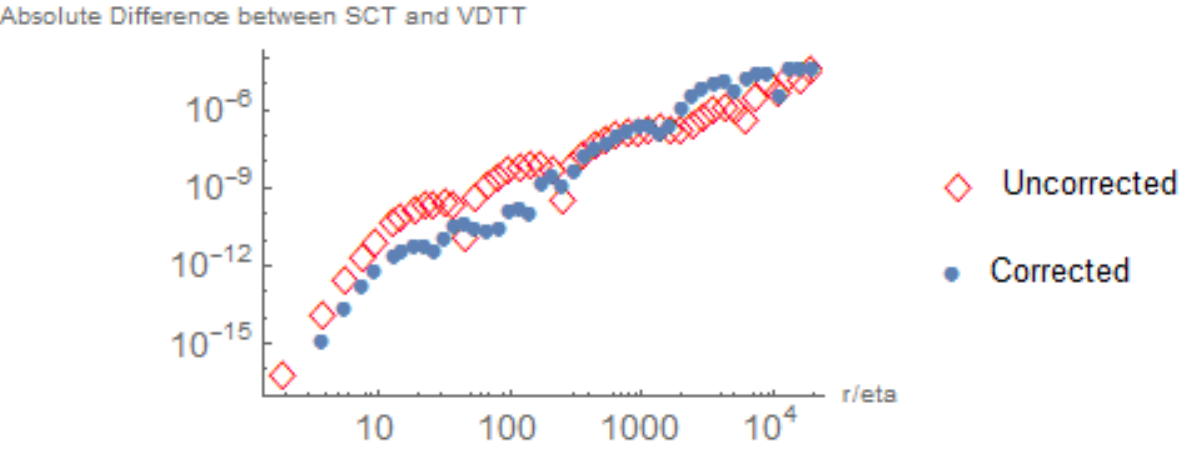}
  \caption{Eighth-Order Structure Function Error}
  \label{fig:508error8}
\end{subfigure}
\caption{Error for Taylor Reynolds Number 508.  Note that the plots are made on a log-log scale.  The blue dots are fits to the structure function formula featuring the Reynolds correction whereas the red diamonds are fits to the structure function formula without the Reynolds number correction.}
\label{fig:508error}
\end{figure}

\begin{figure}
\centering
\begin{subfigure}{.6\textwidth}
  \centering
  \includegraphics[scale=.6]{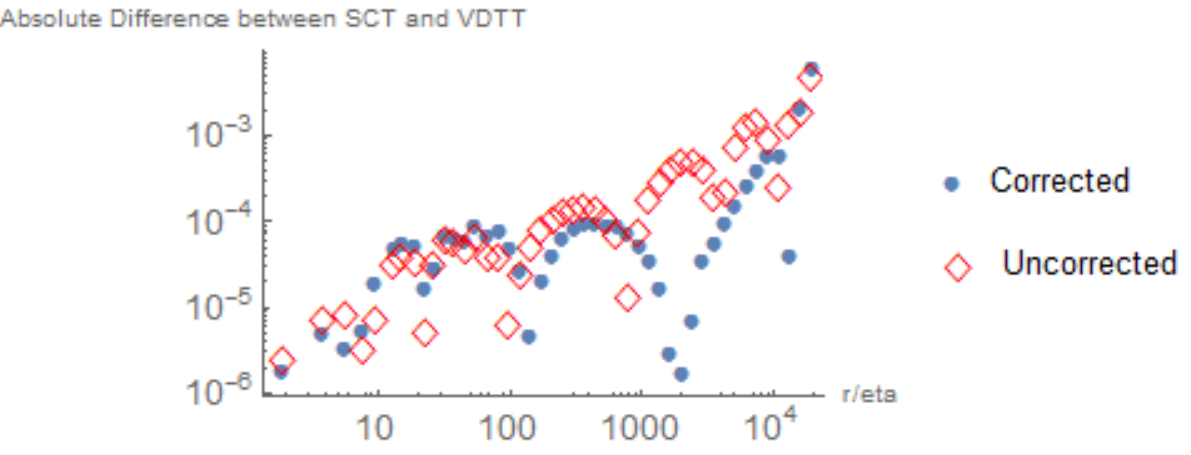}
  \caption{Second-Order Structure Function Error}
  \label{fig:1000error2}
\end{subfigure}%
\begin{subfigure}{.6\textwidth}
  \centering
  \includegraphics[scale=.6]{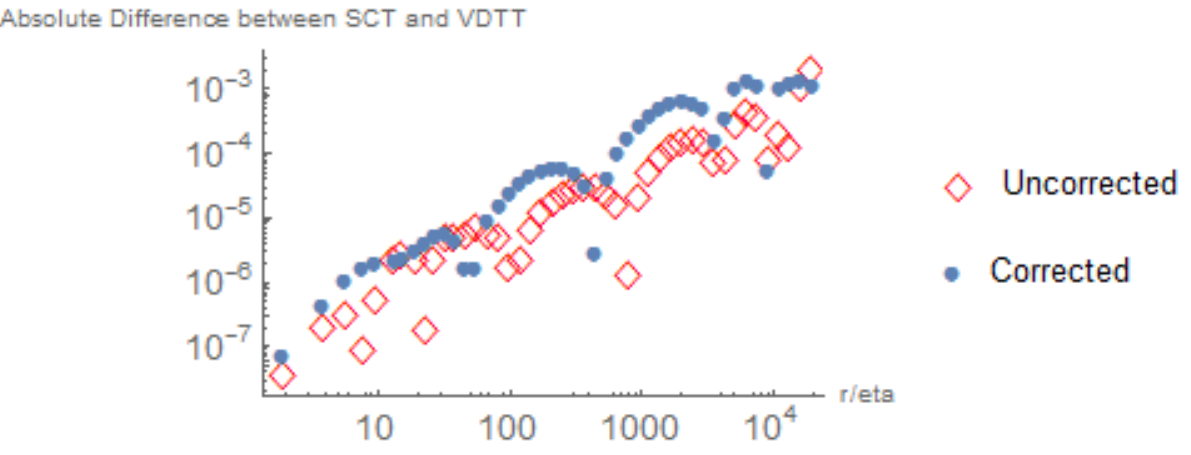}
  \caption{Third-Order Structure Function Error}
  \label{fig:1000error3}
\end{subfigure}\\
\begin{subfigure}{.6\textwidth}
 \centering
 \includegraphics[scale=.6]{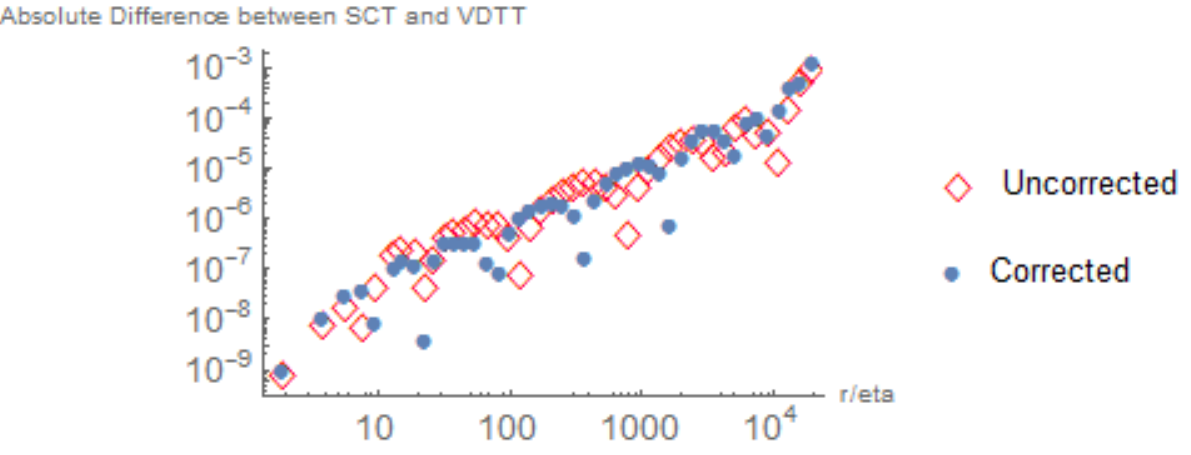}
 \caption{Fourth-Order Structure Function Error}
 \label{fig:1000error4}
\end{subfigure}%
\begin{subfigure}{.6\textwidth}
  \centering
  \includegraphics[scale=.6]{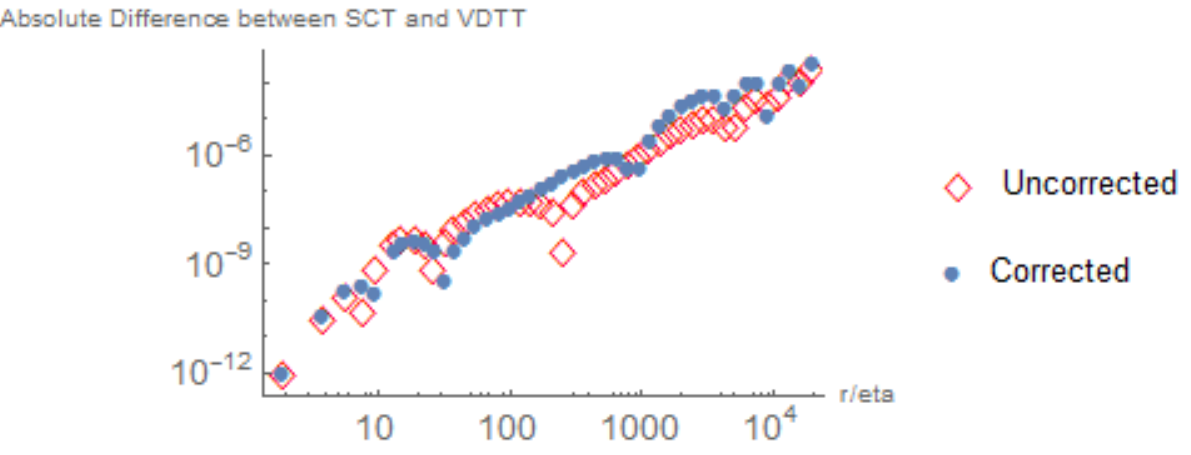}
  \caption{Sixth-Order Structure Function Error}
  \label{fig:1000error6}
\end{subfigure}
\begin{subfigure}{.6\textwidth}
  \centering
  \includegraphics[scale=.6]{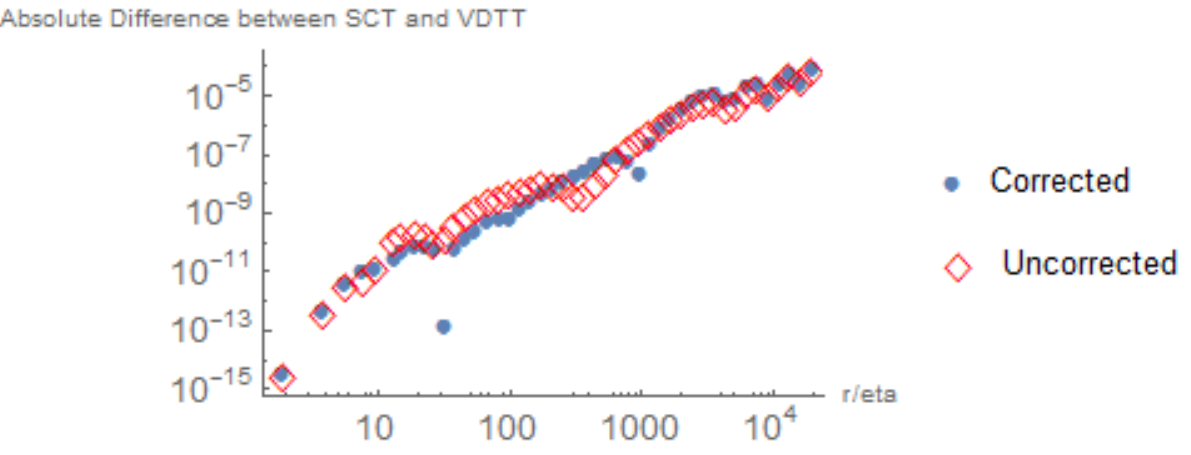}
  \caption{Eighth-Order Structure Function Error}
  \label{fig:1000error8}
\end{subfigure}
\caption{Error for Taylor Reynolds Number 1000.  Note that the plots are made on a log-log scale.  The blue dots are fits to the structure function formula featuring the Reynolds correction whereas the red diamonds are fits to the structure function formula without the Reynolds number correction.}
\label{fig:1000error}
\end{figure}

\begin{figure}
\centering
\begin{subfigure}{.6\textwidth}
  \centering
  \includegraphics[scale=.6]{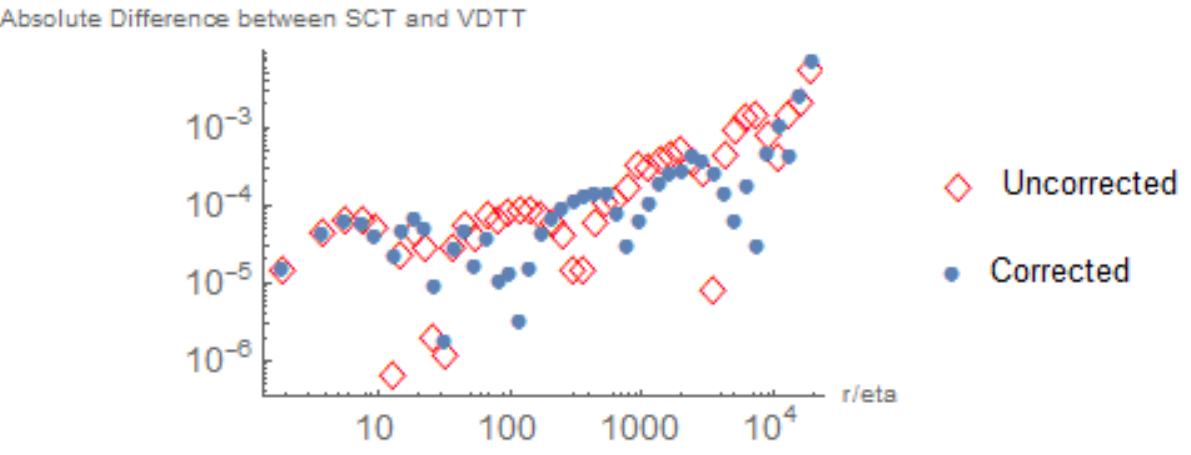}
  \caption{Second-Order Structure Function Error}
  \label{fig:1450error2}
\end{subfigure}%
\begin{subfigure}{.6\textwidth}
  \centering
  \includegraphics[scale=.6]{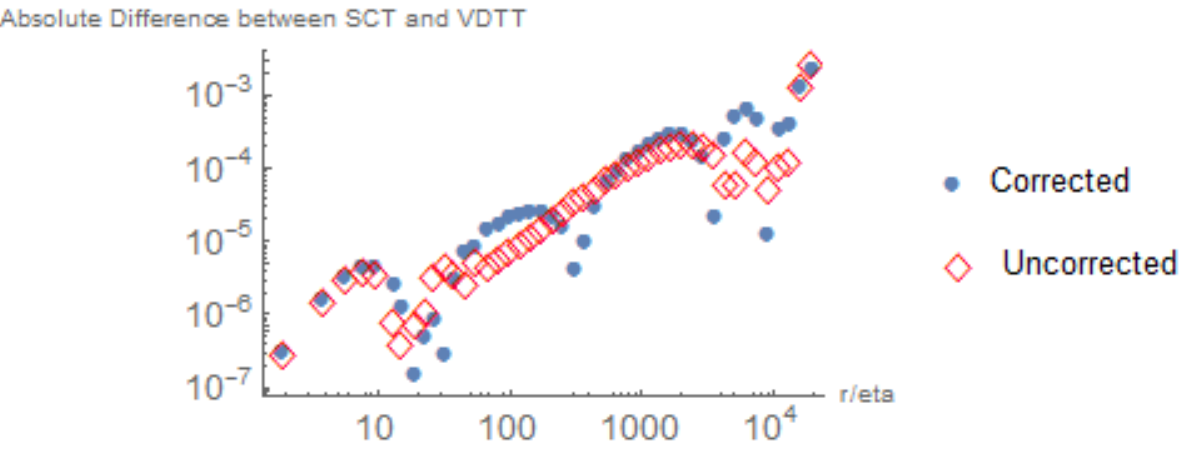}
  \caption{Third-Order Structure Function Error}
  \label{fig:1450error3}
\end{subfigure}\\
\begin{subfigure}{.6\textwidth}
 \centering
 \includegraphics[scale=.6]{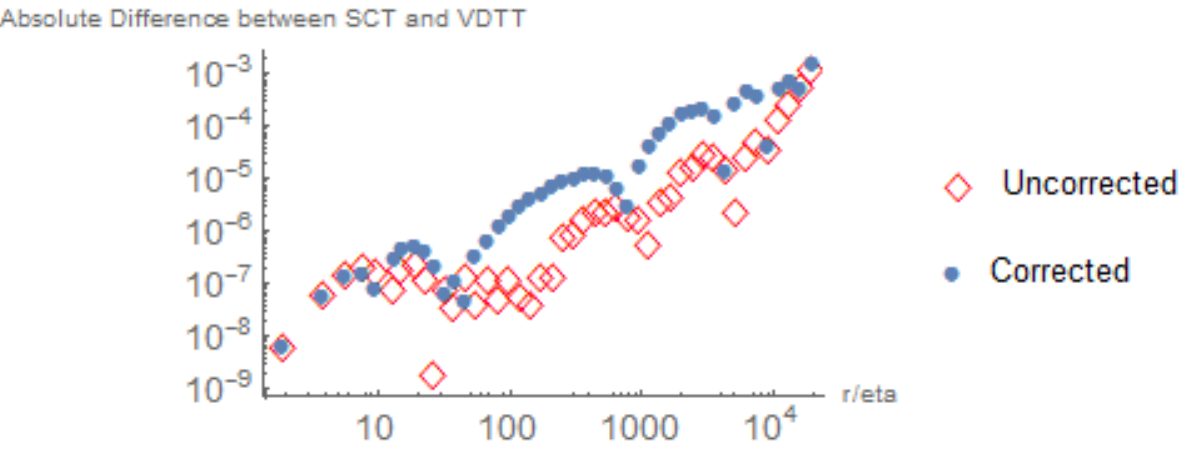}
 \caption{Fourth-Order Structure Function Error}
 \label{fig:1450error4}
\end{subfigure}%
\begin{subfigure}{.6\textwidth}
  \centering
  \includegraphics[scale=.6]{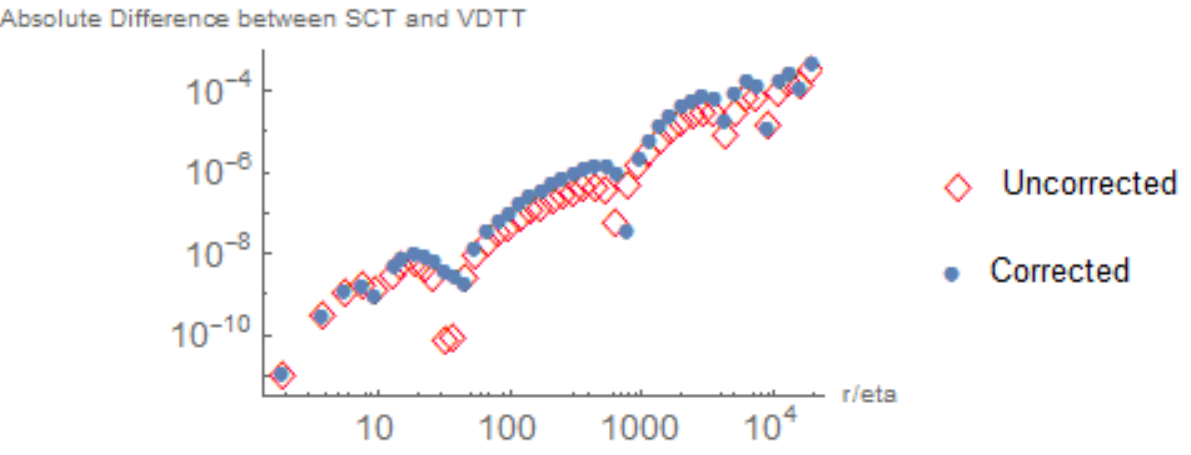}
  \caption{Sixth-Order Structure Function Error}
  \label{fig:1450error6}
\end{subfigure}
\begin{subfigure}{.6\textwidth}
  \centering
  \includegraphics[scale=.6]{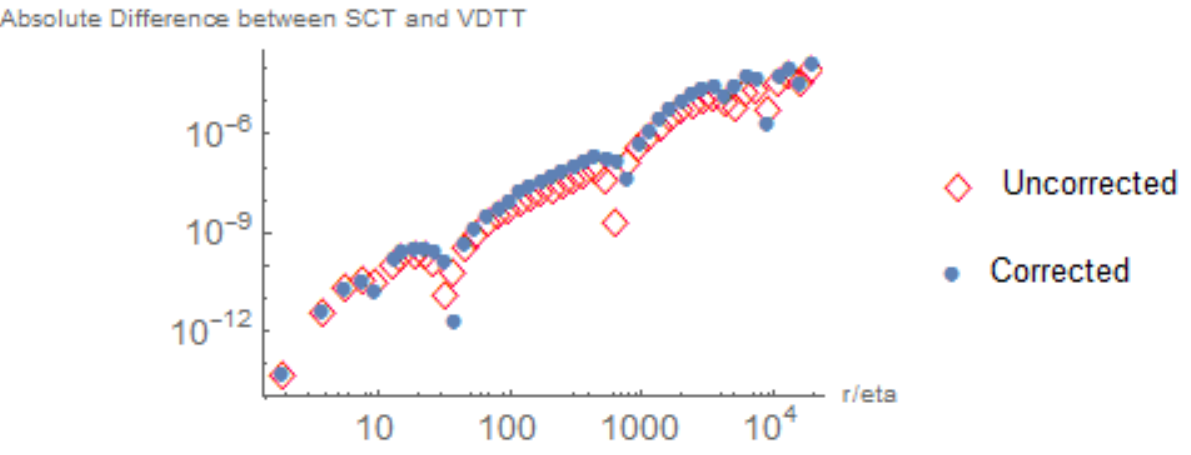}
  \caption{Eighth-Order Structure Function Error}
  \label{fig:1450error8}
\end{subfigure}
\caption{Error for Taylor Reynolds Number 1450.  Note that the plots are made on a log-log scale.  The blue dots are fits to the structure function formula featuring the Reynolds correction whereas the red diamonds are fits to the structure function formula without the Reynolds number correction.}
\label{fig:1450error}
\end{figure}

First we consider the formula for the general $p$-th structure function as given in (\ref{eq:series1}).  One way to let the Reynolds Number go to infinity is to let the viscosity of the fluid go to zero.  
Doing so simplifies the coefficients $A_p$ in $(\ref{eq:Ap})$, so that for $R_\lambda = \infty$, $\nu = 0$, 
\begin{align}
 A_p=\frac{2^\frac{p}{2}\Gamma(\frac{p+1}{2})\sigma_k^p{}_1F_1(-\frac{1}{2}p,\frac{1}{2},-\frac{1}{2}(\frac{M_k}{\sigma_k})^2)}{\vert k\vert ^{\zeta_p}}.
\end{align}
The further denominator terms found in $(26)$ but not above are corrections to the formula to account for the Reynolds number of the flow.  Data fits were also done to the formula without the Reynolds number corrections.  Figures \ref{fig:110error}, \ref{fig:264error}, \ref{fig:508error}, \ref{fig:1000error}, and \ref{fig:1450error} are plots of the error between the formula fits and the data at each data point.  The blue circles are the error to the Reynolds corrected formulas while the red diamonds are the error to the formula without the Reynolds number correction.\\
\indent There are a couple of observations to make about the error plot.  First, for small Taylor Reynolds numbers, it appears that the corrections improve the fitting, especially for the smaller data points.  This improvement erodes as the Taylor Reynolds number increases, until we see very little difference in accuracy for Taylor Reynolds number $1450$.  This makes sense, as the corrections to account for Reynolds number get smaller as the Reynolds number increases, with the formulas becoming the uncorrected version when we let the Reynolds number go to infinity.\\

\begin{figure}
\centering
\begin{subfigure}{.6\textwidth}
  \centering
  \includegraphics[scale=.6]{14506-eps-converted-to}
  \caption{Sixth-Order Structure Function}
  \label{fig:145062}
\end{subfigure}%
\begin{subfigure}{.6\textwidth}
  \centering
  \includegraphics[scale=.6]{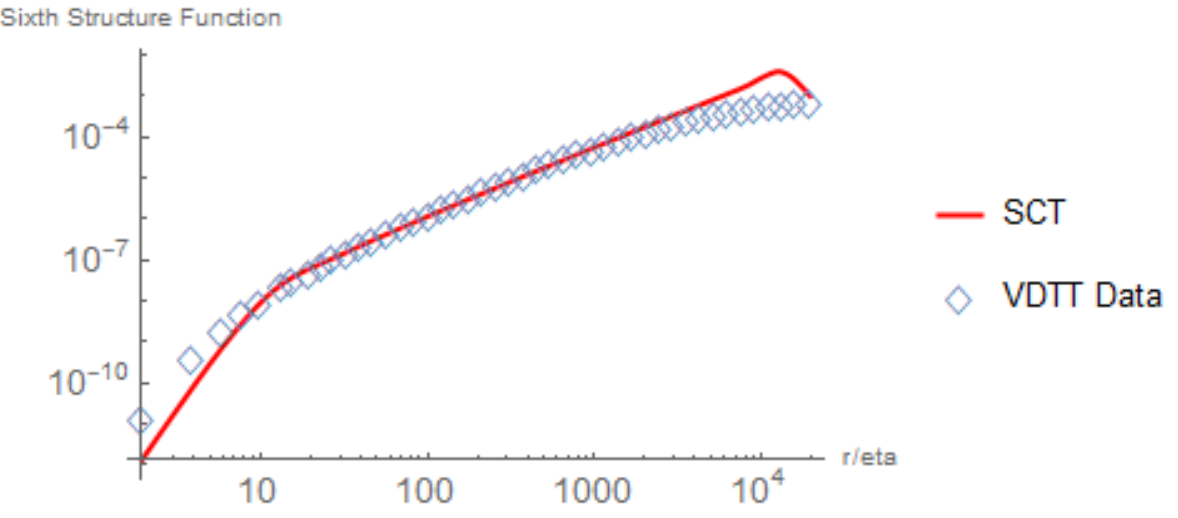}
  \caption{Sixth-Order Structure Function with $A_1$ and $A_2$ given by formula $(\ref{eq:series})$}
  \label{fig:1450ff}
\end{subfigure}\\
\caption{The figure on the left is the original fit to the sixth-order structure function for Reynolds number $1450$.  The right is if we tie $A_1$ and $A_2$ back to the original formula $(\ref{eq:series})$.  Note the effect on the largest values of $r\slash\eta$ and how this creates the wiggles we are seeing.}
\label{fig:1450A}
\end{figure}

\begin{figure}
\centering
\begin{subfigure}{.6\textwidth}
  \centering
  \includegraphics[scale=.6]{14503-eps-converted-to}
  \caption{Third-Order Structure Function}
  \label{fig:145032}
\end{subfigure}%
\begin{subfigure}{.6\textwidth}
  \centering
  \includegraphics[scale=.6]{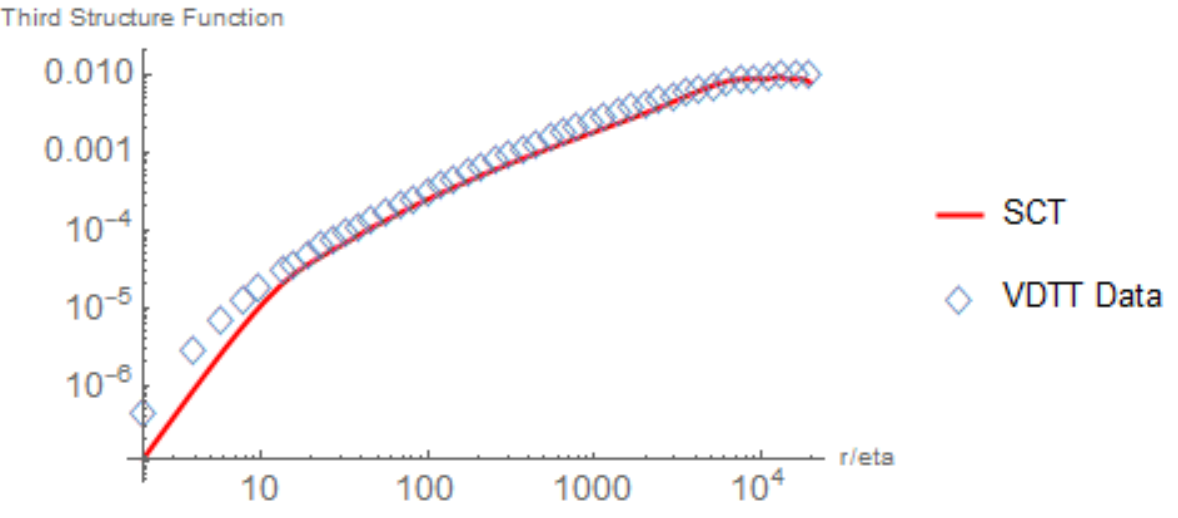}
  \caption{Third-Order Structure Function Error with slight change in parameter $C$.}
  \label{fig:1450C}
\end{subfigure}\\
\begin{subfigure}{.6\textwidth}
\centering
\includegraphics[scale=.6]{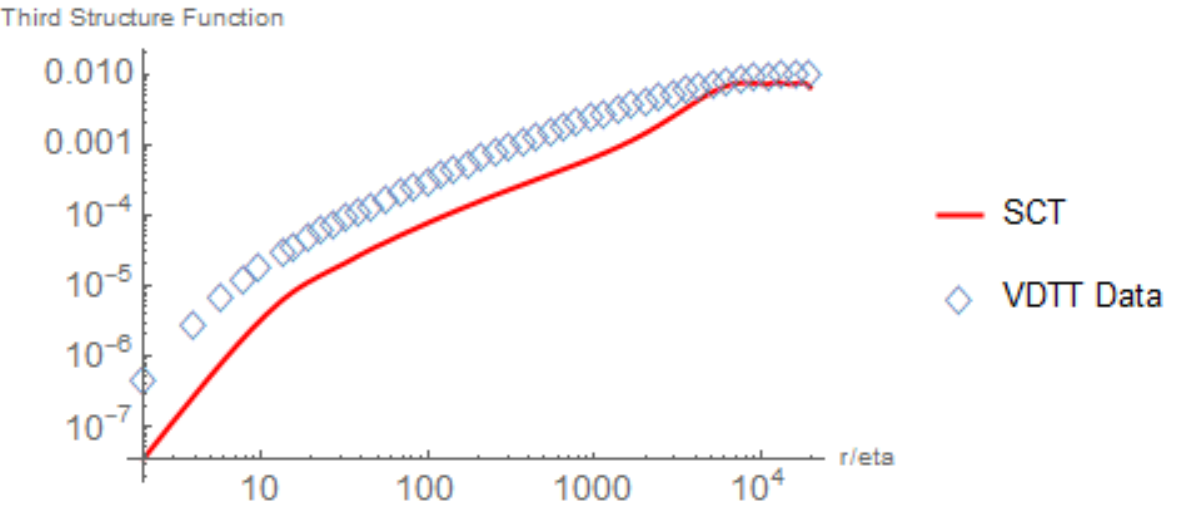}
\caption{Third-Order Structure Function with a greater change in parameter $C$.}
\label{fig:1450C2}
\end{subfigure}
\caption{The figure on the top left is the original fit to the third-order structure function for Reynolds number $1450$.  The top right changes the value of the parameter $C$ from $3.59$ to $4$.  Note the effect here, as the original fit for the dissipative and inertial range are pushed down slightly, and a new transitionary regime is created.  This effect is more pronounced in the bottom figure, when $C$ is increased from $4$ to $6$.}
\label{fig:1450CC}
\end{figure}

\begin{table}
\begin{center}
    \begin{tabular}{| l | l | l | l | l |p{1.5cm}|}
    \hline
    Taylor Reynolds Number&  110 & 264 &508& 1000& 1450 \\ \hline
Second&2.09081&1.49402&1.31448&1.07963&0.984291\\ \hline
Third&1.79012&1.41339&1.05553&0.822192&0.730565\\ \hline
Fourth&1.6408&1.09179&0.920749&0.687336&0.595942\\ \hline
Sixth&1.65727&1.08667&0.91658&0.681818&0.592901\\ \hline
Eighth&1.66164&1.06728&0.901549&0.662111&0.577724\\
    \hline
 \end{tabular}
\caption{The fitted values for $m$ for the uncorrected for Reynolds number effects structure function fits}   
\end{center}
\end{table}
\indent We also see an issue in fitting the smallest data points for solely for Reynolds Number $1450$.  This issue appears to be connected to the system length, as seen in Figure \ref{fig:1450Length}.  A second fit to the fourth structure function for this Reynolds number was found with $D=.921$.  This does improve the fitting for the smaller data points.  However, $D$ being this small causes an issue at the larger data points, namely the sine curve wants to return to zero before the last data point.  Since there are relatively few data points at small values of $r\slash\eta$, we set $D=1.3$.  The value of $1.3$ was chosen as it the smallest number needed to fully capture the larger data points.  Figure \ref{fig:1450Length} also illustrates the effect $D$ has on the fits, serving to place the transition from the dissipative range into the inertial range.\\
\indent One potential point of concern with the fitting result was the probe size.  The size of the probe could influence the fit and a different probe size could produce different result.  To check for this, fits were redone with a reduced number of data points.  In particular, for every Taylor Reynolds number and every structure function, fits were redone without including the first, the first two, and the first three data points respectively.  We saw minimal change in the main parameters. the greatest being a difference of one in the third significant digit.  The robust test for Reynolds Number $508$ are included in figure \ref{fig:robust}.  As we can see, there is not a significant change in the value of $m$ when removing the first couple of data points.  However, the removal of fifteen or more data points removes the entire dissipative range and so we would expect the changes to be significant.  As a result, we are convinced the fits are unaffected by the probe size.\\
\indent Finally, we show the effect parameters $A_1$, $A_2$, and $C$ have on the fits.  Figure \ref{fig:1450A} shows the effects of $A_1$ and $A_2$.  Note that these effects show up for all Reynolds number and all order of structure functions, although they become negligible as the order of the structure function increases.  These two parameters also created the wiggles we see at the largest values of $r\slash\eta$.  Figure \ref{fig:1450CC} shows the effect of parameter $C$.  This parameter places the vertical location of the transition out of the inertial range.

\section{Conclusion}
\indent We started by following Kolmogorov's method as described in Section $2$ to close the Navier Stokes equations 
that describe fully developed turbulence.  
We did this by introducing a stochastic forcing term to account for the small scales, see \cite{BB211}.  
Having closed the model, we then compute a sine series representation for the structure functions of turbulence, with Reynolds number corrections.  
These formulas were then fitted to data generated from the Variable Density Turbulence Tunnel at the Max Planck Institute for Dynamics and Self-Organization.  The fits proved to be good with seven parameters.  However, only three of these parameters $a,b,m$ were active over the entire range of T-R numbers in the experiment, although one more parameter $C$ measures the mean fluctuation velocity and increases over the range of T-R numbers in the experiment.  Of the other four $D, A_1, A_2$ and $C$, one $D$ is active only for the transition from the dissipative range into the inertial range, whereas the other three are active for the transition out of the inertial range.\\
\indent We also compared fits to the formula with a correction to account for the Reynolds number to fits without that correction.  We see that the Reynolds correction formulas generate better fits as the Reynolds number increases for lower structure functions but have little impact on the fits for the higher structure functions.\\

\section{Acknowledgements}
The experimental data presented in this paper was taken during the doctoral studies of Michael Sinhuber and time of John Kaminsky at the Max Planck Institute for Dynamics and Self-Organization.  We are grateful to Eberhard Bodenschatz for fruitful discussions and the possibility to utilize the data.

\bibliography{referencef}
\bibliographystyle{plain}

\end{document}